\documentclass[sn-apa,iicol]{sn-jnl}

\usepackage{times}
\usepackage{epsfig}
\usepackage{graphicx}
\usepackage{comment}
\usepackage{amsmath,amssymb} 
\usepackage{color}
\usepackage{subfigure}
\usepackage{booktabs}
\usepackage{verbatim}
\usepackage{algorithmicx,algorithm}
\usepackage{url}

\newcommand{\clP}{\mathcal{P}}
\newcommand{\clX}{\mathcal{X}}
\newcommand{\clY}{\mathcal{Y}}
\newcommand{\clZ}{\mathcal{Z}}
\newcommand{\bbE}{\mathbb{E}}
\newcommand{\bbR}{\mathbb{R}}
\newcommand{\Lrecon}{L_\mathrm{recon}}
\newcommand{\Lguide}{L_\mathrm{guide}}
\newcommand{\Ldistr}{L_\mathrm{distr}}
\newcommand{\Lpdistr}{L'_\mathrm{distr}}
\newcommand{\Lpercp}{L_\mathrm{percp}}
\newcommand{\yguide}{y_\mathrm{guide}}
\newcommand{\KL}{\mathrm{KL}}

\jyear{2021}%

\theoremstyle{thmstyleone}%
\theoremstyle{thmstyletwo}%

\theoremstyle{thmstylethree}%

\begin{document}

\title[IRN and Extensions]{Invertible Rescaling Network and Its Extensions}


\author[1]{Mingqing Xiao}\email{mingqing\_xiao@pku.edu.cn.com}

\author[2]{Shuxin Zheng}\email{Shuxin.Zheng@microsoft.com}

\author*[2]{Chang Liu}\email{Chang.Liu@microsoft.com}

\author*[1,3]{Zhouchen Lin}\email{zlin@pku.edu.cn}

\author[2]{Tie-Yan Liu}\email{Tie-Yan.Liu@microsoft.com}

\affil[1]{\orgdiv{Key Laboratory of Machine Perception (MoE), School of Intelligence Science and Technology}, \orgname{Peking University}, \orgaddress{\city{Beijing}, \postcode{100871}, \country{P.R. China}}}

\affil[2]{\orgdiv{Machine Learning Group}, \orgname{Microsoft Research Asia}, \orgaddress{\city{Beijing}, \country{P.R. China}}}

\affil[3]{\orgname{Peng Cheng Laboratory}, \orgaddress{\country{P.R. China}}}


\abstract{
Image rescaling is a commonly used bidirectional operation, which first downscales high-resolution images to fit various display screens or to be storage- and bandwidth-friendly, and afterward upscales the corresponding low-resolution images to recover the original resolution or the details in the zoom-in images. However, the non-injective downscaling mapping discards high-frequency contents, leading to the ill-posed problem for the inverse restoration task. This can be abstracted as a general image degradation-restoration problem with information loss. In this work, we propose a novel invertible framework to handle this general problem, which models the bidirectional degradation and restoration from a new perspective, i.e. invertible bijective transformation. The invertibility enables the framework to model the information loss of pre-degradation in the form of distribution, which could mitigate the ill-posed problem during post-restoration. To be specific, we develop invertible models to generate valid degraded images and meanwhile transform the distribution of lost contents to the fixed distribution of a latent variable during the forward degradation. Then restoration is made tractable by applying the inverse transformation on the generated degraded image together with a randomly-drawn latent variable. We start from image rescaling and instantiate the model as Invertible Rescaling Network (IRN), which can be easily extended to the similar decolorization-colorization task. We further propose to combine the invertible framework with existing degradation methods such as image compression for wider applications. 
Experimental results demonstrate the significant improvement of our model over existing methods in terms of both quantitative and qualitative evaluations of upscaling and colorizing reconstruction from downscaled and decolorized images, and rate-distortion of image compression. Code is available at \url{https://github.com/pkuxmq/Invertible-Image-Rescaling}.
}




\maketitle

\section{Introduction}\label{sec1}

Image rescaling is becoming increasingly important in the age of high-resolution (HR) images/videos explosion on the Internet. For efficient storage, transmission, and sharing, such large-sized data are usually downscaled to significantly reduce the size and become bandwidth-friendly~\citep{bruckstein2003down,lin2006adaptive,wu2009low,shen2011down,li2018learning}, while visually valid contents are maintained~\citep{kim2018task,sun2020learned} for previewing or fitting for screens with different resolutions. On the other hand, the inverse restoration task is required by user demands, which aims to upscale the downscaled low-resolution (LR) images to a higher resolution or the original size~\citep{yeo2017will,yeo2018neural,schulter2015fast,giachetti2011real} so that vivid details could be presented. 
However, the non-injective downscaling would cause information loss, as high-frequency contents are lost during downscaling according to the Nyquist-Shannon sampling theorem~\citep{shannon1949communication}. 
Such information loss leads to an intractable ill-posed problem of the inverse tasks~\citep{irani2009super,yang2010image,dong2015image}, since the same downscaled LR image may correspond to multiple possible HR images, and therefore poses great challenges for recovery.

This can be abstracted as a general image degradation-restoration problem with information loss due to dimension reduction. Similar examples also include image decolorization-colorization~\citep{xia2018invertible,ye2020invertible} and image compression. In the following, we first focus on this general problem and then consider the specific instantiation examples. 

There have been many efforts attempting to mitigate this ill-posed problem with machine learning algorithms. For instance, many works consider dealing with the unidirectional restoration task, e.g. for image rescaling, they choose super-resolution (SR) methods to upscale LR images by imposing or learning a prior, i.e. a preference on all possible HR images corresponding to a given LR image, for this inverse task. 
However, mainstream SR algorithms~\citep{dong2015image,lim2017enhanced,zhang2018residual,zhang2018image,dai2019second,wang2018esrgan} leverage a predefined and non-adjustable downscaling method, such as Bicubic interpolation, to guide the learning of upscaling, which omits the compatibility between these two mutually-inverse operations. 
Therefore, simply applying unidirectional restoration methods, e.g. SR, cannot fully leverage the bidirectional nature of the task, resulting in unsatisfactory recoveries.

Some recent works attempt to unify these bidirectional operations through an encoder-decoder framework rather than separating them as two independent tasks. In these methods for image rescaling, an encoder, which serves as a learning-based upscaling-optimal downscaling module, is jointly trained with an upscaling decoder~\citep{kim2018task} or an existing SR module~\citep{li2018learning,sun2020learned}. This encoder-decoder framework is also applied in similar degradation-restoration tasks~\citep{xia2018invertible,ye2020invertible}. 
Taking the bidirectional nature into consideration, such an integrated training method can largely improve the quality of image reconstruction. However, these efforts simply link the two operations through training objectives without any attempt to fully leverage the reciprocal nature of the tasks or capture features of lost contents. So the results cannot meet the expectation as well.

In this paper, we propose a novel invertible framework to largely mitigate this ill-posed problem through invertible bijective transformation. 
With inspiration from the reciprocal nature of this pair of tasks, keeping the knowledge of lost information in the forward procedure, e.g. high-frequency contents in the image rescaling task, would greatly help the inverse recovery. However, it is not acceptable to store or transfer all lost contents to enable an exact recovery. To well address this challenge, we instead deal with these contents in the form of distribution, with the assumption that reasonable lost contents follows a distribution. We develop a novel invertible model to capture the knowledge of distribution in the form of distribution transformation function. 
Specifically, in the forward procedure, our invertible models will transform the original image $x$ into a degraded image $y$ and an auxiliary latent variable $z$ by an invertible transformation. $y$  belongs to a target set of valid degraded images, e.g. the set of visually-pleasing LR images given the HR image $x$ for the image rescaling tasks, and $z$ is a random variable following a fixed pre-specified distribution $p(z)$ (e.g. isotropic Gaussian). The joint distribution of $y$ and $z$ is bijectively transformed from the distribution of $x$ and therefore the random variable $z$ holds the lost ``information'' of $y$ from the perspective of statistical modeling\footnote{Note that the term ``information'' in this sentence means ``uncertainty'' of random variables from the definition of information theory, which does not imply that specific lost contents are ``encoded'' in $z$. The knowledge about lost contents is in our invertible model in the form of the bijective transformation between $x$ and $(y, z)$.}. Learning this bijective transformation enables our model to capture the knowledge of lost contents. 
Then during the inverse restoration procedure, a random sample of $z$ from the pre-specified distribution, together with the degraded image $y$, could recover most contents for the original image through the inverse function of the model. 
We consider two instantiation examples of the this bidirectional problem, i.e. image rescaling and image decolorization-colorization. 
As for the specific architectures, we start from image rescaling and develop Invertible Rescaling Network (IRN), which can be easily extended and adapted to decolorization-colorization.

To realize this invertible framework, several challenges should be tackled during training. Our basic targets include reconstructing original images with high quality and generating degraded images belonging to a target set, e.g. the set of visually-pleasing LR images. A further objective is to accomplish the restoration with an image-agnostic $z$, i.e., $z\sim p(z)$ instead of an image-specific $z \sim p(z\vert y)$, because it is easier for statistical modeling and sampling the independent $p(z)$ without the effort of handling conditions $y$. This is achievable since for any random vector with a density (i.e. $z'\sim p(z'\vert y)$), there exists a bijection $f_y$ such that $f_y(z')\sim N(0,I)$ \citep{HYVARINEN1999}.\footnote{This can be viewed as transferring the dependence of $z$ on $y$ into the process of our model that bijectively transforms mixed $y$ and $z$ into $x$. This treatment avoids the manual allocation of model capacity between capturing the $y$-dependency of the process from $z$ to $x$ and the $y$-dependency of the distribution of $z$, and make it easier for statistical modeling and sampling the random variable $z$. The restoration process, i.e. the inverse transformation of our model with inputs $y$ and $z$, is still dependent on the image content $y$.} For these purposes, we combine a reconstruction loss, a guidance loss, and a distribution matching loss to formulate a novel compact and effective objective function. Note that the last component aims at aligning recovered images with the true original image manifold as well as enforcing $z$ to follow the image-agnostic distribution $p(z)$, which cannot be simply achieved by conventional generative adversarial networks (GANs) nor the maximum likelihood estimation (MLE) method. This is because our invertible model does not give a marginal distribution on the data (it is not a simple generative model), and these conventional methods do not guide the distribution in the latent space for degraded image generation. We formulate the distribution on $y$ as the pushed-forward empirical distribution of $x$, which would inversely pass our invertible model in company with an independent distribution $p(z)$, to recover the distribution of $x$. Therefore, our distribution matching focuses on this recovered one and the data distribution of $x$, and we minimize the JS divergence between them in practice, as other alternative methods such as sample-based maximum mean discrepancy (MMD) method~\citep{ardizzone2019analyzing} could poorly handle the high-dimensional data in our task. Moreover, we show that once the distribution matching on $x$ is achieved, the matching also holds on the $(y,z)$ space with $z$ being image-agnostic, according to the invertible nature of our model.

Furthermore, we propose the combination between our invertible framework and existing degradation methods, and instantiate it by the combination of image rescaling and image compression. Since parts of degradation operations are not always available for adaption with restoration, e.g. image compression has common formats with general standards for convenient and wide applications, we study this combination to enable more applications. We demonstrate the effectiveness to combine our invertible framework with restoration from such degradation.
We note that there could be many other generalized applications of the invertible framework and model as well, such as image steganography, video rescaling, image denoising, etc. Please refer to recent works that adapt the invertible framework and model into various tasks since the publication of our preliminary version of this work\textsuperscript{\ref{footnote1}} for more details~\citep{lu2021large,jing2021hinet,huang2021video,tian2021self,liu2021invertible,xing2021invertible,zhao2021invertible,xie2021enhanced,cheng2021iicnet}. 
Our contributions are concluded as follows:

\begin{itemize}
    \item To our best knowledge, we are the first to model mutually-inverse image degradation and restoration with an invertible bijective transformation.\footnote{The preliminary version of this work has been accepted by \mbox{ECCV~2020} as oral presentation~\citep{xiao2020invertible}.\label{footnote1}} The deliberately designed invertibility enables the framework to model the information loss, which can mitigate the ill-posed nature in this bidirectional problem.
    \item We propose a novel model design and efficient training objectives to realize this framework. It enforces the latent variable $z$ to obey a simple image-agnostic distribution, which enables efficient inverse upscaling based on a sample from the distribution. 
    We develop IRN with deliberately designed architecture for the image rescaling task and demonstrate the easy adaptation to the similar image decolorization-colorization task.
    \item The proposed IRN and its scale-flexible and efficient variants achieve significant performance improvement of reconstructed HR images from the downscaled LR images, compared with state-of-the-art downscaling-SR and encoder-decoder methods. Meanwhile, the largely reduced parameters of IRN compared with these methods indicate the lightweight property and high efficiency of our model.
    \item We further propose the combination between our invertible framework and restoration from existing degradation methods, e.g. combination of image rescaling and compression, for more general applications. Experiments show improvements in these scenarios as well.
\end{itemize}

\section{Related Work}

\subsection{Image Upscaling after Downscaling}

When only the unidirectional upscaling task is considered, image super-resolution (SR) is a widely adopted method with promising results in low-resolution (LR) image upscaling. SR works focus on mitigating the inherent ill-posed problem by learning strong prior information by example-based strategy~\citep{freedman2011image,glasner2009super,schulter2015fast,kim2010single} or deep learning models~\citep{dong2015image,lim2017enhanced,zhang2018residual,zhang2018image,zhong2018joint,dai2019second,wang2018esrgan,guo2020closed,lugmayr2020srflow}. 
The state-of-the-art SR models are to train a neural network with elaborately designed architecture to reconstruct high-resolution (HR) images from the LR counterparts, which are usually generated by Bicubic interpolation from the HR images. However, when it comes to the bidirectional task of image rescaling, considering the image downscaling method would largely benefit the upscaling reconstruction.

Traditional image downscaling methods sub-sample images by a low-pass filter with frequency-based kernels, such as Bilinear, Bicubic, etc.~\citep{mitchell1988reconstruction}. 
For perceptual quality, several detail- or structure-preserving downscaling methods were proposed recently~\citep{kopf2013content,oeztireli2015perceptually,wang2004image,weber2016rapid,liu2017l_} to mitigate the over-smoothness of generated LR images. 
When the potential mutual reinforcement between downscaling and the inverse upscaling task is considered, the upscaling-optimal downscaling methods, which aim to learn the optimal downscaling model for the post-upscaling operation, gain increasing attention and efforts. 
For example, \cite{kim2018task} proposed a task-aware downscaling model based on an auto-encoder framework, which jointly trains the downscaling encoder and upscaling decoder as a united task. 
Similarly, \cite{li2018learning} used a CNN to estimate downscaled images while a learned or specified SR model is adopted for HR image recovery. 
Recently, \cite{sun2020learned} proposed a new content-adaptive-resampler-based image downscaling method, which is jointly trained with existing differentiable upscaling (SR) models. And \cite{chen2020hrnet} proposed a downscaling network based on the discretization of Hamiltonian System, which is trained jointly with SR models as well.
Although these efforts take the bidirectional nature of image rescaling into consideration, they simply link downscaling and upscaling through training objectives while ignoring the lost information during downscaling that leads to the ill-posed problem they suffer from.
In this paper, we propose to model the bidirectional downscaling and upscaling processes with invertible transformation based on invertible neural networks, which could model the lost information and largely mitigate the ill-posed problem.

\textbf{Difference from Super-Resolution.} Please note that the task of image rescaling is different from super-resolution. In our scenario, ground-truth HR images are available at the beginning but we have to use the LR version (e.g. for transmission or preview) instead. We would generate LR images and hope to recover the HR ones afterward from them. While for SR, the target is to generate new HR images for any given LR images.

\subsection{Image Decolorization-Colorization}

Image decolorization methods convert color images to grayscale, which enables applications like aesthetic photography, backward compatibility for legacy display, etc.~\citep{xia2018invertible}, while colorization methods aim to colorize grayscale images. Reconstructing original color images from the decolorized ones is also a bidirectional task with information loss, as color information is lost during decolorization and needs to be recovered, which can be viewed as ``downscaling'' and ``upscaling'' in the color channel dimension. 

Image colorization methods could be used to colorize decolorized images, and existing methods usually requires user-hints~\citep{levin2004colorization,zhang2017real} or learning strong priors by deep learning models~\citep{zhang2016colorful,deshpande2017learning,ardizzone2019guided} to generate color for grayscale images. When it comes to precisely recovering the original color of decolorized images, taking decolorization methods into consideration would help reconstruction as well. 

The most commonly used image decolorization method is to only take the luminance channel and discard color information in color space. Later, several methods have been proposed to preserve the color contrast or structural information which is easily lost during color-to-gray conversion~\citep{bala2004spatial,liu2015gcsdecolor}.
Taking decolorization and colorization as a joint task, \cite{xia2018invertible} first proposed invertible grayscale, which leverages an encoder-decoder architecture of deep learning models to learn to generate grayscale images that is helpful for colorization reconstruction. \cite{ye2020invertible} further improved the network design under this architecture. \cite{kim2018task} also demonstrates the extension of their image rescaling method for this task. However, these methods do not explicitly model the lost information and still significantly suffer from the ill-posed problem. In this work, we demonstrate that our proposed invertible framework could adapt to this bidirectional task well.

\subsection{Image Compression}

Image compression is a kind of data compression on digital images, which can be lossy (e.g. JPEG, BPG) or lossless (e.g. PNG, BMP). Traditional lossy image compression usually involves quantization in the frequency domain and optimal coding rules, while recently image compression methods based on deep learning show promising results of compression ratio and image quality~\citep{balle2016end,rippel2017real,balle2018variational,agustsson2019generative,minnen2018joint,wang2020modeling,cheng2020learned,li2020learning}. As image compression is only for storage saving, it will not change the resolution of images and there is no visually meaningful low-resolution image but only bit-stream output. 
Therefore image compression is different from image rescaling and their methods are usually different.

Despite this, image rescaling is orthogonal to image compression: they can be combined naturally and be applied together in many real applications \citep{2013Overview}. On one hand, the downscaled low-resolution images could be encoded by advanced lossless compression methods; on the other hand, first downscaling images and then compressing them is a common method for larger compression rate~\citep{bruckstein2003down}. Direct image compression methods perform poorly under extremely large compression rate, and are always combined with image rescaling for high compression rate of high-resolution images. In this work, we demonstrate the combination between IRN and lossless as well as lossy compression methods for better lossy compression performance.

\subsection{Invertible Neural Network}
The invertible neural network (INN)~\citep{dinh2015nice,dinh2017density,kingma2018glow,kumar2019videoflow,grathwohl2019ffjord,behrmann2019invertible,chen2019residual,kobyzev2020normalizing} is usually used for generative models. The invertible transformation of INN $f_\theta$ specifies the generative process $x = f_\theta (z)$ given a latent variable $z$, while the inverse mapping $f^{-1}_\theta$ enables explicit computation for the density of the model distribution, i.e. $p_X(x)=p_Z(f^{-1}(x))\left\vert\det Jf^{-1}(x)\right\vert$. Therefore, it is possible to use the maximum likelihood method for stable training of INN. The flexibility for modeling distributions allows INN to be applied in many variational inference tasks as well~\citep{rezende2015variational,kingma2016improved,berg2018sylvester}. Also, due to the strict invertibility, INN has been used to learn representations without information loss~\citep{jacobsen2018revnet}, which has been applied in the super-resolution task as a feature embedding module~\citep{li2019multi,zhu2019residual}.

Several prior works apply INN for tasks with paired data $(x,y)$. For example, \cite{ardizzone2019analyzing} deal with real-world inverse problems from medicine and astrophysics with INN. And \cite{asim2020invertible} leverage INN as effective priors at inverse problems including denoising, compressive sensing, and inpainting. \cite{NEURIPS2020_007ff380} further analyze INN as deep inverse models for generic inverse problems with four benchmarking tasks. Besides, conditional generation with INN, where the invertible modeling between $x$ and $z$ is conditioned on $y$, has also been explored and analyzed, such as in the task of image colorization~\citep{ardizzone2019guided} and super resolution~\citep{lugmayr2020srflow}. Different from these tasks considering unidirectional generation, image degradation-restoration is bidirectional, i.e. both generating $y$ given $x$ and the inverse reconstruction of $x$ are required. Therefore these models are unsuitable for our task, and we propose to model information loss in this task with INN. On the other hand, INN has been applied to conduct image-to-image translation~\citep{van2019reversible}. They consider the paired domain $(X,Y)$ rather than paired data, which is also different from our scenario.

The computational architecture of INN is specially designed to enable invertibility. For example, the mainstream architecture of INN is composed of coupling layers proposed in~\citep{dinh2015nice,dinh2017density}. In this architecture, INN consists of several invertible blocks. For the computation of the $l$-th block, different from conventional neural networks that directly apply neural network transformation on the input $h^l$ as $f(h^l)$, the input $h^l\in\mathbb{R}^{N\times H\times W\times C}$ is first split into $h_1^l, h_2^l$, usually along the channel axis so that $h_1^l\in\mathbb{R}^{N\times H\times W\times C_1}, h_2^l\in\mathbb{R}^{N\times H\times W\times C_2}, C_1+C_2=C$, and the following additive transformations are applied~\citep{dinh2015nice}:
\begin{eqnarray}
    \begin{aligned}
        &h_1^{l+1} = h_1^l + \phi(h_2^l),\\
        &h_2^{l+1} = h_2^l + \eta(h_1^{l+1}),
    \end{aligned}
    \label{eq:invblock}
\end{eqnarray}
where $\phi, \eta$ are functions parameterized by neural networks, e.g. convolutional neural networks. There is no restriction for $\phi, \eta$. The output of the block is the concatenation of the two parts, i.e. $[h_1^{l+1}, h_2^{l+1}]$, which will be the input to the $(l+1)$-th block. The strictly inverse transformation is easily computed given the output:
\begin{eqnarray}
    \begin{aligned}
        &h_2^l = h_2^{l+1} - \eta(h_1^{l+1}),\\
        &h_1^l = h_1^{l+1} - \phi(h_2^l),
    \end{aligned}
    \label{eq:invblock_inv}
\end{eqnarray}
For stronger expression ability, the following computation is often leveraged~\citep{dinh2017density}:
\begin{eqnarray}
    \begin{aligned}
        &h_1^{l+1} = h_1^l \odot \exp(\psi(h_2^l)) + \phi(h_2^l),\\
        &h_2^{l+1} = h_2^l \odot \exp(\rho(h_1^{l+1})) + \eta(h_1^{l+1}),\\
        &h_2^l = (h_2^{l+1} - \eta(h_1^{l+1})) \odot \exp(-\rho(h_1^{l+1})),\\
        &h_1^l = (h_1^{l+1} - \phi(h_2^l)) \odot \exp(-\psi(h_2^l)).
    \end{aligned}
    \label{eq:invblockexp}
\end{eqnarray}

This is the basic component of mainstream INNs that enforces the invertibility of the computation, and the expressive ability of such kind of architecture has been theoretically studied~\citep{NEURIPS2020_2290a738}. There are also other choices for INN architectures. For example, \citet{behrmann2019invertible,chen2019residual} prove that for the commonly used residual neural network architecture $y=f_{\theta}(x)+x$, when the spectral norm of the residual function $f_{\theta}$ is restricted under $1$, this computation is invertible and therefore can be used as a kind of INN. On the other hand, \citet{lu2021implicit} further proposes implicit normalizing flows, in which the computation of INN is implicitly defined by solving an equation. We will design our invertible architecture based on the typical coupling-layer-based invertible blocks, i.e. Eqs.~(\ref{eq:invblock},\ref{eq:invblockexp}), and task-related considerations in Section~\ref{sec: inv arch}.

\section{Methods}\label{sec:methods}

In this section, we first formally present the general mathematical formulation of the image degradation-restoration problem in Section~\ref{sec: math form}. Then we describe the invertible modeling framework of this bidirectional problem in Section~\ref{sec:model-spec}. As for the specific model, we start from image rescaling and elaborate on the specific invertible architecture and training methods for IRN in Section~\ref{sec:image rescaling}. Then we show the adaptation of IRN to the similar decolorization-colorization task in Section~\ref{sec:decolorization-colorization}. Finally, we propose to combine the invertible framework with existing degradation methods with an instantiation of the combination between image rescaling and compression in Section~\ref{sec:combine compression}.

\subsection{Mathematical Formulation of Image Degradation-Restoration}\label{sec: math form}
The basic formulation of the image degradation-restoration problem can be described as:
\begin{equation}
\begin{aligned}
    \min_{\theta}\quad &\sum_{x}\mathcal{L}\left(x, \mathcal{U}\left(\mathcal{D}(x; \theta); \theta\right)\right),\\
    \mathrm{s.t.}\quad & y=\mathcal{D}(x; \theta)\in Y(x), \forall x,
    \label{math form original}
\end{aligned}
\end{equation}
where $x$ is the original image, e.g. HR image for the image rescaling task, $\mathcal{D}$ and $\mathcal{U}$ are respectively the degradation and restoration models parameterized by $\theta$, e.g. downscaling and upscaling of image rescaling, $\mathcal{L}$ is a criterion justifying the quality of recovered images, $y=\mathcal{D}(x; \theta)$ is the model-degraded image, and $Y(x)$ denotes the target set of valid degraded images given $x$, e.g. visually valid LR images given the HR image $x$ for the image rescaling task.
When $\mathcal{D}$ is a given mapping without parameters to optimize, the problem of learning $\mathcal{U}$ only resorts to a typical restoration problem, e.g. image super-resolution. In contrast, in the degradation-restoration problem, $\mathcal{D}$ is also learned and contributes to a better restoration. 

In many tasks, although we do not have the explicit expression of $Y(x)$, it is much easier to obtain a valid degraded image in this set. For example, typical interpolation methods (e.g. Bicubic) could produce visually valid LR images for the image rescaling tasks. As for the rescaling and decolorization-colorization tasks in this paper, we instantiate the constraint in (\ref{math form original}) by narrowing the set around a given sample. 
Specifically, let $y_{\text{guide}}(x)$ denote an available degraded image, e.g. an LR image downscaled by a typical interpolation method which well demonstrates what is a visually valid LR image as a sample in $Y(x)$. 
We instantiate $Y(x)$ by $Y_\text{guide}(x)=\{y\,\vert\, \lVert y - y_\text{guide}(x) \rVert < \epsilon\}$. So in practice only one valid degraded image $y_{\text{guide}}(x)$ is required and the original problem turns into:
\begin{equation}
\begin{aligned}
    \min_{\theta}\quad &\sum_x\mathcal{L}\left(x, \mathcal{U}\left(\mathcal{D}(x; \theta); \theta\right)\right),\\
    \mathrm{s.t.}\quad & \lVert \mathcal{D}(x; \theta)-y_{\text{guide}}(x) \rVert < \epsilon.
\end{aligned}
\end{equation}
In Section~\ref{sec:train-method}, this constraint will be further relaxed and formulate a guidance loss in practice.

\begin{figure*} [ht]
    \centering
    \includegraphics[scale=0.35]{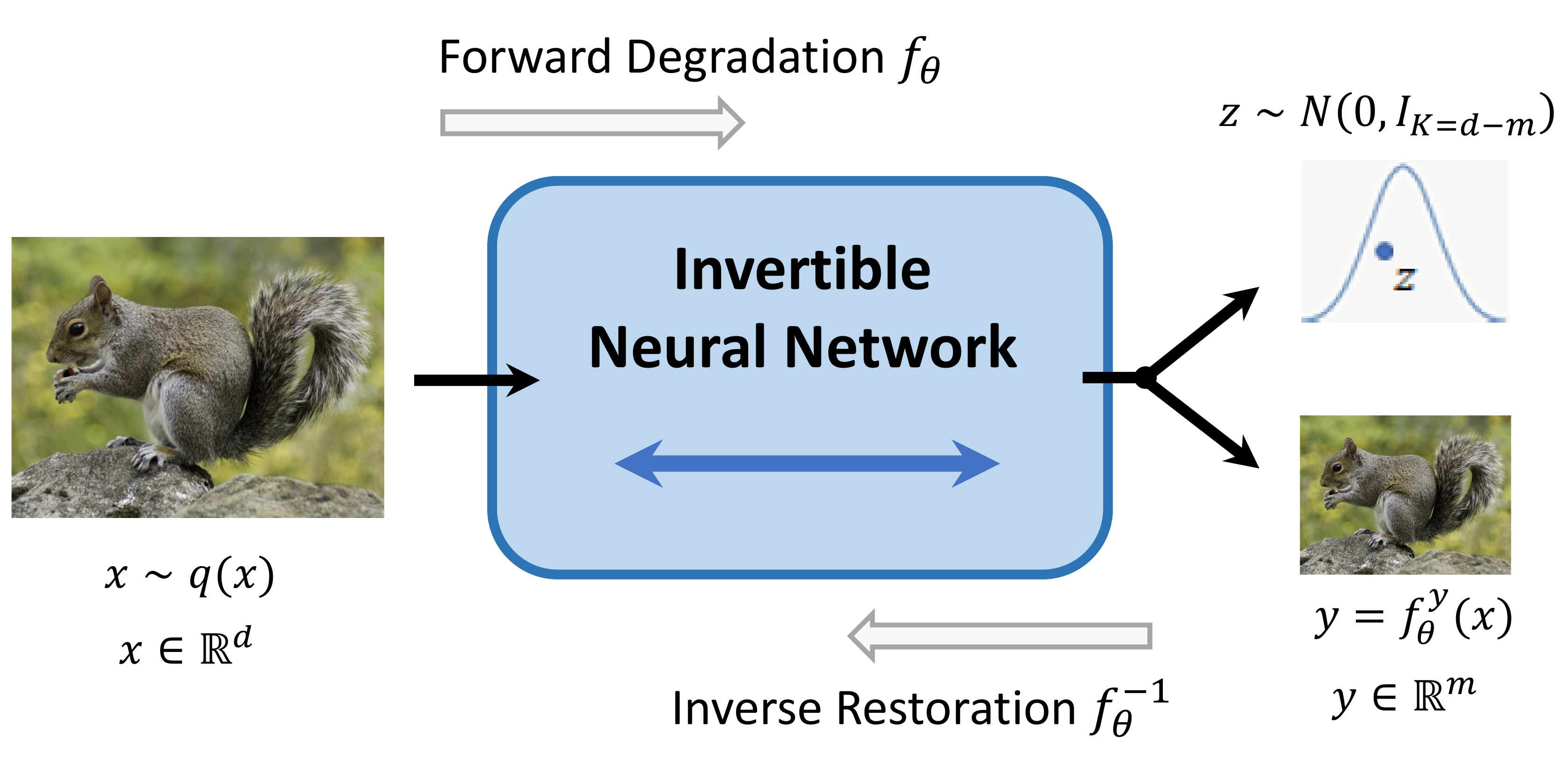}
    \caption{Illustration of the invertible modeling framework for the degradation-restoration problem. In the forward degradation procedure, the image $x$ is transformed to a valid degraded image $y$ and an image-agnostic latent variable $z$ through a parameterized invertible function $f_{\theta}(\cdot)$; in the inverse upscaling procedure, a randomly drawn $z$ combined with $y$ are transformed to restore image $x$ through the inverse function $f_{\theta}^{-1}(\cdot)$.}
    \label{fig:problem formulation}
\end{figure*}

Now we have described the basic settings of image degradation-restoration. The problem formulation under our invertible framework will be illustrated in the following sections. 

\subsection{Specification of Invertible Modeling}\label{sec:model-spec}
\subsubsection{Formulation of Invertible Framework}

As described in the Introduction, we model the bidirectional degradation and restoration from the perspective of invertible bijective transformation. 
Fig.~\ref{fig:problem formulation} illustrates the sketch of our invertible framework. To model lost information during degradation, we introduce an auxiliary latent random variable $z$, and leverage an invertible neural network to bijectively transform the distribution of $x$ to the joint distribution of a pre-specified distribution $p(z)$ and the distribution of model-degraded image $y$. Then the distribution of lost contents is transformed to $p(z)$ together with the generation of $y$. As described in the introduction, we note that for any random vector with a density (i.e. $z'\sim p(z'\vert y)$), there exists a bijection $f_y$ such that $f_y(z')\sim N(0,I)$ \citep{HYVARINEN1999}; therefore for easier modeling and sampling of $p(z)$ without handling conditions, we choose image-agnostic $z\sim p(z)$ as an additional desideratum, which will be enforced by distribution matching. In this way, the distribution of lost contents is captured by our model without preserving image-specific lost contents or $z$, and a random sample of $z'$ from $p(z)$ in company with the degraded image $y$ could reconstruct a image $x'$ with reasonable lost contents by the inverse function of our invertible model. Let $f_{\theta}$ denote the parameterized bijective transformation. Then the degradation procedure of our model is expressed as $(y, z)=f_{\theta}(x)$, where $y$ is the output degraded image. Correspondingly, the restoration procedure is $x'=f_{\theta}^{-1}(y, z')$, where $z' \sim p(z)$. As $z'$ is random, the restored image $x'$ is also random. This defines the restoration distribution $p_\theta(x\vert y)$, representing the uncertainty over all possible original images that could yield $y$. The randomness of $z$ corresponds to the randomness of reasonable $x$ in $p_\theta(x\vert y)$. 
Note that this inverse transformation will mix $y$ and $z'$ so that the generation process is still dependent on the image-specific information. 

The invertible modeling framework is particularly suitable for the degradation-restoration problem under a measure-theoretic point of view, in that it has the unique advantage of being \emph{cyclically compatible} \citep[Def.~2.1]{liu2021generative}.
This means the model-defined restoration distribution $p_\theta(x\vert y)$ and degradation distribution $p_\theta(y\vert x)$ always come from the same joint distribution of $(x,y)$. Since the degradation distribution $p_\theta(y\vert x) = \delta_{f_\theta^y(x)} (y)$ ($f_\theta^y(x)$ denotes the $y$-part of the output of $(y,z) = f_\theta(x)$) is a Dirac delta distribution, the restoration distribution $p_\theta(x\vert y)$ is compatible with it if and only if it is supported within the preimage set of the degradation transformation $f_\theta^y$, i.e. $(f_\theta^y)^{-1}(\{y\}) := \{x \mid f_\theta^y(x) = y\}$ \citep[Thm.~2.6]{liu2021generative}.
Due to the invertibility of $f_\theta$, for any $z' \in \mathbb{R}^K$, the restored image $f_\theta^{-1}(y,z')$ is always in the preimage set since $f_\theta^y(f_\theta^{-1}(y,z')) = y$. In this way, the model only needs to focus on learning the distribution over all possible original images without worrying about conflicting with the degradation process.

With invertible modeling, the problem formulation is described as:
\begin{equation}
\begin{aligned}
    \min_{\theta}\quad &\sum_x\mathbb{E}_{z\sim p(z)}\left[\mathcal{L}\left(x, f_{\theta}^{-1}\left([f_{\theta}^y(x), z]\right)\right)\right],\\
    \mathrm{s.t.}\quad & \lVert f_{\theta}^y(x)-y_{\text{guide}} \rVert < \epsilon,\\
    & \{f_\theta^z(x)\}_x \sim p(z),
\end{aligned}
\label{mathematical formulation}
\end{equation}
where $f_{\theta}^y$ and $f_{\theta}^z$ denote the transformations whose outputs correspond to $y$ and $z$ of the output of $f_{\theta}(x)$ respectively.
In Section~\ref{sec:train-method}, the constraint regarding distributions will be relaxed and formulate a distribution loss in practice.

\subsubsection{Realization of Invertible Framework}\label{sec:train-method}

Our invertible framework specifies a correspondence between the distributions of the original image $x$ and the degraded image $y$, as well as the image-agnostic distribution $p(z)$ of the latent variable $z$. To realize this framework, we should train the invertible model denoted by $f_{\theta}$. This subsection introduces the general training objectives for our invertible models, while some adaptions will be detailed for specific tasks in Sections~\ref{sec:image rescaling} and~\ref{sec:decolorization-colorization}. The training objectives are to drive the above relations and match our requirements, i.e. solve (\ref{mathematical formulation}). 
We will make the constrained optimization problem (\ref{mathematical formulation}) practical by reforming it as jointly optimizing three objective terms as introduced below.

{\textbf{Reconstruction}}\quad
As described in Section~\ref{sec:model-spec}, our invertible framework is under the context of distribution. Therefore it is not for the correspondence between the point $x$ and $y$ if $z$ is not specified. Given a image $x^{(n)}$, the model-degraded image $f_\theta^y (x^{(n)})$ will be restored by our model with the image-agnostic latent variable $z\sim p(z)$, resulting in $f_\theta^{-1} (f_\theta^y (x^{(n)}), z)$ which also follows a distribution. We hope to restrict this distribution around the original image so that the image can be validly recovered by the model using any sample of $z$ from $p(z)$. This arbitrariness would inversely encourage the disentanglement between $z$ and $y$ in the forward process as well. To achieve this, we encourage the reconstructed image with any random sample $z$ to match the original $x^{(n)}$, leading to the reconstruction loss which minimizes the expected difference over all original images:
\begin{align}
    \Lrecon(\theta) := \sum_{n=1}^N \bbE_{z\sim p(z)} [ \ell_\clX ( x^{(n)}, f_\theta^{-1} (f_\theta^y (x^{(n)}), z) ) ],
    \label{eq.recon loss}
\end{align}
where $\ell_\clX$ is a difference metric on $\clX$, e.g. the $L_1$ or $L_2$ loss. We estimate the expectation w.r.t $z$ by one random sample from $p(z)$ each evaluation in practice. This loss corresponds to the objective in (\ref{mathematical formulation}).

{\textbf{Guidance}}\quad
As described in Section~\ref{sec: math form}, we hope to generate a valid degraded image belonging to a target set, whose expression is not explicitly known, but we can instantiate it as a constraint w.r.t. the distance to guiding degraded images. We relax this constraint as a loss added in the objective, which encourages the model-degraded images to resemble guiding images. Let $\yguide^{(n)}$ denote this guiding image (for example, an LR image generated by the Bicubic interpolation for the image rescaling task). The guidance loss is expressed as:
\begin{align}
    \Lguide(\theta) := \sum_{n=1}^N \ell_\clY (\yguide^{(n)}, f_\theta^y (x^{(n)})),
    \label{eq.guide loss}
\end{align}
where $\ell_\clY$ is a difference metric on $\clY$, e.g. the $L_1$ or $L_2$ loss.
This kind of objective was also adopted in literatures~\citep{kim2018task,sun2020learned}.

{\textbf{Distribution Matching}}\quad
The third part of the training objective is to match the distribution of latent variable $z$ and original images. We first describe our notations for the distributions. We denote the data distribution of original images as $q(x)$, which is available through the sample cloud $\{x^{(n)}\}_{n=1}^N$. Note that when traversing over this sample cloud, $\{y^{(n)}\}_{n=1}^N$ generated by our model also form a sample cloud of a distribution. We use the push-forward distribution ${f_\theta^y}_\# [q] (y)$ to denote this distribution of $y$, which represents the distribution of the transformed random variable $y = f_\theta^y (x)$ with $x \sim q(x)$. We define the push forward distribution ${f_{\theta}^z}_{\#}[q](z)$ in the same way. 
Similarly, the inversely reconstructed images compose a sample cloud $\{f_\theta^{-1} (y^{(n)}, z^{(n)})\}_{n=1}^N$ following a distribution, where $z^{(n)}\sim p(z)$ is a randomly drawn latent variable. As $z\sim p(z)$ is to be independent from $y$, we have $(y^{(n)}, z^{(n)}) \sim {f_\theta^y}_\# [q] (y) \, p(z)$. Therefore, we can denote the distribution of reconstructed images as ${f_\theta^{-1}}_\# \big[ {f_\theta^y}_\# [q] (y) \, p(z) \big] (x)$.

Our model should enforce $z\sim p(z)$ to be image-agnostic and match the model-reconstructed distribution towards data distribution. 
This corresponds to the constraint on the distribution in (\ref{mathematical formulation}).  
Therefore we relax the constraint as a loss added in the objective as well, and introduce the distribution matching loss to achieve these two goals:
\begin{align}
    \Ldistr(\theta) := L_\clP \big( {f_\theta^{-1}}_\# \big[ {f_\theta^y}_\# [q](y) \, p(z) \big] (x), q(x) \big),
    \label{eq.distr loss}
\end{align}
where $L_\clP$ is a difference metric of distributions. The distribution matching loss directly pushes the model-reconstructed images to lie on the manifold of true original images, which matches the distribution and enables the recovered images to be more realistic (note that the reconstruction loss only restrict them around the original images). At the same time, it drives the independence of $z\sim p(z)$ from $y$ in the forward transformation. This is because if $f_\theta$ is invertible, the distribution matching holds on $\clX$ if and only if it holds on $\clY\times\clZ$ in the asymptotic case, i.e. ${f_\theta^{-1}}_\# \big[ {f_\theta^y}_\# [q] (y) \, p(z) \big] (x) = q(x)$ is equivalent to ${f_\theta^y}_\# [q] (y) \, p(z) = {f_\theta}_\# [q] (y,z)$. In this way, the loss also drives the coupled distribution ${f_\theta}_\# [q] (y,z)$ from the forward transformation towards the decoupled distribution ${f_\theta^y}_\# [q] (y) \, p(z)$, realizing the matching of independent $z \sim p(z)$.

As for the probability metric $L_\clP$, we can employ the JS divergence due to the high-dimensionality and unknown density function in our problem. We estimate the loss as:
\begin{align}
	& \Ldistr(\theta) = \mathrm{JS}( {f_\theta^{-1}}_\# \big[ {f_\theta^y}_\# [q](y) \, p(z) \big] (x), q(x) ) \notag \\
	={} & \frac{1}{2} \max_T \Big\{ \bbE_{q(x)} \left[ \log \sigma(T(x)) \right] \notag \\
	    & {} + \bbE_{x' \sim {f_\theta^{-1}}_\# \big[ {f_\theta^y}_\# [q](y) \, p(z) \big] (x')} \left[ \log \left( 1 - \sigma(T(x')) \right) \right] \Big\} \notag\\ & {} + \log 2 \notag \\
	={} & \frac{1}{2} \max_T \big\{ \bbE_{q(x)} \left[ \log \sigma(T(x)) \right] \notag \\
	    & {} + \bbE_{ (y, z) \sim {f_\theta^y}_\# [q](y) \, p(z) } \left[ \log \left( 1 - \sigma(T( f_\theta^{-1} (y, z) )) \right) \right] \big\} \notag\\ & {} + \log 2 \notag \\
	\approx{} & \frac{1}{2 N} \max_T \sum_n \Big\{ \log \sigma(T (x^{(n)})) \notag \\
	    & {} + \log \left( 1 - \sigma(T( f_\theta^{-1} ( f_\theta^y (x^{(n)}), z^{(n)} ) )) \right) \Big\} + \log 2,
	\label{eqn:bwddismat}
\end{align}
where $\sigma$ is the sigmoid function, $T: \clX \to \bbR$ is a function on $\clX$ and $\sigma(T(\cdot))$ is regarded as the discriminator in GAN literatures~\citep{goodfellow2014generative}. The ``$\approx$'' is due to Monte Carlo estimation: $\{z^{(n)}\}_{n=1}^N$ are i.i.d. samples from $p(z)$ and $\{x^{(n)}\}_{n=1}^N \sim q(x)$.
In practice, we can parameterize the function $T$ with a neural network $T_\phi$, and thus $\max_T$ amounts to $\max_\phi$.
We can follow the same way as GANs to optimize $\theta$ and $\phi$ so that the JS divergence is minimized.

\begin{figure*}[ht]
	\centering
	\includegraphics[width=\textwidth]{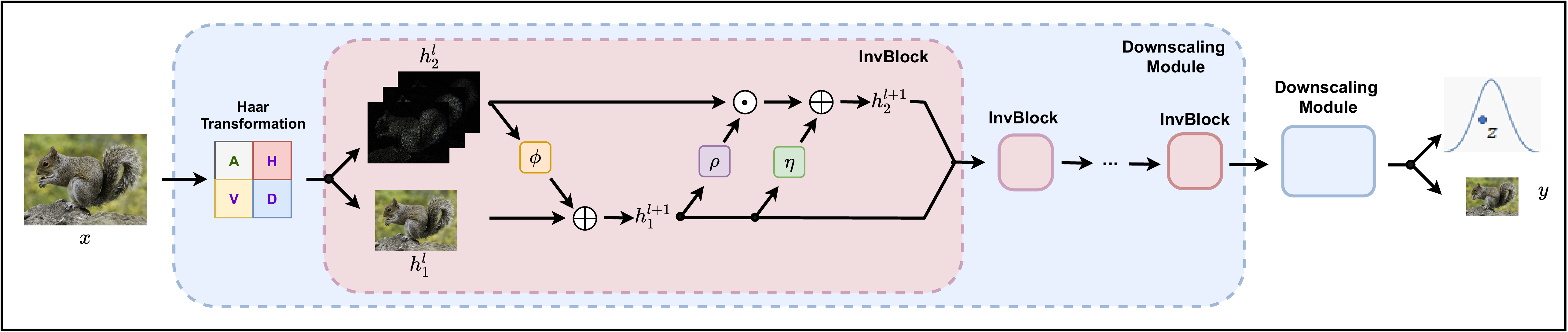}
	\caption{Illustration of our Invertible Rescaling Network (IRN) as the instantiation model of our invertible modeling framework. The invertible architecture is composed of Downscaling Modules, in which InvBlocks are stacked after a Haar Transformation. Each Downscaling Module reduces the spatial resolution by 2$\times$. The $\exp(\cdot)$ of $\rho$ is omit.}
	\label{fig:framework}
\end{figure*}

\subsection{Model for Image Rescaling}\label{sec:image rescaling}

As for specific models, we start from image rescaling in this section. 
We develop Invertible Rescaling Network (IRN) as the instantiation model of our inverible modeling framework for image rescaling, and we will describe the specific invertible architecture and training methods of IRN. 
We also present the algorithms for downscaling and upscaling in our IRN model in Algorithms~\ref{downscaling},~\ref{upscaling} as an example to better demonstrate the input, output, and procedure of our invertible framework. Note that in practice the HR image $x$ and LR image $y$ will be quantized to 8-bit representation, as will be indicated in Section~\ref{sec: inv arch}. We omit this detail in the algorithm description and treat the domain as $\mathbb{R}$.

\begin{algorithm}[h]
\caption{Downscaling of IRN}
\hspace*{0.02in} {\bf Input:}
HR image $x\in \mathbb{R}^{H\times W\times C}$, scale size $s$, model $f_{\theta, s}$\\
\hspace*{0.02in} {\bf Output:}
LR image $y\in \mathbb{R}^{\frac{H}{s}\times\frac{W}{s}\times C}$
\begin{algorithmic}[1]
\State Calculate $(y, z) = f_{\theta, s}(x)$
\State \Return $y$
\end{algorithmic}
\label{downscaling}
\end{algorithm}
\begin{algorithm}[h]
\caption{Upscaling of IRN}
\hspace*{0.02in} {\bf Input:}
LR image $y\in \mathbb{R}^{H\times W\times C}$, scale size $s$, model $f_{\theta, s}$\\
\hspace*{0.02in} {\bf Output:}
HR image $x\in \mathbb{R}^{sH\times sW\times C}$
\begin{algorithmic}[1]
\State Randomly sample $z\sim p(z), z\in \mathbb{R}^{H\times W\times (s^2-1)C}$
\State Calculate $x = f_{\theta, s}^{-1}(y, z)$
\State \Return $x$
\end{algorithmic}
\label{upscaling}
\end{algorithm}

\subsubsection{Invertible Architecture}\label{sec: inv arch}

Fig.~\ref{fig:framework} illustrates the architecture of our proposed IRN, which is basically composed of stacked \textit{Downscaling Modules} consisting of one \textit{Haar Transformation} and several \textit{InvBlocks}. Each \textit{Downscaling Module} will reduce the spatial resolution by $2\times$. The overall architecture is invertible given that each component is invertible.

\textbf{The Haar Transformation}\quad In each \textit{Downscaling Module}, a Haar Transformation is first applied to equip the model with a certain inductive bias for splitting low- and high-frequency contents, which are approximately preserved and lost contents during image downscaling respectively. 
The Haar Transformation, which is an invertible wavelet transformation, will decompose the input into a low-pass representation and three directions of high-frequency coefficients~\citep{ardizzone2019guided}. Specifically, given the input raw image or feature maps with height $H$, width $W$ and channel $C$, a tensor of shape $(\frac{1}{2}H, \frac{1}{2}W, 4C)$ is produced, where the first $C$ slices are the low-pass representation equivalent to the Bilinear interpolation downscaling, and the other three groups of $C$ slices correspond to the high-frequency residual in the vertical, horizontal and diagonal directions respectively. With the help of the Haar Transformation, the model could effectively separate low- and high-frequency information, which benefits the following generation of $y$ and transformation from $x_H$ to $z$. 
And the spatial resolution is reduced by $2\times$ after the Haar Transformation.

\textbf{InvBlock}\quad InvBlocks are the main components for the target invertible transformations. Given that the input has been split into low- and high-frequency components by the Haar Transformation, we introduce InvBlocks based on the coupling layer architecture described in Eqs.~(\ref{eq:invblock},\ref{eq:invblock_inv},\ref{eq:invblockexp}), whose two branches (i.e. the split of $h_1^l$ and $h_2^l$ in Eq.~\eqref{eq:invblock}) correspond to these two components respectively. The transformation would further polish the input representations for the generation of a suitable LR image as well as an independent and properly distributed latent representation for lost information. As for the detailed computation, considering the importance of shortcut connection in image scaling tasks~\citep{lim2017enhanced,wang2018esrgan}, we employ the additive transformation (Eq.~\ref{eq:invblock}) for the low-frequency part $h_1^l$, and the enhanced affine transformation (Eq.~\ref{eq:invblockexp}) for the high-frequency part $h_2^l$ to enhance the model capacity. 
This also equips the model with a certain inductive bias for the generation of LR images with the low-frequency part going straight through, and could stabilize the training of IRN. 
The details of the InvBlock architecture are illustrated in Fig.~\ref{fig:framework}, except that the $\exp(\cdot)$ operation after function $\rho$ is omitted here.

We employ a densely connected convolutional block, which has demonstrated its effectiveness for image scaling tasks in~\citep{wang2018esrgan}, to parameterize the transformation functions $\phi(\cdot), \eta(\cdot), \rho(\cdot)$. To avoid numerical explosion due to the $\exp(\cdot)$ function, we employ a centered sigmoid function and a scale term after function $\rho(\cdot)$.

\textbf{Quantization}\quad The outputs of our model are floating-point values, while the common image formats such as RGB are quantized to 8-bit representation. To enable storage compatibility, we adopt a rounding operation as the quantization module on the generated LR image. The quantized LR image is saved by PNG format and used for upscaling. However, the nondifferentiable property of quantization poses challenges for training with back-propagation. To overcome the obstacle, we apply the Straight-Through Estimator \citep{bengio2013estimating} to calculate the gradients for the quantization module. The notation for quantization is omitted in the following for simplicity.

\subsubsection{Scale-flexible and Efficient Implementation}

There could be further improvements over the architecture to adapt IRN to more scales or more computation efficiency. Specifically, we will introduce the learnable downsampling module and improvement on computational efficiency to enable scale-flexible and efficient implementation.

\begin{figure} [ht]
    \centering
    \includegraphics[scale=0.225]{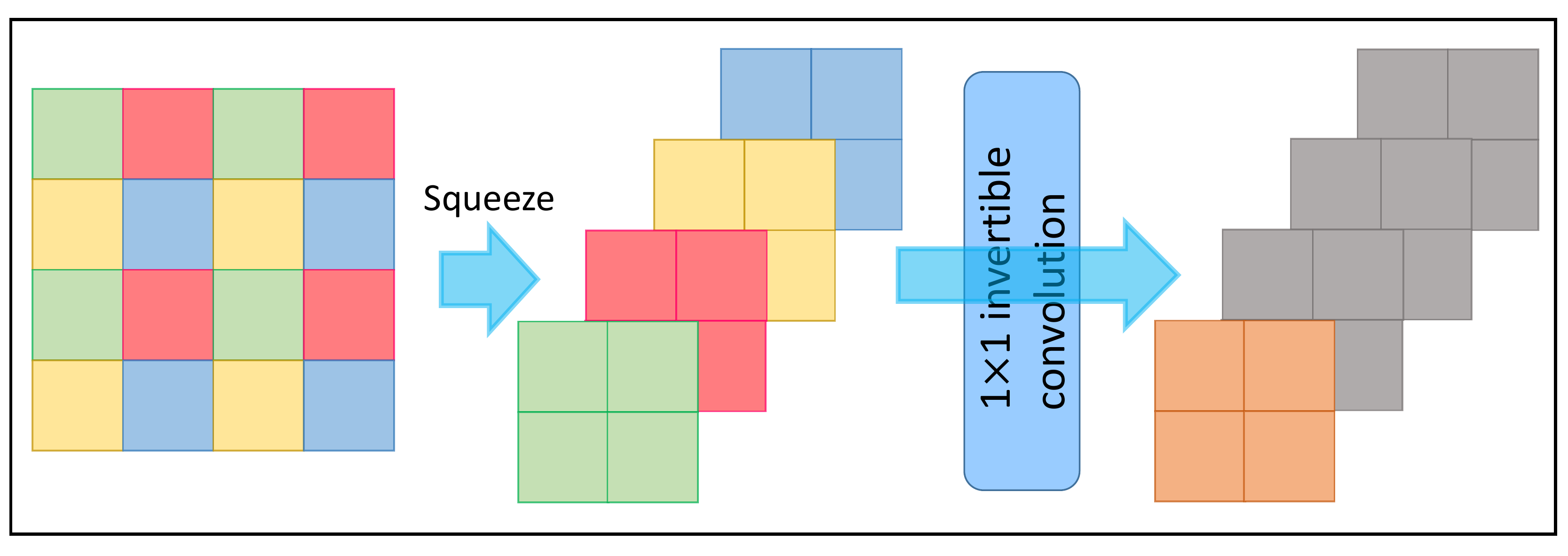}
    \caption{Illustration of the learnable dowmsampling module ($2\times$ example). It consists of a squeeze operation to downscale the spatial resolution by $N$ times and a $1\times1$ invertible convolution to transform the squeezed $N\times N$ elements.}
    \label{fig:learnable downsampling}
\end{figure}

\paragraph{Learnable Downsampling} Although the Haar Wavelet Transformation is able to serve for downsampling and splitting high- and low-frequency contents well, stacking multiple transformations can only rescale images by the scales that are the power of two. This largely restricts the rescaling scope for our model. To enable more scales, such as $3\times$, we propose to leverage a learnable downsampling layer to replace Haar Transformation in the architecture. It consists of a squeeze operation and one $1\times1$ invertible convolution.

As shown in Fig.~\ref{fig:learnable downsampling}, the squeeze operation downscales the spatial resolution for a certain scale $N$ by squeezing spatial elements into channels. Then, a $1\times1$ invertible convolution is applied to transform the squeezed $N\times N$ elements before InvBlocks. $1\times1$ invertible convolution is first proposed in GLOW \citep{kingma2018glow} for channel permutation. Different from their purpose, we expect it to learn to split low- and high-frequency contents under arbitrary scales and adapt the following InvBlocks better. The Haar Transformation can be viewed as a special case of this downsampling module under $2\times$ scale, as it provides a fixed rather than learnable prior. For this module, we provide a prior for extracting low-frequency in initialization by setting parameters of the $1\times1$ invertible convolution in order that the first channel after transformation is the average of $N\times N$ elements, while the other channels are the identity transformation to enable the invertibility.

We denote the IRN model with learnable downsampling as IRN$_\text{LD}$.

\textbf{Fractional scaling factors}\quad In real applications, there would be fractional scaling factors. We can deal with them by combining IRN and traditional interpolation methods. Specifically, for the scaling factor $s_1$, we choose IRN with scaling factor $s_2=[s_1]$ and rescale HR images with interpolation (e.g. Bicubic) by scale $\frac{s_2}{s_1}$ and $\frac{s_1}{s_2}$ before and after passing them into IRN respectively. This has been demonstrated in recent work as well~\citep{xing2022scale}.

\paragraph{Improving Computation Efficiency}\label{sec:computation efficiency}

We note that the architecture that stacks multiple Downscaling Modules containing one downsampling module and multiple InvBlocks suffers from much-increased FLOPs during computation. This is because InvBlocks in the previous Downscaling Modules other than the last one will apply convolution operations on tensors with larger spatial resolution, which significantly increases computational cost. To further improve computation efficiency, we propose to modify the architecture to first apply downsampling modules (e.g. multiple Haar Transformation or learnable downsampling) and then go through multiple InvBlocks. This enables the convolution operations to be applied on smaller resolutions, which could largely reduce the FLOPs and runtime under a similar amount of parameters.

We denote the IRN model under this architecture as IRN$_\text{E}$. It differs from IRN only when IRN stacks multiple Downscaling Modules.

\subsubsection{Training Objectives}\label{sec:train-method-irn}

The training objectives of IRN mainly follow the reconstruction (Eq.(\ref{eq.recon loss})), guidance (Eq.(\ref{eq.guide loss})), and distribution matching (Eq.(\ref{eq.distr loss})) to realize the invertible framework as described in Section~\ref{sec:train-method}. For image rescaling, the reconstruction and guidance is adapted as HR reconstruction and LR guidance correspondingly, which means calculating $L_{recon}$ between reconstructed and original HR images and calculating $L_{guide}$ between model-generated LR images and LR images generated by the Bicubic interpolation methods, respectively. 
Based on the above objectives, we can optimize our IRN model by minimizing the combination of the three losses, which relaxes the constrained problem~(\ref{mathematical formulation}) into an unconstrained one. However, as an issue in practice, we find it difficult to directly do the optimization due to the unstable training process of GANs~\citep{arjovsky2017towards}. Therefore, we propose to adopt a weakened but more stable surrogate loss for the distribution matching as a pre-training stage, forming a two-stage training procedure.

As explained in Section~\ref{sec:train-method}, the distribution matching on $\clX$ has the same asymptotic effect as on $\clY\times\clZ$, i.e. $L_\clP( {f_\theta^y}_\# [q] (y) \, p(z), {f_\theta}_\# [q] (y,z) )$. Our surrogate loss considers partial distribution matching on $\clZ$, i.e. $L_\clP( p(z), {f_\theta^z}_\# [q] (z) )$, which is more flexible as the density function of $p(z)$ is available. We choose cross entropy (CE) as a more stable distribution metric for minimization:
\begin{align}
    & \Lpdistr(\theta) := \mathrm{CE}( {f_\theta^z}_\# [q] (z), p(z) ) \notag \\
    =& - \! \bbE_{{f_\theta^z}_\# [q](z)} [\log p(z)] = - \bbE_{q(x)} [\log p( z \!=\! f_\theta^z (x) )].
    \label{eq.distr loss'}
\end{align}

Note that the maximum likelihood estimation (MLE) $\max_\theta \bbE_{q(x)} [ \log {f_\theta^{-1}}_\# [p_{y,z}] (x) ]$ commonly used in related INN-based generative models~\citep{dinh2015nice,dinh2017density,kingma2018glow,ardizzone2019guided}
is however not applicable to our model,
since it requires a joint distribution $p(y,z)$ with tractable density function on the $(y,z)$ end,
while we only have a distribution $p(z)$ on $z$.\footnote{MLEs corresponding to minimizing $\KL( q(x\vert y), {f_\theta^{-1}(y, \cdot)}_\# [p_z] (x) )$ or $\KL\Big( q(x), \Big( \bbE_{{f_\theta^y}_\# [q](y)} [f_\theta^{-1}(y, \cdot)] \Big)_\# [p_z] (x) \Big)$ are also impossible, since the pushed-forward distributions are only supported on a lower-dimensional manifold (dimension of $z$) in $\clX$ so their densities are not well-defined (i.e., the densities are a.e. zero in $\clX$ and are infinite on the manifold).} Therefore we can only leverage a stable but weakened surrogate loss.

Our pre-training stage will minimize the following total objective, and we call IRN as this trained model:
\begin{align}
    L_\mathrm{IRN} := \lambda_1 \Lrecon + \lambda_2 \Lguide + \lambda_3 \Lpdistr,
    \label{eq.IRN objective function}
\end{align}
where $\lambda_1, \lambda_2, \lambda_3$ are coefficients for balancing different loss terms.

After the pre-training, we adopt the trained model as the initialization and restore the full distribution matching loss $\Ldistr$ based on JS divergence for the training objective. Additionally, as $\Ldistr$ encourages reconstructed HR images to be more realistic, we also add a perceptual loss ~\citep{johnson2016perceptual} $\Lpercp$ on $\clX$ to further enhance the perceptual quality. Instead of pixels, the perceptual loss measures the difference between two images on their semantic features, which are extracted by pre-trained deep learning models (e.g. VGG).
There are several variants of the perceptual loss which mainly differ from the feature positions~\citep{ledig2017photo,wang2018esrgan}, and we adopt the variant proposed in~\cite{wang2018esrgan}.

Therefore, the second stage minimizes the following total objective, and we call the model as IRN+:
\begin{align}
    L_\mathrm{IRN+} := \lambda_1 \Lrecon + \lambda_2 \Lguide + \lambda_3 \Ldistr + \lambda_4 \Lpercp.
    \label{eq.IRN+ objective function}
\end{align}

\subsection{Model for Image Decolorization-Colorization}\label{sec:decolorization-colorization}

Image decolorization-colorization is a commonly seen task~\citep{xia2018invertible,ye2020invertible} and is another instantiation of bidirectional degradation-restoration problem, in which color information in the channel dimension is lost. The core idea of our problem formulation is the same as Fig.~\ref{fig:problem formulation}, which transforms the distribution of image-specific lost information into an image-agnostic Gaussian distribution. Some adaptation of the specific model to fit this task is illustrated as the following.

\subsubsection{Architecture}

The basic architecture is similar to Fig.~\ref{fig:framework}. Different from splitting low- and high-frequency contents as image rescaling, we need to split grayscale and color information, and produce a grayscale image while capturing the distribution of color information here. Therefore, we need to replace the downsampling module with a graying module. We directly leverage the YCbCr color space representation of the image to split the information in the channel. Then these two branch of information (i.e. Y and CbCr) go through InvBlocks as introduced previously. We denote this model as IRN$_\text{color}$.

\subsubsection{Training Objectives}

We also leverage the three components for the objective, i.e. guidance loss  (Eq.(\ref{eq.guide loss})), reconstruction loss (Eq.(\ref{eq.recon loss})), and distribution matching loss (Eq.(\ref{eq.distr loss})). In particular, for the guidance loss, we adapt it as a Grayscale Guidance, in which the Y channel under YCbCr representation of the image is leveraged as the guidance. The reconstruction loss is to compute the difference between reconstructed images and original ones. For distribution matching, we choose the stable cross entropy introduced in Section~\ref{sec:train-method-irn} here, because the human perception of color is less sensitive and the unstable perceptual-driven loss is not necessary for good results. Besides, because colorization has more diverse results than upscaling, to stabilize and improve our training for the reconstruction of original color images, we will consider an alternative choice to only encourage the most probable point of latent variable $z$ in its distribution rather than the whole distribution to perfectly reconstruct original images. That is, when $z$ follows the standard Gaussian distribution, we set $z=0$ rather than a random sample in the inverse computation. 
For more discussion about this please refer to Section 4.2.5.

\subsection{Combination of Image Rescaling and Compression}\label{sec:combine compression}

Our invertible framework jointly models degradation and restoration as an invertible bijective transformation. In real applications, some parts of degradation operations are not always available to adapt with restoration, e.g. for convenience. For example, the widely used image compression follows general standards, and formats such as PNG and JPEG are the most commonly used ones with well-established support in most digital devices. Therefore, we propose the combination of our invertible framework and restoration from existing degradation methods for wider applications.

Specifically, we consider the instantiation of the combination between image rescaling and compression, which is also a common method for a higher compression rate of high-resolution images~\citep{bruckstein2003down}, because direct image compression methods perform poorly under an extremely large compression rate. In this work, we demonstrate the combination between IRN and lossless as well as lossy compression methods for better lossy compression performance.

Note that it is also possible to directly generalize the invertible framework for image compression with some additional efforts. Please refer to \citep{wang2020modeling} for the preliminary attempt.

\subsubsection{Methods}

For lossless image compression methods, LR images can be encoded without information loss, therefore IRN can be directly combined with them, i.e. directly compress the downscaled LR images generated by IRN. 

For existing lossy image compression methods, there would be inevitable information loss during encoding, i.e. additional degradation caused by the lossy compression. So directly combining IRN with them , e.g. first compress LR images of IRN and then directly pass compressed images to IRN, may go against the principle of modeling lost information in the proposed invertible framework. Additional restoration for such degradation is required for good performance.

To mitigate this problem, we propose to leverage an additional module to partially restore the lost information by lossy compression methods. 
Specifically, downscaled images of IRN will first be compressed by lossy compression methods, e.g. JPEG, and the compressed image will go through a Compression Restore Module (CRM) before being passed to IRN. CRM is taken as a neural network model, whose input is the compressed LR image with degradation and output is the LR image restored from the degradation caused by lossy compression. This module is trained to restore lost information of the given compression method, which is similar to many methods considering the unidirectional restoration task. We will elaborate on the detailed architecture and evaluate the compression performance in the next section. The combination of IRN and CRM is the instantiation model of our proposed combination of invertible framework and restoration from existing degradation methods.

\begin{table*} [ht]
	\centering
	\footnotesize
	\tabcolsep=2mm
	\caption{Quantitative evaluation results (PSNR / SSIM) of different downscaling and upscaling methods for image reconstruction on benchmark datasets: Set5, Set14, BSD100, Urban100, and DIV2K validation set. For our method, differences on average PSNR / SSIM from different $z$ samples are less than 0.02. We report the mean result over 5 draws.}
	\begin{tabular}{c|c|c|c|c|c|c|c}
		\hline
		Downscaling \& Upscaling  &  Scale & Param  &  Set5  &  Set14  &  BSD100  &  Urban100  &  DIV2K \\
		
		\hline  
		\hline
		Bicubic \& Bicubic & 2$\times$ & / & 33.66 / 0.9299 & 30.24 / 0.8688 & 29.56 / 0.8431 & 26.88 / 0.8403 & 31.01 / 0.9393 \\
		
		\hline
		Bicubic \& SRCNN \citep{dong2015image} & 2$\times$ & 57.3K &  36.66 / 0.9542  &  32.45 / 0.9067  &  31.36 / 0.8879  &  29.50 / 0.8946 &   35.60 / 0.9663  \\
		
		\hline
		Bicubic \& EDSR \citep{lim2017enhanced} & 2$\times$ &  40.7M & 38.20 / 0.9606  &  34.02 / 0.9204  &  32.37 / 0.9018  &  33.10 / 0.9363 &  35.12 / 0.9699 \\
		
		\hline
		Bicubic \& RDN \citep{zhang2018residual} & 2$\times$ & 22.1M &  38.24 / 0.9614  &  34.01 / 0.9212  &  32.34 / 0.9017  &  32.89 / 0.9353 &   --  \\
		
		\hline
		Bicubic \& RCAN \citep{zhang2018image} & 2$\times$ & 15.4M &  38.27 / 0.9614  &  34.12 / 0.9216  &  32.41 / 0.9027  &  33.34 / 0.9384  &  --  \\
		
		\hline
		Bicubic \& SAN \citep{dai2019second} & 2$\times$ & 15.7M &  38.31 / 0.9620  &  34.07 / 0.9213  &  32.42 / 0.9028  &  33.10 / 0.9370  &  --  \\
		
		\hline
		TAD \& TAU \citep{kim2018task} & 2$\times$ & -- & 38.46 /  --   & 35.52 /  --  & 36.68 /  --   & 35.03 /  --  & 39.01 /  --  \\
		
		\hline
		CNN-CR \& CNN-SR \citep{li2018learning} & 2$\times$ & -- & 38.88 / -- & 35.40 / -- & 33.92 / -- & 33.68 / -- & --\\
		
		\hline
		CAR \& EDSR \citep{sun2020learned} & 2$\times$ & 51.1M & 38.94 / 0.9658 & 35.61 / 0.9404 & 33.83 / 0.9262 & 35.24 / 0.9572 & 38.26 / 0.9599 \\
		
		\hline
		IRN (ours) & 2$\times$ & 1.66M & \textcolor{red}{43.99} / \textcolor{red}{0.9871} & \textcolor{red}{40.79} / \textcolor{red}{0.9778} & \textcolor{red}{41.32} / \textcolor{red}{0.9876} & \textcolor{red}{39.92} / \textcolor{red}{0.9865} & \textcolor{red}{44.32} / \textcolor{red}{0.9908} \\

		\hline  
		\hline
		Bicubic \& Bicubic & 4$\times$ & / & 28.42 / 0.8104 & 26.00 / 0.7027 & 25.96 / 0.6675 & 23.14 / 0.6577 & 26.66 / 0.8521 \\
		
		\hline
		Bicubic \& SRCNN \citep{dong2015image} & 4$\times$ & 57.3K &  30.48 / 0.8628  &  27.50 / 0.7513  &  26.90 / 0.7101  &  24.52 / 0.7221 &   --  \\
		
		\hline
		Bicubic \& EDSR \citep{lim2017enhanced} & 4$\times$ & 43.1M & 32.62 / 0.8984 & 28.94 / 0.7901 & 27.79 / 0.7437 & 26.86 / 0.8080 & 29.38 / 0.9032 \\

		\hline
		Bicubic \& RDN \citep{zhang2018residual} & 4$\times$ & 22.3M &  32.47 / 0.8990  &  28.81 / 0.7871  &  27.72 / 0.7419  &  26.61 / 0.8028 &   --  \\
		
		\hline
		Bicubic \& RCAN \citep{zhang2018image} & 4$\times$ & 15.6M & 32.63 / 0.9002 & 28.87 / 0.7889 & 27.77 / 0.7436 & 26.82 / 0.8087 & 30.77 / 0.8460 \\
		
		\hline
		Bicubic \& ESRGAN \citep{wang2018esrgan} & 4$\times$ & 16.3M & 32.74 / 0.9012 & 29.00 / 0.7915 & 27.84 / 0.7455 & 27.03 / 0.8152 & 30.92 / 0.8486 \\
		
		\hline
		Bicubic \& SAN \citep{dai2019second} & 4$\times$ & 15.7M &  32.64 / 0.9003  &  28.92 / 0.7888  &  27.78 / 0.7436  &  26.79 / 0.8068  &  --  \\
		
		\hline
		TAD \& TAU \citep{kim2018task} & 4$\times$ & -- & 31.81 /  --  & 28.63 /  --   & 28.51 /  --   & 26.63 /  --   & 31.16 /  --   \\
		
		\hline
		CAR \& EDSR \citep{sun2020learned} & 4$\times$ & 52.8M & 33.88 / 0.9174 & 30.31 / 0.8382 & 29.15 / 0.8001 & 29.28 / 0.8711 & 32.82 / 0.8837 \\
		
		\hline
		IRN (ours) & 4$\times$ & 4.35M & \textcolor{red}{36.19} / \textcolor{red}{0.9451} & \textcolor{red}{32.67} / \textcolor{red}{0.9015} & \textcolor{red}{31.64} / \textcolor{red}{0.8826} & \textcolor{red}{31.41} / \textcolor{red}{0.9157} & \textcolor{red}{35.07} / \textcolor{red}{0.9318} \\

        \hline
        \hline
		Bicubic \& Bicubic & 8$\times$ & / & 24.40 / 0.6580 & 23.10 / 0.5660 & 23.67 / 0.5480 & 20.74 / 0.5160 & 23.70 / 0.6387 \\
		
		\hline
		Bicubic \& SRCNN \citep{dong2015image} & 8$\times$ & 57.3K &  25.33 / 0.6900  &  23.76 / 0.5910  &  24.13 / 0.5660  &  21.29 / 0.5440 &   --  \\
		
		\hline
		Bicubic \& EDSR \citep{lim2017enhanced} & 8$\times$ & -- & 26.96 / 0.7762 & 24.91 / 0.6420 & 24.81 / 0.5985 & 22.51 / 0.6221 & 25.50 / -- \\

		\hline
		Bicubic \& RCAN \citep{zhang2018image} & 8$\times$ & 15.8M & 27.31 / 0.7878 & 25.23 / 0.6511 & 24.98 / 0.6058 & 23.00 / 0.6452 & -- \\
		
		\hline
		Bicubic \& SAN \citep{dai2019second} & 8$\times$ & 15.8M &  27.22 / 0.7829  &  25.14 / 0.6476  &  24.88 / 0.6011  &  22.70 / 0.6314  &  --  \\
		
		\hline
		TAD \& TAU \citep{kim2018task} & 8$\times$ & -- & -- & --  & --  & -- & 26.77 /  --   \\
		
		\hline
		IRN (ours) & 8$\times$ & 11.1M & \textcolor{red}{31.20} / \textcolor{red}{0.8736} & \textcolor{red}{28.40} / \textcolor{red}{0.7698} & \textcolor{red}{27.49} / \textcolor{red}{0.7239} & \textcolor{red}{26.67} / \textcolor{red}{0.7947} & \textcolor{red}{30.29} / \textcolor{red}{0.8280} \\

        \hline
	    
	\end{tabular}
	
    \label{quantitative results}
\end{table*}

\section{Experiments}\label{sec:exp}

\subsection{Datasets and Settings}\label{sec:exp setting}

Our experiments include three parts: image rescaling, image decolorization-colorization, as well as the combination between image rescaling and compression. For the training of all tasks, we employ the widely used DIV2K~\citep{agustsson2017ntire} image restoration dataset to train our models. It contains 800 high-quality 2K resolution training images and 100 validation images. Besides, for the first two tasks, we evaluate our model on 4 additional standard datasets, i.e. the Set5~\citep{bevilacqua2012low}, Set14~\citep{zeyde2010single}, BSD100~\citep{martin2001database}, and Urban100~\citep{huang2015single}; and for the third task, we also evaluate our model on the widely used Kodak dataset~\citep{franzen1999kodak}. For image rescaling, following the setting in~\citep{lim2017enhanced}, we quantitatively evaluate the peak noise-signal ratio (PSNR) and SSIM~\citep{wang2004image} on the Y channel of images represented in the YCbCr (Y, Cb, Cr) color space. 
We also evaluate LPIPS~\citep{zhang2018perceptual}, PI~\citep{blau20182018}, and FID~\citep{heusel2017gans} as quantitative metrics of perceptual evaluation. 
For the other two tasks, we evaluate PSNR and SSIM on the RGB three-channel color space. 

\begin{table*} [ht]
	\centering
	\footnotesize
	\tabcolsep=2mm
	\caption{Quantitative evaluation results (PSNR / SSIM) of different 3$\times$ image downscaling and upscaling methods on benchmark datasets: Set5, Set14, BSD100, Urban100, and DIV2K validation set. For our model, differences on average PSNR / SSIM of different samples for z are less than 0.02. We report the mean result.}
	\begin{tabular}{c|c|c|c|c|c|c|c}
		
		\hline
		Downscaling \& Upscaling  &  Scale & Param  &  Set5  &  Set14  &  BSD100  &  Urban100  &  DIV2K \\
	
		\hline  
		\hline
		Bicubic \& Bicubic & 3$\times$ & / & 30.39 / 0.8682 & 27.55 / 0.7742 & 27.21 / 0.7385 & 24.46 / 0.7349 & 26.95 / 0.8556 \\
		
		\hline
		Bicubic \& SRCNN \citep{dong2015image} & 3$\times$ & 57.3K &  32.75 / 0.9090  &  29.30 / 0.8215  &  28.41 / 0.7863  &  26.24 / 0.7989 &   30.48 / 0.9117  \\
		
		\hline
		Bicubic \& EDSR \citep{lim2017enhanced} & 3$\times$ & 43.7M & 34.65 / 0.9280 & 30.52 / 0.8462 & 29.25 / 0.8093 & 28.80 / 0.8653 & 34.17 / 0.9476 \\

		\hline
		Bicubic \& RDN \citep{zhang2018residual} & 3$\times$ & 22.3M &  34.71 / 0.9296  &  30.57 / 0.8468  &  29.26 / 0.8093  &  28.80 / 0.8653 &   34.13 / 0.9484  \\
		
		\hline
		Bicubic \& RCAN \citep{zhang2018image} & 3$\times$ & 15.6M & 34.74 / 0.9299 & 30.65 / 0.8482 & 29.32 / 0.8111 & 29.09 / 0.8702 & 34.44 / 0.9499 \\
		
		\hline
		Bicubic \& SAN \citep{dai2019second} & 3$\times$ & 15.7M &  34.75 / 0.9300  &  30.59 / 0.8476  &  29.33 / 0.8112  &  28.93 / 0.8671  &  34.30 / 0.9494  \\

		\hline
		IRN$_\text{LD}$ (ours) & 3$\times$ & 3.14M & \textcolor{red}{37.94} / \textcolor{red}{0.9586} & \textcolor{red}{34.64} / \textcolor{red}{0.9313} & \textcolor{red}{33.80} / \textcolor{red}{0.9306} & \textcolor{red}{33.45} / \textcolor{red}{0.9470} & \textcolor{red}{37.33} / \textcolor{red}{0.9586} \\

        \hline
	    
	\end{tabular}
	
    \label{quantitative results of IRN_LD}
    
\end{table*}

\begin{table*} [ht]
	\centering
	\footnotesize
	\caption{Quantitative perceptual evaluation results of different 4$\times$ image downscaling and upscaling methods on benchmark datasets: Set5, Set14, BSD100, Urban100, and DIV2K validation set. For LPIPS, PI, and FID, lower is better. The best result is in red, and the second best result is in blue.}
	\begin{tabular}{c|c|c|c|c|c}
		
		\hline
		\multirow{2}*{(Metrics)} & PSNR / SSIM & PSNR / SSIM & PSNR / SSIM & PSNR / SSIM & PSNR / SSIM \\
		& LPIPS / PI / FID & LPIPS / PI / FID & LPIPS / PI / FID & LPIPS / PI / FID & LPIPS / PI / FID \\
	
		\hline  
		Downscaling \& Upscaling  &  Set5  &  Set14  &  BSD100  &  Urban100  &  DIV2K \\
		\hline
		\hline
		\multirow{2}*{Bicubic \& ESRGAN} & 32.74 / 0.9012 & 29.00 / 0.7915 & 27.84 / 0.7455 & 27.03 / 0.8152 & 30.92 / 0.8486 \\
		
		& 0.169 / 6.095 / 53.87 & 0.273 / 5.342 / 74.75 & 0.358 / 5.190 / 93.1 & 0.198 / 5.041 / 24.41 & 0.256 / 5.274 / 15.91 \\

		\hline
		\multirow{2}*{Bicubic \& ESRGAN+} &  30.57 / 0.8561  &  26.39 / 0.7054  &  25.52 / 0.6618  &  24.48 / 0.7420  &  28.17 / 0.7759  \\
		
		& \textcolor{blue}{0.076} / \textcolor{blue}{3.842} / \textcolor{blue}{27.61} & 0.133 / \textcolor{red}{2.944} / 55.17 & \textcolor{blue}{0.165} / \textcolor{blue}{2.494} / 49.00 & 0.126 / \textcolor{blue}{3.740} / 20.75 & \textcolor{blue}{0.115} / \textcolor{red}{3.202} / 13.56 \\

		\hline
		\hline
		\multirow{2}*{IRN (ours)} & \textcolor{red}{36.19} / \textcolor{red}{0.9451} & \textcolor{red}{32.67} / \textcolor{red}{0.9015} & \textcolor{red}{31.64} / \textcolor{red}{0.8826} & \textcolor{red}{31.41} / \textcolor{red}{0.9157} & \textcolor{red}{35.07} / \textcolor{red}{0.9318} \\
		
		& 0.078 / 4.195 / 33.88 & \textcolor{blue}{0.123} / 3.635 / \textcolor{blue}{35.96} & 0.166 / 3.069 / \textcolor{blue}{42.11} & \textcolor{blue}{0.084} / 4.021 / \textcolor{red}{9.13} & 0.119 / 3.804 / \textcolor{red}{5.78} \\
		
        \hline
        
		\multirow{2}*{IRN+ (ours)} & \textcolor{blue}{33.59} / \textcolor{blue}{0.9147} & \textcolor{blue}{29.97} / \textcolor{blue}{0.8444} & \textcolor{blue}{28.94} / \textcolor{blue}{0.8189} & \textcolor{blue}{28.24} / \textcolor{blue}{0.8684} & \textcolor{blue}{32.24} / \textcolor{blue}{0.8921} \\
		
		& \textcolor{red}{0.031} / \textcolor{red}{3.382} / \textcolor{red}{11.15} & \textcolor{red}{0.067} / \textcolor{blue}{2.952} / \textcolor{red}{32.38} & \textcolor{red}{0.074} / \textcolor{red}{2.398} / \textcolor{red}{22.06} & \textcolor{red}{0.055} / \textcolor{red}{3.541} / \textcolor{blue}{13.00} & \textcolor{red}{0.054} / \textcolor{blue}{3.240} / \textcolor{blue}{7.90} \\
        \hline
	    
	\end{tabular}
	
    \label{perceptual accessment results}
    
\end{table*}

For image rescaling, we train and compare our IRN model in $2\times$, $4\times$ and $8\times$ downscaling scale with 1, 2, and 3 downscaling modules respectively. Each downscaling module has 8 InvBlocks and downscales the original image by $2\times$. The transformation functions $\phi(\cdot), \eta(\cdot), \rho(\cdot)$ in InvBlocks are parameterized by a densely connected convolutional block, which is referred to as Dense Block in~\cite{wang2018esrgan}. For experiments of IRN$_\text{LD}$ model in $3\times$ scale, we use one downscaling module with learnable downsampling and 12 InvBlocks. For experiments of IRN$_\text{E}$ model in $4\times$ scale, we use one downscaling module with 16 InvBlocks (downscaling first). We use Adam optimizer~\citep{kingma2014adam} with $\beta_1=0.9, \beta_2=0.999$ to train our model. The mini-batch size is set to 16. The input HR image is randomly cropped into $144 \times 144$ and augmented by applying random horizontal and vertical flips. In the pre-training stage, the total number of iteration is $500K$, and the learning rate is initialized as $2\times10^{-4}$ where halved at $[100k, 200k, 300k, 400k]$ mini-batch updates. The hyper-parameters in Eq.~(\ref{eq.IRN objective function}) are set as $\lambda_1=1,\lambda_2=s^2,\lambda_3=1$, where $s$ denotes the scale. After pre-training, we finetune our model for another $200K$ iterations as described in Section~\ref{sec:train-method-irn}. The learning rate is initialized as $1\times10^{-4}$ and halved at $[50k, 100k]$ iterations. We set $\lambda_1=0.01,\lambda_2=s^2, \lambda_3=1, \lambda_4=0.01$ in Eq.~(\ref{eq.IRN+ objective function}) and pre-train the discriminator for 5000 iterations. The discriminator is similar to \cite{ledig2017photo}, which contains eight convolutional layers with $3\times3$ kernels, whose numbers increase from 64 to 512 by a factor of 2 every two layers, and two dense layers with 100 hidden units.

For image decolorization-colorization, the graying module has 8 InvBlocks. The hyper-parameters are set as $\lambda_1=1,\lambda_2=9,\lambda_3=1$. Other optimizers and iteration settings are the same as image rescaling.

For combination with image compression, we leverage the IRN$_{2\times}$ model trained in image rescaling task and further finetune it for $100K$ iterations in the rescaling task by adding a random noise on the generated LR images during upscaling in training, in order to make the model more robust to possible changes on LR images due to compression and restoration. The model for Kodak is additionally finetuned for $2.5K$ iterations on Kodak. We train a compression restore module (CRM) for each compression ratio of JPEG. The CRM contains 8 residual in residual dense blocks (RRDB) proposed in~\citep{wang2018esrgan}, and is trained by a $L_2$ loss on reconstructed LR images and LR images before compression. The optimizer and iteration settings are the same as IRN.

\subsection{Image Rescaling}
\subsubsection{Evaluation on Reconstructed HR Images}

\begin{figure*} [ht]
    \centering
    \includegraphics[scale=0.115]{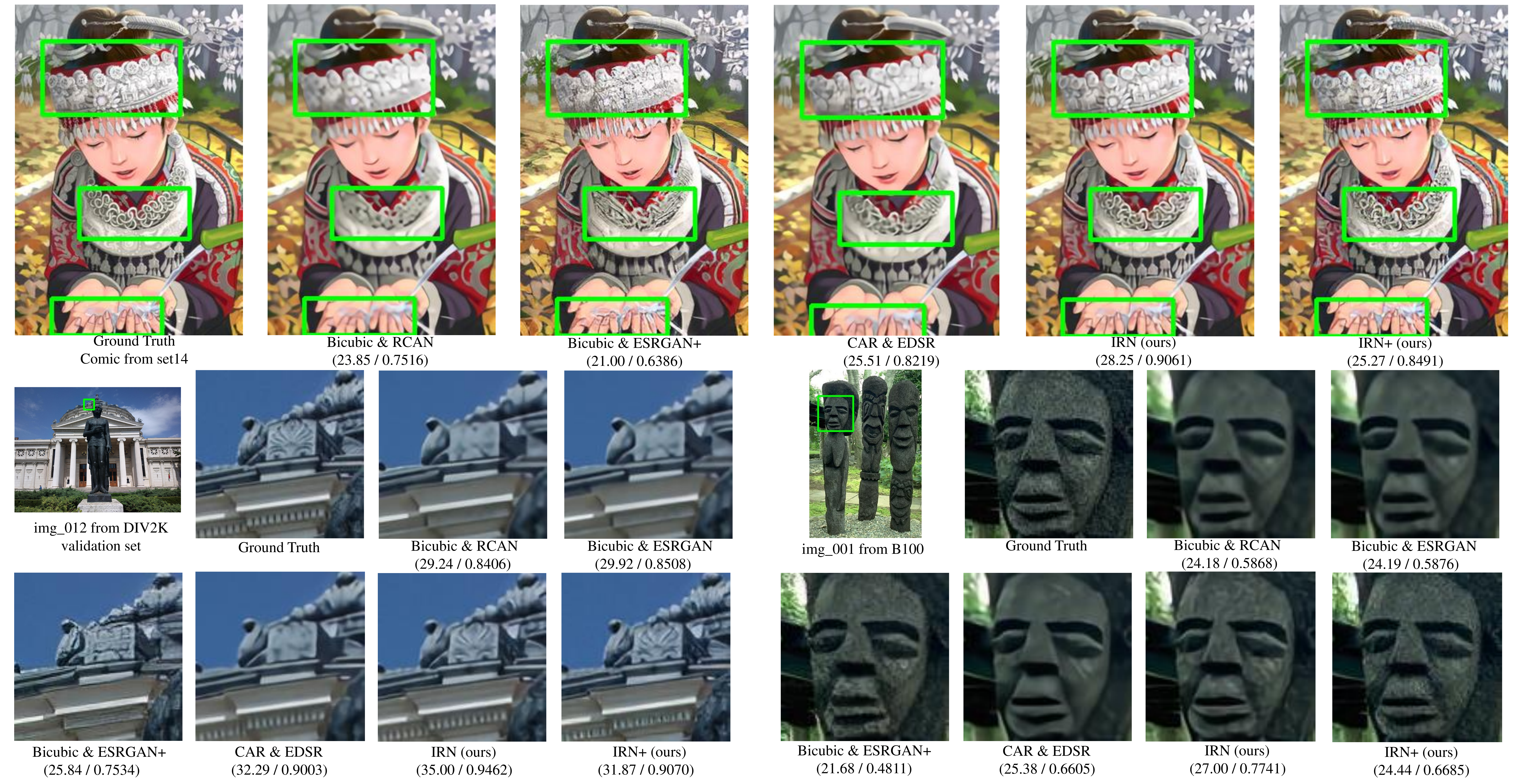}
    \caption{Qualitative results of upscaling the $4\times$ downscaled images. IRN recovers rich details, leading to both visually pleasing performance and high similarity to the original images. IRN+ produces even sharper and more realistic details. See the appendix for more results.}
    \label{fig:qualitative results}
\end{figure*}

\begin{figure*} [ht]
    \centering
    \subfigure[]{
    \includegraphics[scale=0.05]{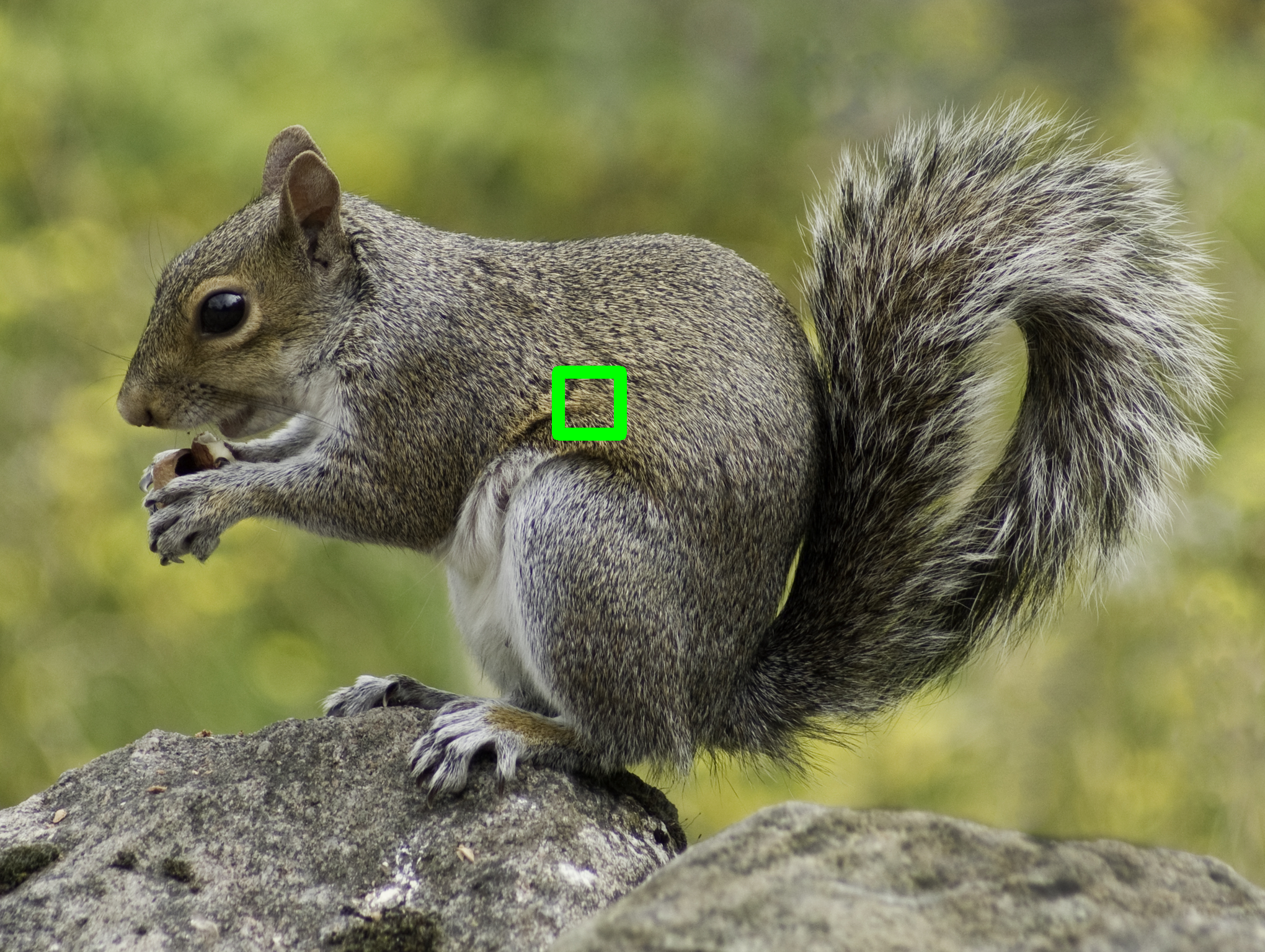}
    \label{ground truth}
    }
    \subfigure[]{
    \includegraphics[scale=0.775]{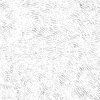}
    \label{detail1}
    }
    \subfigure[]{
    \includegraphics[scale=0.775]{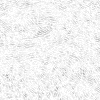}
    \label{detail2}
    }
    \subfigure[]{
    \includegraphics[scale=0.775]{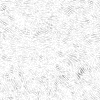}
    \label{detail3}
    }

    \caption{Visualisation of the difference of upscaled HR images from multiple draws of $z$. (a): original image;
    (b-d): HR image differences of three $z$ samples from another common $z$ sample. Darker color means larger difference. It shows that the differences are high-frequency noises in high-frequency regions without a typical texture.
    }
    \label{fig:different samples}
\end{figure*}

In this section, we present the quantitative and qualitative performance of HR images reconstructed by our model and other downscaling and upscaling methods. Two kinds of baselines are considered: (1) downscaling with Bicubic interpolation and upscaling with state-of-the-art SR models trained with this downscaling kernel~\citep{dong2015image,lim2017enhanced,zhang2018residual,zhang2018image,wang2018esrgan,dai2019second}; (2) downscaling with upscaling-optimal models~\citep{kim2018task,li2018learning,sun2020learned} and upscaling with corresponding SR models. For the notations, we identify the downscaling and upscaling methods respectively for baselines while use IRN or IRN+ as a whole to denote our invertible model for the bidirectional tasks; and following our notation, we use ESRGAN to represent the pre-trained PSNR-driven model of~\cite{wang2018esrgan} while ESRGAN+ for their GAN-based perceptual-driven model. In addition, the influence of different samples of $z$ on our reconstructed HR images and the effectiveness of different types of loss in the pre-training stage are investigated.

\textbf{Quantitative Results}\quad
As shown in table~\ref{quantitative results}, IRN significantly outperforms the state-of-the-art baseline models regarding quantitative evaluation PSNR and SSIM in all datasets. Although upscaling-optimal downscaling methods largely enhance the reconstruction performance of SR models compared with Bicubic interpolation due to the unification of bidirectional tasks, they still suffer from the ill-posed problem caused by information loss and therefore the results are hardly satisfying. Contrarily, by modeling the lost information with invertibility, IRN significantly boosts the PSNR with about 4-5 dB, 2-3 dB, and 3-4 dB on each dataset under $2\times$, $4\times$, and $8\times$ scale respectively compared with the state-of-the-art results, where the improvement is up to 5.94 dB. The PSNR results indicate an exponential reduction of information loss due to its logarithmic computation, which is consistent with the significant improvement of SSIM. The results of IRN+ are in the appendix because it is visual perception oriented. IRN$_\text{LD}$ extends IRN to more flexible downscaling and upscaling scales. Table~\ref{quantitative results of IRN_LD} demonstrates the significant improvement of IRN$_\text{LD}$ on $3\times$ scale as well, with about 3-5 dB improvement on the PSNR metric compared with other methods.

It is noteworthy that IRN achieved the best results with a relatively small amount of parameters. When upscaling with SR models, it requires more than 15M parameters for better results, while the model sizes of our IRN are only 1.66M, 4.35M, and 11.1M in the scale $2\times$, $4\times$, and $8\times$. It indicates the lightweight property and high efficiency of our proposed invertible model.

We also quantitatively evaluate the perceptual performance as shown in Table~\ref{perceptual accessment results}. LPIPS and PI are full-reference and no-reference methods for perceptual evaluation of each image respectively, and FID is the metric for the perceptual similarity between two groups of images. We compare IRN and IRN+ with the representative PSNR-driven model ESRGAN and perceptual-driven model ESRGAN+, and the results demonstrate significant improvements of our models. Particularly, IRN+ with full distribution matching and perceptual loss achieves the best result considering both PSNR/SSIM and perceptual indexes, which also accords with the qualitative results below.

\begin{figure*} [ht]
    \centering
    \subfigure[IRN]{
        \includegraphics[scale=0.4]{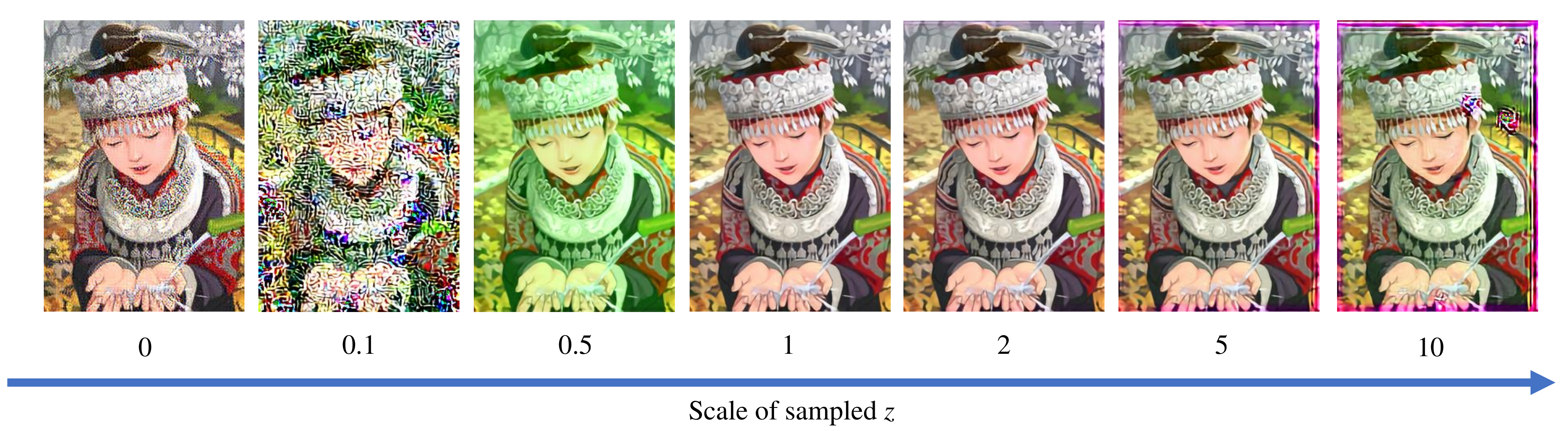}
    }
    \subfigure[IRN+]{
        \includegraphics[scale=0.4]{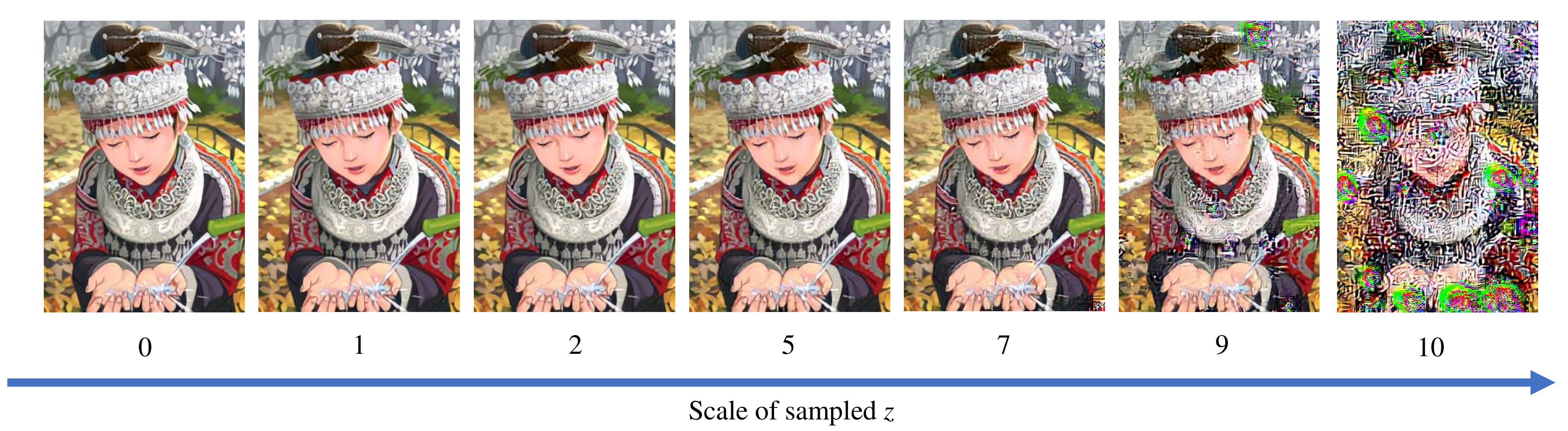}
    }
    \caption{Results of HR images by IRN and IRN+ with out-of-distribution samples of $z$. We train $z$ with an isotropic Gaussian distribution, and illustrate upscaling results when scaling $z$ sampled from the isotropic Gaussian distribution.}
    \label{fig:scale results}
\end{figure*}

\begin{table*} [ht]
	\centering
	\small
	\caption{Analysis results (PSNR/SSIM) of training IRN with $L_1$ or $L_2$ LR guide and HR reconstruction loss, with/without partial distribution matching loss, on Set5, Set14, BSD100, Urban100 and DIV2K validation sets with scale 4$\times$.}
	\begin{tabular}{c|c|c|c|c|c|c|c}
		\hline
		$L_{guide}$ & $L_{recon}$ & $L_{distr'}$ & Set5 & Set14 & BSD100 & Urban100 & DIV2K \\
		\hline
		\hline
		$L_1$ & $L_1$ & Yes & 34.75 / 0.9296 & 31.42 / 0.8716 & 30.42 / 0.8451 & 30.11 / 0.8903 & 33.64 / 0.9079 \\
		\hline
		$L_1$ & $L_2$ & Yes & 34.93 / 0.9296 & 31.76 / 0.8776 & 31.01 / 0.8562 & 30.79 / 0.8986 & 34.11 / 0.9116 \\
		\hline
		$L_2$ & $L_1$ & Yes & \textcolor{red}{36.19} / 0.9451 & \textcolor{red}{32.67} / \textcolor{red}{0.9015} & \textcolor{red}{31.64} / \textcolor{red}{0.8826} & \textcolor{red}{31.41} / \textcolor{red}{0.9157} & \textcolor{red}{35.07} / \textcolor{red}{0.9318} \\
		\hline
		$L_2$ & $L_2$ & Yes & 35.93 / 0.9402 & 32.51 / 0.8937 & 31.64 / 0.8742 & 31.40 / 0.9105 & 34.90 / 0.9308 \\
		\hline
		\hline
		$L_2$ & $L_1$ & No &  36.12 / \textcolor{red}{0.9455} & 32.18 / 0.8995 & 31.49 / 0.8808 & 30.91 / 0.9102 & 34.90 / 0.9308 \\
		\hline
		
	\end{tabular}
	\label{ablation_loss}
\end{table*}
    
\begin{table*} [ht]
	\centering
	\small
	\caption{Analysis results (PSNR/SSIM) of training IRN with different loss weights for HR reconstruction and LR guidance loss, for image reconstruction on Set5, Set14, BSD100, Urban100 and DIV2K validation sets with scale 4$\times$.}
	\begin{tabular}{c|c|c|c|c|c|c}
		\hline
		$\lambda_1$ & $\lambda_2$ & Set5 & Set14 & BSD100 & Urban100 & DIV2K \\
		\hline
		\hline
		$1$ & $16$ & \textcolor{red}{36.19} / \textcolor{red}{0.9451} & \textcolor{red}{32.67} / \textcolor{red}{0.9015} & \textcolor{red}{31.64} / \textcolor{red}{0.8826} & \textcolor{red}{31.41} / \textcolor{red}{0.9157} & \textcolor{red}{35.07} / \textcolor{red}{0.9318} \\
		\hline
		$1$ & $160$ & 35.94 / 0.9439 & 32.32 / 0.8961 & 31.40 / 0.8757 & 31.26 / 0.9121 & 34.81 / 0.9276 \\
		\hline
		$1$ & $1.6$ & 35.72 / 0.9391 & 32.06 / 0.8863 & 31.14 / 0.8676 & 30.52 / 0.8992 & 34.47 / 0.9221 \\
		\hline
		
	\end{tabular}
	\label{analysis loss weight}
\end{table*}
    
\begin{table*} [ht]
	\centering
	\small
	\caption{Analysis results (PSNR/SSIM) between the LR images downscaled by IRN trained by different loss weights and by Bicubic on Set5, Set14, BSD100, Urban100 and DIV2K validation sets with scale 4$\times$.}
	\begin{tabular}{c|c|c|c|c|c|c}
		\hline
		$\lambda_1$ & $\lambda_2$ & Set5 & Set14 & BSD100 & Urban100 & DIV2K \\
		\hline
		\hline
		$1$ & $16$ & 44.60 / 0.9964 & 42.47 / 0.9928 & 43.24 / 0.9923 & 41.28 / 0.9916 & 44.37 / 0.9933 \\
		\hline
		$1$ & $160$ & 50.14 / 0.9988 & 47.57 / 0.9977 & 48.62 / 0.9976 & 47.46 / 0.9977 & 50.06 / 0.9980 \\
		\hline
		$1$ & $1.6$ & 34.25 / 0.9820 & 34.00 / 0.9764 & 35.59 / 0.9755 & 33.40 / 0.9720 & 35.59 / 0.9782 \\
		\hline
		
	\end{tabular}
	\label{analysis loss weight LR}
\end{table*}

\textbf{Qualitative Results}\quad
When it comes to qualitative evaluation, we visually demonstrate the details of the upscaled images by different methods. Fig.~\ref{fig:qualitative results} demonstrates the better visual quality and fidelity of our IRN and IRN+ model compared with previous state-of-the-art methods. IRN could recover richer details, while IRN+ further produces sharper and more realistic images, leading to their pleasing visual quality. For instance, IRN and IRN+ are the only models that are able to reconstruct the 'Comic' image with the complicated textures on the headwear and necklace, as well as the sharp and realistic fingers. Previous perceptual-driven models such as ESRGAN+, however, would produce unreasonable and unpleasing details, leading to great dissimilarity. The better results of our models owe to the modeling of information loss, as well as the distribution matching and perceptual loss for IRN+. More visual results are in the appendix.

\textbf{Visualisation on the Influence of $z$}\quad
We further investigate the influence of random $z$. As described in Section~\ref{sec:model-spec}, different samples of $z\sim p(z)$ aim to only focus on the randomness of reasonable high-frequency contents. Visually, we calculate and visualize the difference between different draws of $z$ in Fig.~\ref{fig:different samples}. It shows that only a tiny noisy distinction without typical textures is observed in high-frequency regions, which are almost imperceptible if combined with low-frequency contents. Quantitatively, different samples of $z$ result in the PSNR difference that is less than 0.02 dB for each image, which also indicates that the randomness mainly lies in high-frequency noise. These results indicate that our models have learned the knowledge to restore meaningful lost high-frequency contents while embedding imperceptible noises into the randomness of distribution.

Additionally, we test our model with out-of-distribution samples to verify its effectiveness and sensitivity. Our models are trained with $p(z)$ being an isotropic Gaussian distribution, and we test IRN and IRN+ by inversely passing $(y, \alpha z)$ to obtain $x_\alpha$ with the control of the scale $\alpha$ of sampled $z\sim p(z)$. Note that the probability density for samples with $\alpha<1$ is still large for the Gaussian distribution, e.g. the point of $z=0$ has the largest probability density, and therefore the reconstruction should still be valid if distribution matching is fully realized. As shown in Fig.~\ref{fig:scale results}, IRN+ could validly reconstruct HR images when the sampled $z$ lie in areas with a large probability density or with small disturbance, and more noisy textures and degradations would appear when there is a larger deviation from the original distribution. This indicates that IRN+ fully realizes the distribution matching for $p(z)$ and is robust to mild deviation. On the other hand, IRN without the full distribution matching objective fails to validly reconstruct HR images when the scale $\alpha\neq 1$, which indicates that it only learns to validly reconstruct images by $z$ around the areas with a large density of training samples rather than the full distribution. This demonstrates the effectiveness of our full distribution matching objective.

\textbf{Analysis on the Losses}\quad
We also conduct analysis experiments for the losses of Eqs.~(\ref{eq.guide loss}, \ref{eq.recon loss}, \ref{eq.distr loss'}), which is shown in Table~\ref{ablation_loss}, Table~\ref{analysis loss weight} and Table~\ref{analysis loss weight LR}. We can see from Table~\ref{ablation_loss} that when the LR guidance takes the $L_2$ loss while the HR reconstruction is the $L_1$ loss, IRN gets the best training performance. The underlying explanation is that our forward procedure aims to learn a valid downscaling transformation that is beneficial to the inverse upscaling, rather than exactly the Bicubic downscaling, so the $L_2$ loss that is less sensitive to minor changes from the guidance would be more suitable; while the goal of our inverse procedure is to accurately reconstruct the original HR image, thus the $L_1$ loss encouraging more pixel-wise similarity is profitable. The results also demonstrate the improvement brought by our surrogate partial distribution matching loss (Eq.~\eqref{eq.distr loss'}), which acts on the marginal distribution on $\clZ$ to encourage the forward distribution learning.

As described in Section~\ref{sec:exp setting}, our default weights for HR reconstruction and LR guidance loss are $\lambda_1=1$ and $\lambda_2=s^2$ in order to keep the losses on the same scale. To further justify the choice, we study the weights with different scales of ratios. We conduct analysis experiments with IRN in $4\times$ scale. The original weights are $\lambda_1=1$, $\lambda_2=16$, we largely increase or decrease the weight for LR guidance, i.e. $\lambda_2=160$ or $\lambda_2=1.6$. The evaluation results on image reconstruction are shown in Table~\ref{analysis loss weight}. It shows that the reconstruction quality is quite robust to the ratio between HR reconstruction and LR guidance, and the original weights that keep the losses on the same scale achieve the best results. We also compare the images downscaled by IRN trained by different loss weights with those downscaled by Bicubic to verify the validity of LR images. The results are in Table~\ref{analysis loss weight LR}. It shows that the LR similarity is strongly correlated with the ratio of LR guidance loss, and the larger the loss is, the more similar LR images are. When $\lambda_2=16$, it is enough to keep the LR images valid due to the strong similarity (PSNR$>$40, SSIM$>$0.99), and setting $\lambda_2=160$ could improve the LR similarity but not HR reconstruction quality. When $\lambda_2=1.6$, however, the LR similarity is significantly dropped, and there could be slight artifacts on the LR images on the validation datasets, which hamper the HR reconstruction. As a result, the reconstruction performance of $\lambda_2=1.6$ is the worst. Therefore, keeping the losses on the same scale as the original setting is the best choice for our model.

\begin{table} [ht]
	\centering
	\small
	\tabcolsep=1mm
	\caption{SSIM results between the images downscaled by IRN and by Bicubic on the Set5, Set14, BSD100, Urban100 and DIV2K validation sets.}
	\begin{tabular}{c|c|c|c|c|c}
		\hline
		Scale & Set5 & Set14 & BSD100 & Urban100 & DIV2K \\
		\hline
		\hline
		$2\times$ & 0.9957 & 0.9936 & 0.9936 & 0.9941 & 0.9945 \\
		\hline
		$4\times$ & 0.9964 & 0.9927 & 0.9923 & 0.9916 & 0.9933 \\
		\hline
		$8\times$ & 0.9958 & 0.9926 & 0.9918 & 0.9879 & 0.9919 \\
		\hline
		
	\end{tabular}
	\label{lr_ssim}
\end{table}

\begin{table*} [ht]
	\centering
	\small
	\tabcolsep=1mm
	\caption{Ablation study on the invertibility. Quantitative results (PSNR/SSIM) for $4\times$ scale on the Set5, Set14, BSD100, Urban100 and DIV2K validation sets are reported.}
	\begin{tabular}{c|c|c|c|c|c|c}
		\hline
		Downscaling \& Upscaling & Param & Set5 & Set14 & BSD100 & Urban100 & DIV2K \\
		\hline
		\hline
		IRN & 4.35M & 36.19 / 0.9451 & 32.67 / 0.9015 & 31.64 / 0.8826 & 31.41 / 0.9157 & 35.07 / 0.9318 \\
		\hline
		Bicubic \& IRN-U & 4.35M & 32.03 / 0.8930 & 28.54 / 0.7800 & 27.52 / 0.7336 & 25.97 / 0.7801 & 30.37 / 0.8358 \\
		\hline
		IRN-D* \& ESRGAN$_s$ & 4.35+4.47M & 35.14 / 0.9365 & 31.47 / 0.8807 & 30.61 / 0.8588 & 29.62 / 0.8903 & 33.71 / 0.9150 \\
		\hline
		IRN-D* \& ESRGAN & 4.35+16.3M & 35.87 / 0.9432 & 32.31 / 0.8963 & 31.37 / 0.8775 & 30.98 / 0.9116 & 34.75 / 0.9288 \\
		\hline
		\midrule
		\hline
		IRN-D \& IRN-U (tiny) & 1.09M & 34.87 / 0.9283 & 31.34 / 0.8721 & 30.47 / 0.8510 & 29.39 / 0.8790 & 33.49 / 0.9061 \\
		\hline
		IRN (tiny) & 1.09M & \textbf{35.64} / \textbf{0.9402} & \textbf{32.00} / \textbf{0.8891} & \textbf{31.12} / \textbf{0.8698} & \textbf{30.36} / \textbf{0.8994} & \textbf{34.41} / \textbf{0.9230} \\
		\hline
		\hline
		IRN-D \& IRN-U (small) & 2.18M & 35.88 / 0.9432 & 32.31 / 0.8959 & 31.31 / 0.8755 & 30.65 / 0.9060 & 34.63 / 0.9267 \\
		\hline
		IRN (small) & 2.18M & \textbf{36.04} / \textbf{0.9432} & \textbf{32.49} / \textbf{0.8955} & \textbf{31.45} / \textbf{0.8764} & \textbf{31.13} / \textbf{0.9102} & \textbf{34.84} / \textbf{0.9279} \\
		\hline
		\hline
		IRN-D \& IRN-U & 4.35M & 35.93 / 0.9418 & 32.57 / 0.8974 & 31.41 / 0.8750 & 31.31 / 0.9124 & 34.77 / 0.9265 \\
		\hline
		IRN & 4.35M & \textbf{36.19} / \textbf{0.9451} & \textbf{32.67} / \textbf{0.9015} & \textbf{31.64} / \textbf{0.8826} & \textbf{31.41} / \textbf{0.9157} & \textbf{35.07} / \textbf{0.9318} \\
		\hline
		\hline
		IRN-D \& IRN-U (large) & 8.70M & 36.21 / 0.9450 & 32.84 / 0.9008 & 31.57 / 0.8772 & 31.59 / 0.9169 & 35.05 / 0.9297 \\
		\hline
		IRN (large) & 8.70M & \textbf{36.32} / \textbf{0.9461} & \textbf{32.86} / \textbf{0.9032} & \textbf{31.74} / \textbf{0.8845} & \textbf{31.59} / \textbf{0.9179} & \textbf{35.18} / \textbf{0.9330} \\
		\hline
		
	\end{tabular}
	\label{ablation on invertibility}
\end{table*}

\begin{table*} [ht]
	\centering
	\footnotesize
	\tabcolsep=1mm
	\caption{Computation efficiency results of different methods for downscaling or upscaling images by different scales, with the HR image size 1920$\times$1080.}
	\begin{tabular}{c|c|c|c|c|c|c}
		\hline
		Downscaling \& Upscaling Method & Scale & Param (Down+Up) & FLOPs (Down) & FLOPS (Up) & RunTime (ms) (Down) & RunTime (ms) (Up) \\
		\hline
		\hline
		Bicubic \& RCAN~\citep{zhang2018image} & 2$\times$ & 15.4M & / & 7.96$\times 10^{12}$ & / & 2188 \\
		\hline
		Bicubic \& ESRGAN~\citep{wang2018esrgan} & 2$\times$ & 16.7M & / & 9.31$\times 10^{12}$ & / & 2251 \\
		\hline
		CAR \& EDSR~\citep{sun2020learned} & 2$\times$ & 10.7M + 40.73M & 2.12$\times 10^{12}$ & 2.11$\times 10^{13}$ & 228 & 2476 \\
		\hline
		IRN (ours) & 2$\times$ & 1.67M & 8.66$\times 10^{11}$ & 8.66$\times 10^{11}$ & 344 & 347 \\
		\hline
		\hline
		Bicubic \& RCAN~\citep{zhang2018image} & 4$\times$ & 15.6M & / & 2.07$\times 10^{12}$ & / & 633 \\
		\hline
		Bicubic \& ESRGAN~\citep{wang2018esrgan} & 4$\times$ & 16.7M & / & 2.33$\times 10^{12}$ & / & 593 \\
		\hline
		CAR \& EDSR~\citep{sun2020learned} & 4$\times$ & 9.89M + 43.09M & 8.97$\times 10^{11}$ & 6.52$\times 10^{12}$ & 107 & 706 \\
		\hline
		IRN (ours) & 4$\times$ & 4.36M & 1.21$\times 10^{12}$ & 1.21$\times 10^{12}$ & 515 & 521 \\
		\hline
		IRN$_\text{E}$ (ours) & 4$\times$ & 5.37M & 6.97$\times 10^{11}$ & 6.97$\times 10^{11}$ & 264 & 269 \\
		\hline
		
	\end{tabular}
	\label{computation efficiency}
\end{table*}

\begin{table*} [ht]
	\centering
	\small
	\tabcolsep=1mm
	\caption{Quantitative results (PSNR/SSIM) of IRN and IRN$_\text{E}$ for $4\times$ scale on the Set5, Set14, BSD100, Urban100 and DIV2K validation sets.}
	\begin{tabular}{c|c|c|c|c|c|c}
		\hline
		Downscaling \& Upscaling & Param & Set5 & Set14 & BSD100 & Urban100 & DIV2K \\
		\hline
		\hline
		IRN & 4.35M & 36.19 / 0.9451 & 32.67 / 0.9015 & 31.64 / 0.8826 & 31.41 / 0.9157 & 35.07 / 0.9318 \\
		\hline
		IRN$_\text{E}$ & 5.37M & 35.52 / 0.9393 & 32.14 / 0.8935 & 31.17 / 0.8777 & 30.65 / 0.9107 & 34.53 / 0.9282 \\
		\hline
		
	\end{tabular}
	\label{irn and irn_e}
\end{table*}

\subsubsection{Evaluation on Downscaled LR Images}

To verify the validity of our downscaling, we evaluate the quality of IRN-downscaled LR images. Table~\ref{lr_ssim} demonstrates the similarity index SSIM between our LR images and Bicubic-based LR images. It quantitatively shows that the images are extremely similar to each other. More figures in the appendix illustrate the visual similarity between the images, demonstrating the proper and valid visual perception of our LR images similar to Bicubic-based ones. Therefore, the downscaling of IRN can perform as well and valid as the guidance Bicubic interpolation.

\subsubsection{Ablation on Invertibility}

To further demonstrate the effectiveness of the proposed invertible framework, we conduct ablation comparisons by simply leveraging IRN architecture to upscale Bicubic-downscaled images (we denote the model as IRN-U), by training existing SR models to upscale IRN-downscaled images (IRN model is pre-trained and we denote it as IRN-D* here), and by joint training separate IRN-D and IRN-U models in an encoder-decoder framework.

For the first experiment, we pad $z$ by 0 to keep the dimension in order to train the model. As shown in Table~\ref{ablation on invertibility}, simply training the architecture of IRN on Bicubic-downscaled images fails to reach a satisfactory performance. This illustrates that our improvement is not from network architecture or capacity.

For the second experiment, we train the ESRGAN model~\citep{wang2018esrgan} (one of the state-of-the-art SR models with codes, we use its PSNR-driven model) on LR images downscaled by pre-trained IRN. We train a small model with similar parameters with IRN (we denote it as ESRGAN$_s$), and a model with original capacity. As shown in Table~\ref{ablation on invertibility}, without our invertible framework, the performance will drop much even if more parameters are used.

For the third experiment, we train IRN-D \& IRN-U and IRN under different amount of parameters. As shown in Table~\ref{ablation on invertibility}, without invertibility, separate IRN-D \& IRN-U models achieve much lower performance, especially when the amount of parameters is small. This illustrates the improvement by our invertible framework, as well as the highly efficient utilization of parameters that enables lightweight models.

\subsubsection{Computation Efficiency}

The previous results demonstrate the lightweight property of IRN considering parameters. We further compare detailed computation efficiency between IRN and other methods with available open-source code. We demonstrate the results of 2$\times$ and 4$\times$ here. 

We calculate the FLOPs and RunTime for models to downscale or upscale images, setting the size of high-resolution images as $1920\times 1080$, and running on one Tesla-P100 GPU. All methods are implemented in PyTorch, except CAR~\citep{sun2020learned} which is partially in CUDA code. 
As shown in Table~\ref{computation efficiency}, IRN demonstrates overall computation efficiency.

IRN$_\text{E}$ could improve computation efficiency for larger scales that require multiple downscaling modules in IRN. As shown in Table~\ref{computation efficiency}, in $4\times$ scale, IRN$_\text{E}$ could reduce about $50\%$ of FLOPS and RunTime. Table~\ref{irn and irn_e} shows the performance of IRN$_\text{E}$. There might exists a balance between computation efficiency and performance.

\subsubsection{Discussion on Randomness of $z$}

\begin{table*} [ht]
	\centering
	\footnotesize
	\caption{Quantitative evaluation results (PSNR / SSIM) of IRN and IRN ($z=0$) on benchmark datasets: Set5, Set14, BSD100, Urban100, and DIV2K validation set.}
	\begin{tabular}{c|c|c|c|c|c|c|c}
		
		\hline
		Downscaling \& Upscaling  &  Scale & Param  &  Set5  &  Set14  &  BSD100  &  Urban100  &  DIV2K \\
	
		\hline  
		\hline
	
		IRN & 4$\times$ & 4.35M & 36.19 / 0.9451 & 32.67 / 0.9015 & 31.64 / 0.8826 & 31.41 / 0.9157 & 35.07 / 0.9318 \\
        
		\hline
		IRN ($z=0$) & 4$\times$ & 4.35M & 36.23 / 0.9463 & 32.70 / 0.9019 & 31.63 / 0.8832 & 31.22 / 0.9137 & 35.04 / 0.9321 \\
		
        \hline
	    
	\end{tabular}
	
    \label{discussion on randomness of z}
    
\end{table*}

In this subsection, we would like to have some discussions on the randomness of $z$ and the current implementation of our model. 

First, when there is information loss, restoration would certainly contain randomness due to the uncertainty. To fully model the information loss from the perspective of statistical modeling, we have to leverage a random latent variable $z$ and learn the bijective distribution transformation between the distribution of $x$ and the joint distribution of $y$ and $z$, and the randomness of $z$ corresponds to randomness of reasonable lost contents.

As for our IRN model, which is in the pre-training stage without the full distribution matching objective and is different from IRN+, it does not fully model the full distribution, but only around the density of training samples of $z$ (see the paragraph \textbf{Visualisation on the Influence of $z$} in Section 4.2.1). So for this model, an alternative to not consider the randomness, e.g. taking $z=0$ which has the largest probability density in the Gaussian distribution, may be still valid considering the density on this point, as shown in Table~\ref{discussion on randomness of z}. Note that this only encourages the point with the largest probability density to recover an HR image, and it degrades the bijective transformation between two distributions into the bijective transformation between two points (i.e. it does not model the distribution or consider randomness by choosing only one preferred point in the distribution). In this setting, the losses for IRN may correspond to the losses to match data points. The results show that our invertible model is valid for this degraded condition as well.

However, our general goal is to model the full distribution as IRN+, which is a more general case and has more potential. For example, the reconstructed HR images should have many different possible realistic high-frequency details, and our general framework has the potential to model such diversity according to the randomness of $z$. 

In our current experiments, because the training dataset does not contain enough such diversity information, e.g. different perceptible high-frequency textures of similar low-frequency contents, and one of our main training objectives during pre-training is to encourage the pixel level similarity of reconstructed and original HR images, the diversity with different $z$ mainly lies in the randomness of imperceptible high-frequency details, and the PSNR scores are similar. In potential future applications, it is possible for realistic diversities with proper datasets.

In this work, we present our general invertible framework that can model the full distribution of lost information, which may have more potential future applications.

\subsection{Invertible Image Decolorization-Colorization}

As described in Section~\ref{sec:decolorization-colorization}, the proposed invertible framework and model can be extended to other bidirectional tasks, such as image decolorization-colorization. In this section, we present experiments of the extended model under this task, to illustrate the generalization ability of our model.

We compare our model with TAD Gray \& TAU Color~\citep{kim2018task} and invertible grayscale~\citep{xia2018invertible}, which all follow the encoder-decoder framework. Because \cite{xia2018invertible} has different training settings and datasets, we train and test their model under a similar setting as theirs on the DIV2K dataset that is rescaled to 256$\times$256. We also test our model that is trained on the original DIV2K dataset on this rescaled dataset.

\begin{table} [ht]
	\centering
	\footnotesize
	\tabcolsep=0.5mm
	\caption{Quantitative results (PSNR) of different decolorization-colorization methods for image reconstruction on the Set5, Set14, BSD100, Urban100 and DIV2K validation sets.}
	\begin{tabular}{c|c|c|c|c|c}
		\hline
		Method & Set5 & Set14 & BSD100 & Urban100 & DIV2K \\
		\hline
		\hline
		Baseline~\citep{kim2018task} & 19.12 & 21.14 & 24.21 & 23.29 & 21.10 \\
		\hline
		TAD-G \& TAU-C & 35.22 & 32.67 & 32.73 & 30.98 & 36.63 \\
		\hline
		IRN$_\text{color}$ (ours) & \textcolor{red}{40.86} & \textcolor{red}{36.78} & \textcolor{red}{42.43} & \textcolor{red}{38.77} & \textcolor{red}{42.65} \\
		\hline
		
	\end{tabular}
	\label{color1}
\end{table}

\begin{table} [ht]
	\centering
	\small
	\tabcolsep=2mm
	\caption{Quantitative results (PSNR/SSIM) of different decolorization-colorization methods for image reconstruction on the DIV2K validation set that is rescaled to 256$\times$256.}
	\begin{tabular}{c|c|c}
		\hline
		Method & Param & DIV2K\_256$\times$256 \\
		\hline
		\hline
		Invertible Grayscale & 7.42M & 31.52 / 0.9475 \\
		\hline
		IRN$_\text{color}$ (ours) & 1.41M & \textcolor{red}{37.27} / \textcolor{red}{0.9800} \\
		\hline
		
	\end{tabular}
	\label{color2}
\end{table}

As shown in Table~\ref{color1}, IRN$_\text{color}$ can perfectly reconstruct the original color images from grayscale ones, with most RGB PSNR results above 40 dB, which indicates that the reconstructed images are almost the same as original ones. And compared with TAD Gray \& TAD Color~\citep{kim2018task}, IRN$_\text{color}$ demonstrates the significant improvement of the quality of reconstructed images, indicating the advantage of our invertible framework. 

Table~\ref{color2} also demonstrates the significant improvement of IRN$_\text{color}$ compared with \cite{xia2018invertible}. Note that under this test setting, the distribution of images could be inconsistent with training images for IRN$_\text{color}$ due to the degradation by rescaling images to the size 256$\times$256. Despite this, IRN$_\text{color}$ still outperforms \cite{xia2018invertible} by 5.75 dB with much fewer parameters, further indicating the effectiveness and high efficiency of the proposed model.

\begin{figure} [ht]
    \centering
    \includegraphics[scale=0.225]{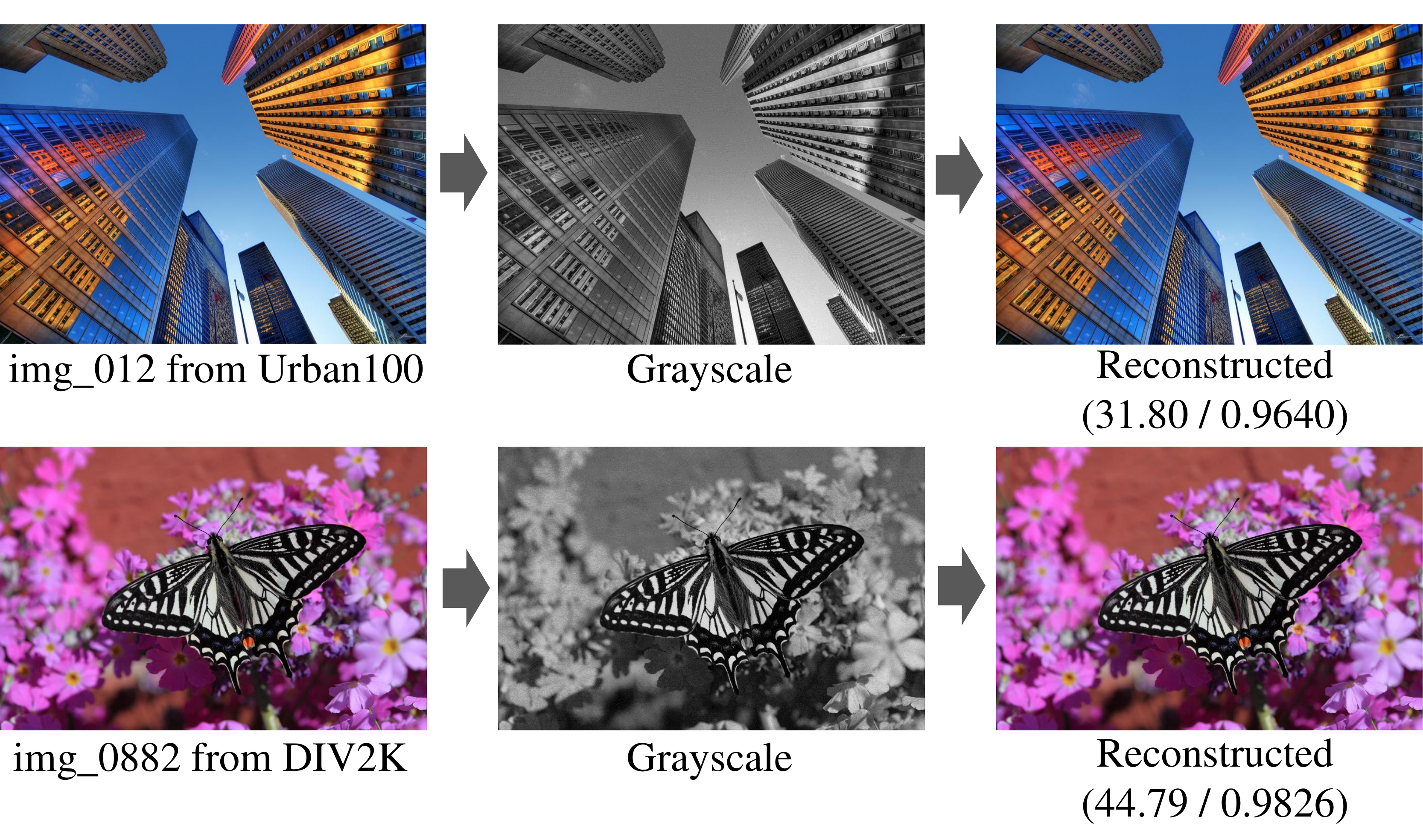}
    \caption{Qualitative demonstration of decolorization-colorization by IRN$_\text{color}$.}
    \label{fig:qualitative color result1}
\end{figure}

\begin{figure} [ht]
    \centering
    \includegraphics[scale=0.225]{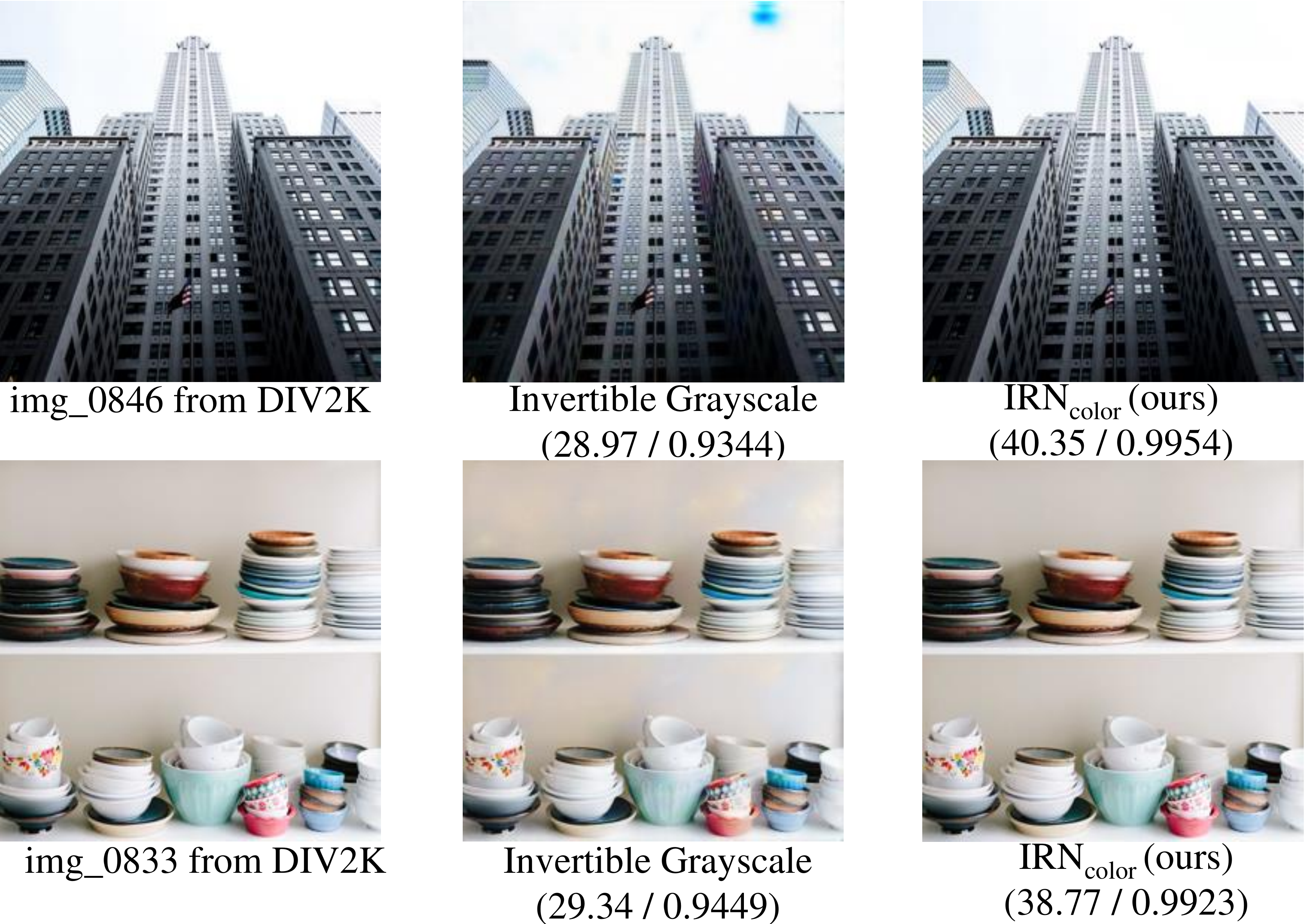}
    \caption{Qualitative comparison of colorization reconstruction for grayscale images between different methods.}
    \label{fig:qualitative color result2}
\end{figure}

Fig.~\ref{fig:qualitative color result1} and Fig.~\ref{fig:qualitative color result2} illustrate the visual quality of the grayscale and reconstructed images, as well as the comparison with other methods. It shows that the reconstructed images could have almost the same perception as the original ones. And compared with \cite{xia2018invertible}, whose reconstructed images may contain some noise or strange variegation, IRN$_\text{color}$ achieves more fidelity and better visual perception.

\subsection{Combination with Image Compression}

\begin{table} [ht]
	\centering
	\footnotesize
	\tabcolsep=2mm
	\caption{Comparison results of combination between image rescaling and lossless image compression methods on average RGB PSNR and total storage size of DIV2K validation set.}
	\begin{tabular}{c|c|c|c}
		\toprule
		Method & Scale & PSNR (dB) & Storage (MB)\\
		\hline
		\hline
		PNG & / & / & 470\\
		\hline
		FLIF & / & / & 294\\
		\hline
		\hline
		JPEG (Q=20) & / & 29.59 & 16.2\\
		\hline
		Bicubic\&ESRGAN+PNG & 4$\times$ & 29.47  & 32.4 (+100.0\%)\\ 
		\hline
		Bicubic\&ESRGAN+FLIF & 4$\times$ & 29.47 & 22.4 (+38.3\%) \\
		\hline
		\hline
		JPEG (Q=32) & / & 31.11 & 21.7\\
		\hline
		CAR\&EDSR+PNG & 4$\times$ & 31.09  & 30.2 (+39.2\%)\\ 
		\hline
		CAR\&EDSR+FLIF & 4$\times$ & 31.09 & 21.3 (-1.8\%) \\
		\hline
		\hline
		JPEG (Q=57) & / & 32.94 & 31.4\\
		\hline
		\textbf{IRN+PNG} & 4$\times$ & 32.95  & 34.9 (+11.1\%)\\ 
		\hline
		\textbf{IRN+FLIF} & 4$\times$ & 32.95 & \textcolor{red}{28.7 (-8.6\%)} \\
		\hline
		\hline
		JPEG (Q=96) & / & 40.70 & 122\\
		\hline
		\textbf{IRN+PNG} & 2$\times$ & 40.87  & 131 (+7.3\%)\\ 
		\hline
		\textbf{IRN+FLIF} & 2$\times$ & 40.87 & \textcolor{red}{108 (-11.5\%)} \\
		\hline
		\hline
		JPEG (Q=14) & / & 28.36 & 13.07\\
		\hline
		\textbf{IRN+PNG} & 8$\times$ & 28.50  & \textcolor{red}{9.16 (-29.9\%)}\\ 
		\hline
		\textbf{IRN+FLIF} & 8$\times$ & 28.50 & \textcolor{red}{7.68 (-41.2\%)} \\
		\bottomrule
		
	\end{tabular}
	\label{compression lossless}
\end{table}

\begin{figure*} [ht]
    \centering
    \subfigure[]{
        \includegraphics[scale=0.5]{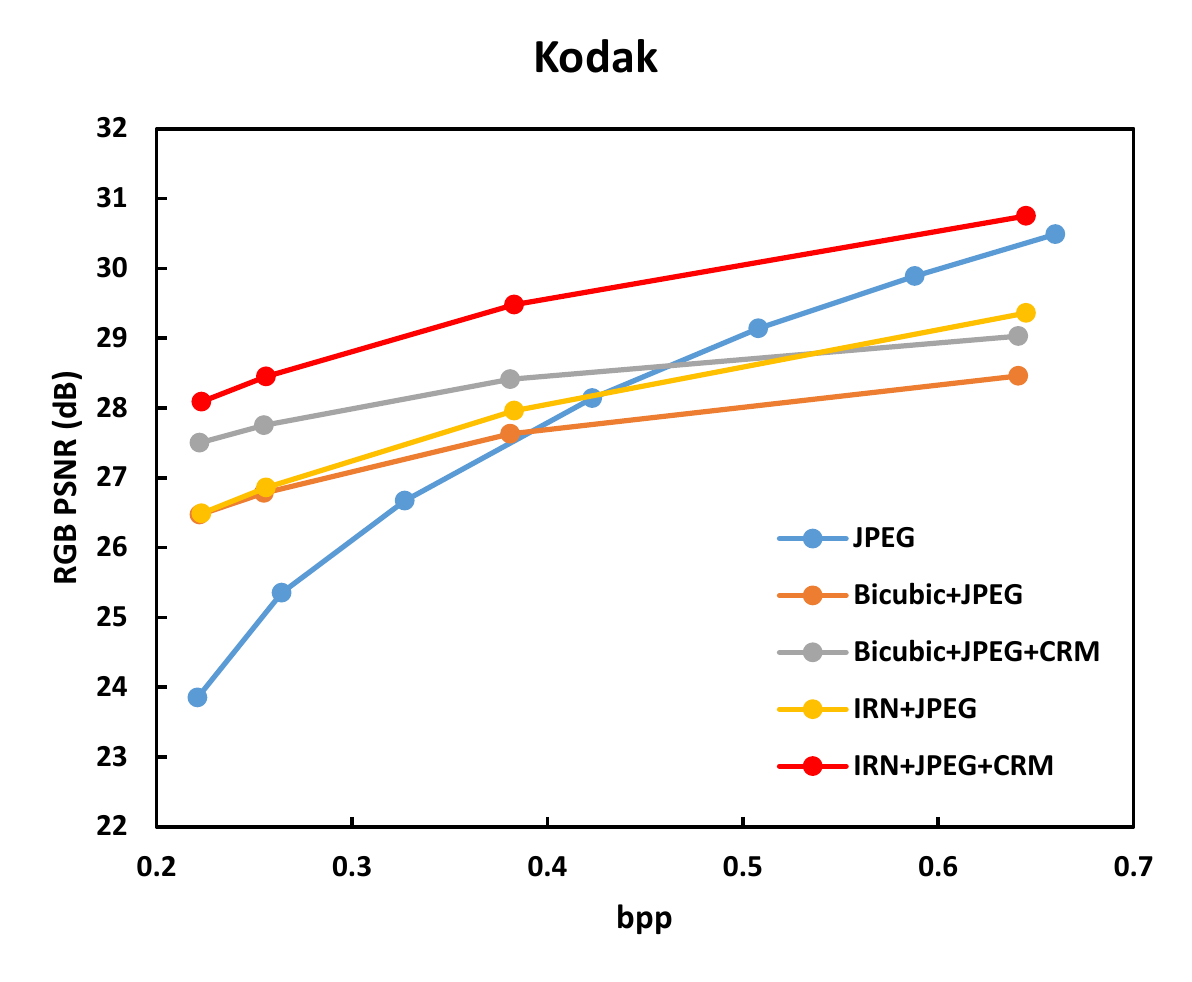}
    }
    \subfigure[]{
        \includegraphics[scale=0.5]{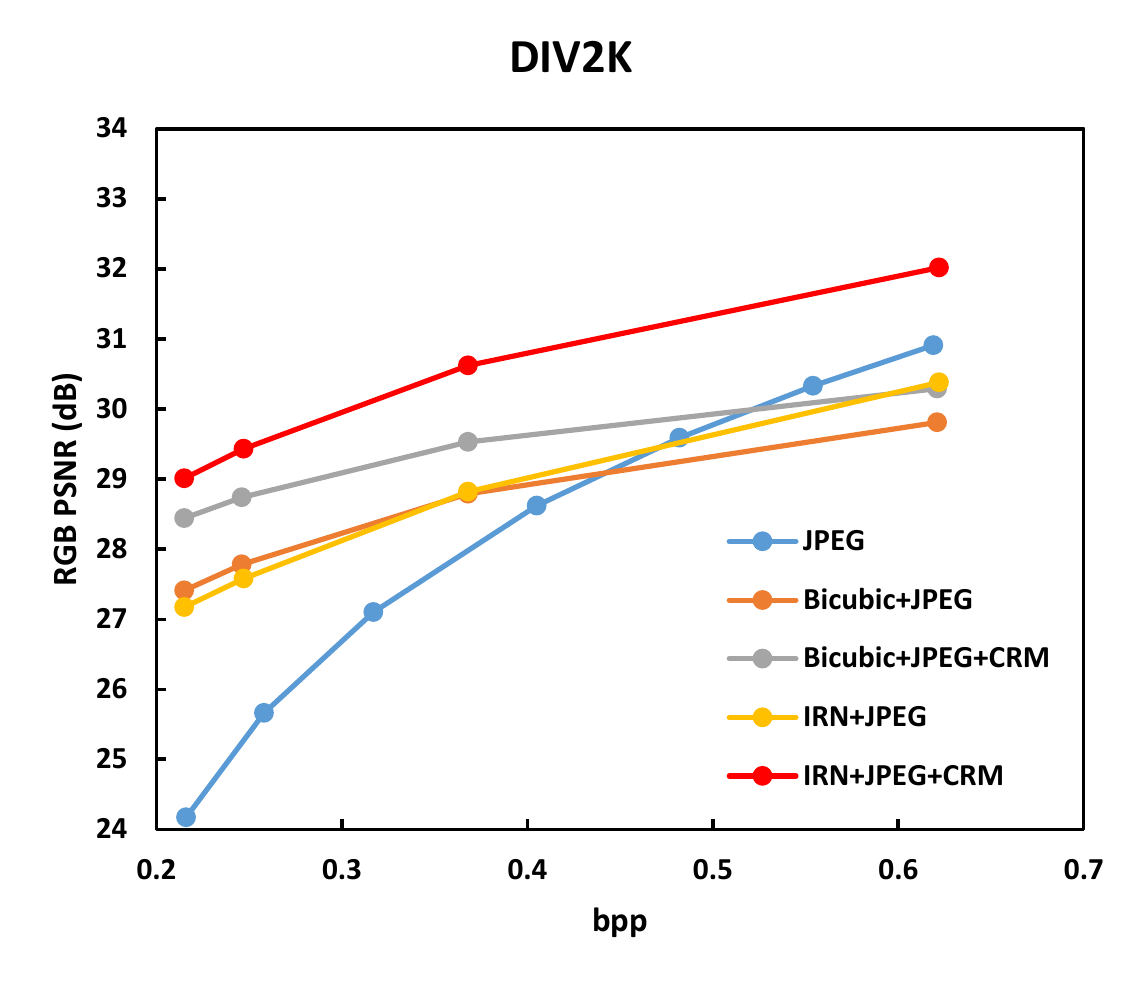}
    }
    \caption{Results of combination between image rescaling and lossy image compression methods on different datasets. The rescaling scale is $2\times$. We tune the quality of JPEG algorithm for different compression ratios. RGB PSNR and bit rate (bit per pixel, bpp) are evaluated.}
    \label{fig:compression lossy}
\end{figure*}

\begin{figure*} [htbp]
    \centering
    \includegraphics[scale=0.15]{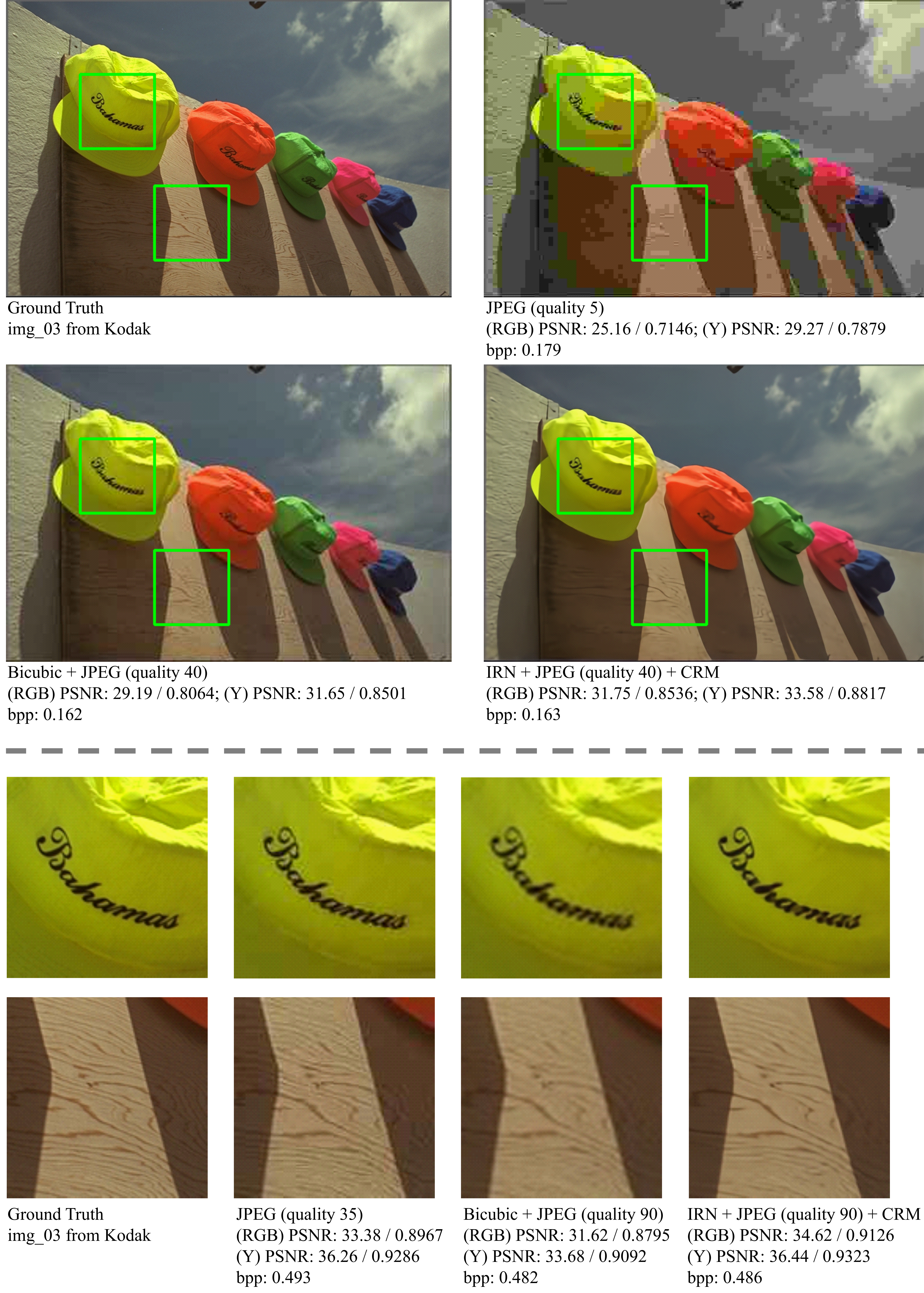}
    \caption{Qualitative results of image compression methods.}
    \label{fig:compression qualitative results}
\end{figure*}

In this section, we evaluate the combination of image rescaling and image compression methods as described in Section~\ref{sec:combine compression}. 

For the combination with lossless image compression, we choose two representative methods, i.e. PNG and FLIF~\citep{sneyers2016flif}, as an example. PNG is a classical lossless image compression algorithm, while FILF is a more recent one based on machine learning algorithms. We choose the popular JPEG lossy image compression method as the comparison standard for the trade-off between compression ratio and image quality. Because there is no hyper-parameter for image rescaling and lossless image compression to control the compression ratio, we tune the quality of JPEG to compare the compression performance with different rescaling methods under similar image quality respectively. We evaluate the total storage size for the DIV2K validation set, which contains 100 images, as compression performance, and average RGB PSNR as image quality.

As shown in Table~\ref{compression lossless}, when compared with other image downscaling and upscaling methods, IRN always shows its advantage in the trade-off between compression ratio and image quality. When compared with classical lossy image compression methods, IRN with advanced lossless compression methods can directly outperform JPEG. IRN could get promising results, especially under the condition that high compression performance is required.

For the combination with lossy image compression, we choose the classical JPEG algorithm as an example. As described in Section~\ref{sec:combine compression}, we train a Compression Restore Module (CRM) to restore the lost information in compression, which is a neural network consisting of eight residual in residual dense blocks (RRDB) introduced in the ESRGAN model~\citep{wang2018esrgan}. We tune the quality of JPEG, and the R-D curves are shown in Fig.~\ref{fig:compression lossy}. 
As explained in Section~\ref{sec:combine compression}, directly combining IRN and JPEG may not perform well because JPEG introduces additional information loss which goes against our invertible framework. This problem is mitigated by CRM. Results demonstrate that IRN combined with JPEG and CRM achieves satisfactory compression performance compared with traditional image rescaling and compression methods. Also, the ablation experiments of Bicubic+JPEG, Bicubic+JPEG+CRM, and IRN+JPEG illustrate that the performance improvement is not majorly owed to CRM, but the effectiveness of our proposed combination between the invertible framework and restoration from existing degradation methods. Additionally, we present qualitative visual results in Fig.~\ref{fig:compression qualitative results}. It demonstrates the improvement of our proposed model for clearer details under similar compression ratios.

\section{Conclusion}\label{sec:conclusion}

In this paper, we propose a novel invertible framework for the bidirectional image degradation-restoration task, which models degradation and restoration from the perspective of invertible transformation to largely mitigate the ill-posed problem. By bijectively transforming the distribution of image-specific lost contents into a pre-specified image-agnostic distribution together with the generation of degraded images, the proposed invertible framework can model lost information and keep the knowledge of distribution transformation in the invertible model. In the inverse restoration, an easily sampled latent variable in company with the generated degraded image is able to reconstruct images through the inverse transformation. Our deliberately designed architecture and effective training objectives enable the proposed IRN model to achieve the goals of the invertible framework in the image rescaling scenario, and it is easily adapted to similar tasks such as image decolorization-colorization. Further, we propose the combination between our invertible framework and restoration from existing degradation methods for wider applications, with an instantiation of the combination of image rescaling and compression. Our extensive experiments demonstrate the significant improvement of our model both quantitatively and qualitatively, as well as the lightweight property and high efficiency of our model. More ablation and extension experiments further provide detailed analysis and illustrate the generalization ability of the proposed method.

\backmatter

\bmhead{Supplementary information}




In supplementary materials, we provide the appendix of the manuscript and the full implementation codes.

\bmhead{Acknowledgments}

The authors would like to thank Yaolong Wang, Di He, Guolin Ke and Jiang Bian for their help on discussions, experiments and writing in the preliminary version of this paper. The authors would also like to thank the reviewers for their valuable suggestions. Z. Lin was supported by the major key project of PCL (grant no. PCL2021A12) and the NSF China (No. 62276004).

\section*{Declarations}

\bmhead{Funding}
Z. Lin was supported by the major key project of PCL (grant no. PCL2021A12) and the NSF China (No. 62276004).

\bmhead{Competing interests}
The authors have no competing interests to declare that are relevant to the content of this article.

\bmhead{Availability of data and materials}
All the datasets used in the paper are publicly available.

\bmhead{Code availability}
Our code is available at \url{https://github.com/pkuxmq/Invertible-Image-Rescaling}. We also provide the full code in the supplementary materials.

\bmhead{Authors' contributions}
M. Xiao, S. Zheng, and C. Liu conceptualized the
work and designed the methodology. M. Xiao and C. Liu formulated the mathematical formulation. M. Xiao conducted the experiments. M. Xiao, S. Zheng, and C. Liu analyzed the results. Z. Lin and TY. Liu supervised the work. All authors wrote and revised the manuscript.

\small
\bibliography{IRN}

\begin{thebibliography}{}
\providecommand{\doi}[1]{\url{https://doi.org/#1}}
\bibcommenthead

\bibitem [\protect \citeauthoryear {%
Agustsson%
\ \BBA {} Timofte%
}{%
Agustsson%
\ \BBA {} Timofte%
}{%
{\protect \APACyear {2017}}%
}]{%
agustsson2017ntire}
\APACinsertmetastar {%
agustsson2017ntire}%
\begin{APACrefauthors}%
Agustsson, E.%
\BCBT {}\ \BBA {} Timofte, R.%
\end{APACrefauthors}%
\unskip\
\newblock
\APACrefYearMonthDay{2017}{}{}.
\newblock
{\BBOQ}\APACrefatitle {Ntire 2017 challenge on single image super-resolution:
  Dataset and study} {Ntire 2017 challenge on single image super-resolution:
  Dataset and study}.{\BBCQ}
\newblock
 \APACrefbtitle {{Proceedings of the IEEE Conference on Computer Vision and
  Pattern Recognition Workshops}.} {{Proceedings of the IEEE Conference on
  Computer Vision and Pattern Recognition Workshops}.}
\PrintBackRefs{\CurrentBib}

\bibitem [\protect \citeauthoryear {%
Agustsson%
, Tschannen%
, Mentzer%
, Timofte%
\BCBL {}\ \BBA {} Gool%
}{%
Agustsson%
\ \protect \BOthers {.}}{%
{\protect \APACyear {2019}}%
}]{%
agustsson2019generative}
\APACinsertmetastar {%
agustsson2019generative}%
\begin{APACrefauthors}%
Agustsson, E.%
, Tschannen, M.%
, Mentzer, F.%
, Timofte, R.%
\BCBL {} Gool, L.V.%
\end{APACrefauthors}%
\unskip\
\newblock
\APACrefYearMonthDay{2019}{}{}.
\newblock
{\BBOQ}\APACrefatitle {Generative adversarial networks for extreme learned
  image compression} {Generative adversarial networks for extreme learned image
  compression}.{\BBCQ}
\newblock
 \APACrefbtitle {{Proceedings of the IEEE International Conference on Computer
  Vision}.} {{Proceedings of the IEEE International Conference on Computer
  Vision}.}
\PrintBackRefs{\CurrentBib}

\bibitem [\protect \citeauthoryear {%
Ardizzone%
, Kruse%
\BCBL {}\ \protect \BOthers {.}}{%
Ardizzone%
, Kruse%
\BCBL {}\ \protect \BOthers {.}}{%
{\protect \APACyear {2019}}%
}]{%
ardizzone2019analyzing}
\APACinsertmetastar {%
ardizzone2019analyzing}%
\begin{APACrefauthors}%
Ardizzone, L.%
, Kruse, J.%
, Wirkert, S.%
, Rahner, D.%
, Pellegrini, E.W.%
, Klessen, R.S.%
\BDBL {}K{\"o}the, U.%
\end{APACrefauthors}%
\unskip\
\newblock
\APACrefYearMonthDay{2019}{}{}.
\newblock
{\BBOQ}\APACrefatitle {Analyzing inverse problems with invertible neural
  networks} {Analyzing inverse problems with invertible neural
  networks}.{\BBCQ}
\newblock
 \APACrefbtitle {{Proceedings of the International Conference on Learning
  Representations}.} {{Proceedings of the International Conference on Learning
  Representations}.}
\PrintBackRefs{\CurrentBib}

\bibitem [\protect \citeauthoryear {%
Ardizzone%
, L{\"u}th%
, Kruse%
, Rother%
\BCBL {}\ \BBA {} K{\"o}the%
}{%
Ardizzone%
, L{\"u}th%
\BCBL {}\ \protect \BOthers {.}}{%
{\protect \APACyear {2019}}%
}]{%
ardizzone2019guided}
\APACinsertmetastar {%
ardizzone2019guided}%
\begin{APACrefauthors}%
Ardizzone, L.%
, L{\"u}th, C.%
, Kruse, J.%
, Rother, C.%
\BCBL {} K{\"o}the, U.%
\end{APACrefauthors}%
\unskip\
\newblock
\APACrefYearMonthDay{2019}{}{}.
\newblock
{\BBOQ}\APACrefatitle {Guided Image Generation with Conditional Invertible
  Neural Networks} {Guided image generation with conditional invertible neural
  networks}.{\BBCQ}
\newblock
\APACjournalVolNumPages{{arXiv preprint arXiv:1907.02392}}{}{}{}.
\newblock

\newblock

\PrintBackRefs{\CurrentBib}

\bibitem [\protect \citeauthoryear {%
Arjovsky%
\ \BBA {} Bottou%
}{%
Arjovsky%
\ \BBA {} Bottou%
}{%
{\protect \APACyear {2017}}%
}]{%
arjovsky2017towards}
\APACinsertmetastar {%
arjovsky2017towards}%
\begin{APACrefauthors}%
Arjovsky, M.%
\BCBT {}\ \BBA {} Bottou, L.%
\end{APACrefauthors}%
\unskip\
\newblock
\APACrefYearMonthDay{2017}{}{}.
\newblock
{\BBOQ}\APACrefatitle {Towards principled methods for training generative
  adversarial networks} {Towards principled methods for training generative
  adversarial networks}.{\BBCQ}
\newblock
 \APACrefbtitle {{Proceedings of the International Conference on Learning
  Representations}.} {{Proceedings of the International Conference on Learning
  Representations}.}
\PrintBackRefs{\CurrentBib}

\bibitem [\protect \citeauthoryear {%
Asim%
, Daniels%
, Leong%
, Ahmed%
\BCBL {}\ \BBA {} Hand%
}{%
Asim%
\ \protect \BOthers {.}}{%
{\protect \APACyear {2020}}%
}]{%
asim2020invertible}
\APACinsertmetastar {%
asim2020invertible}%
\begin{APACrefauthors}%
Asim, M.%
, Daniels, M.%
, Leong, O.%
, Ahmed, A.%
\BCBL {} Hand, P.%
\end{APACrefauthors}%
\unskip\
\newblock
\APACrefYearMonthDay{2020}{}{}.
\newblock
{\BBOQ}\APACrefatitle {Invertible generative models for inverse problems:
  mitigating representation error and dataset bias} {Invertible generative
  models for inverse problems: mitigating representation error and dataset
  bias}.{\BBCQ}
\newblock
 \APACrefbtitle {{Proceedings of the International Conference on Machine
  Learning}.} {{Proceedings of the International Conference on Machine
  Learning}.}
\PrintBackRefs{\CurrentBib}

\bibitem [\protect \citeauthoryear {%
Bala%
\ \BBA {} Eschbach%
}{%
Bala%
\ \BBA {} Eschbach%
}{%
{\protect \APACyear {2004}}%
}]{%
bala2004spatial}
\APACinsertmetastar {%
bala2004spatial}%
\begin{APACrefauthors}%
Bala, R.%
\BCBT {}\ \BBA {} Eschbach, R.%
\end{APACrefauthors}%
\unskip\
\newblock
\APACrefYearMonthDay{2004}{}{}.
\newblock
{\BBOQ}\APACrefatitle {Spatial color-to-grayscale transform preserving
  chrominance edge information} {Spatial color-to-grayscale transform
  preserving chrominance edge information}.{\BBCQ}
\newblock
 \APACrefbtitle {{Color and Imaging Conference}.} {{Color and Imaging
  Conference}.}
\PrintBackRefs{\CurrentBib}

\bibitem [\protect \citeauthoryear {%
Ball{\'e}%
, Laparra%
\BCBL {}\ \BBA {} Simoncelli%
}{%
Ball{\'e}%
\ \protect \BOthers {.}}{%
{\protect \APACyear {2017}}%
}]{%
balle2016end}
\APACinsertmetastar {%
balle2016end}%
\begin{APACrefauthors}%
Ball{\'e}, J.%
, Laparra, V.%
\BCBL {} Simoncelli, E.P.%
\end{APACrefauthors}%
\unskip\
\newblock
\APACrefYearMonthDay{2017}{}{}.
\newblock
{\BBOQ}\APACrefatitle {End-to-end optimized image compression} {End-to-end
  optimized image compression}.{\BBCQ}
\newblock
 \APACrefbtitle {{Proceedings of the International Conference on Learning
  Representations}.} {{Proceedings of the International Conference on Learning
  Representations}.}
\PrintBackRefs{\CurrentBib}

\bibitem [\protect \citeauthoryear {%
Ball{\'e}%
, Minnen%
, Singh%
, Hwang%
\BCBL {}\ \BBA {} Johnston%
}{%
Ball{\'e}%
\ \protect \BOthers {.}}{%
{\protect \APACyear {2018}}%
}]{%
balle2018variational}
\APACinsertmetastar {%
balle2018variational}%
\begin{APACrefauthors}%
Ball{\'e}, J.%
, Minnen, D.%
, Singh, S.%
, Hwang, S.J.%
\BCBL {} Johnston, N.%
\end{APACrefauthors}%
\unskip\
\newblock
\APACrefYearMonthDay{2018}{}{}.
\newblock
{\BBOQ}\APACrefatitle {Variational image compression with a scale hyperprior}
  {Variational image compression with a scale hyperprior}.{\BBCQ}
\newblock
 \APACrefbtitle {{Proceedings of the International Conference on Learning
  Representations}.} {{Proceedings of the International Conference on Learning
  Representations}.}
\PrintBackRefs{\CurrentBib}

\bibitem [\protect \citeauthoryear {%
Behrmann%
, Grathwohl%
, Chen%
, Duvenaud%
\BCBL {}\ \BBA {} Jacobsen%
}{%
Behrmann%
\ \protect \BOthers {.}}{%
{\protect \APACyear {2019}}%
}]{%
behrmann2019invertible}
\APACinsertmetastar {%
behrmann2019invertible}%
\begin{APACrefauthors}%
Behrmann, J.%
, Grathwohl, W.%
, Chen, R.T.%
, Duvenaud, D.%
\BCBL {} Jacobsen, J\BHBI H.%
\end{APACrefauthors}%
\unskip\
\newblock
\APACrefYearMonthDay{2019}{}{}.
\newblock
{\BBOQ}\APACrefatitle {Invertible Residual Networks} {Invertible residual
  networks}.{\BBCQ}
\newblock
 \APACrefbtitle {{Proceedings of the International Conference on Machine
  Learning}.} {{Proceedings of the International Conference on Machine
  Learning}.}
\PrintBackRefs{\CurrentBib}

\bibitem [\protect \citeauthoryear {%
Bengio%
, L{\'e}onard%
\BCBL {}\ \BBA {} Courville%
}{%
Bengio%
\ \protect \BOthers {.}}{%
{\protect \APACyear {2013}}%
}]{%
bengio2013estimating}
\APACinsertmetastar {%
bengio2013estimating}%
\begin{APACrefauthors}%
Bengio, Y.%
, L{\'e}onard, N.%
\BCBL {} Courville, A.%
\end{APACrefauthors}%
\unskip\
\newblock
\APACrefYearMonthDay{2013}{}{}.
\newblock
{\BBOQ}\APACrefatitle {Estimating or propagating gradients through stochastic
  neurons for conditional computation} {Estimating or propagating gradients
  through stochastic neurons for conditional computation}.{\BBCQ}
\newblock
\APACjournalVolNumPages{arXiv preprint arXiv:1308.3432}{}{}{}.
\newblock

\newblock

\PrintBackRefs{\CurrentBib}

\bibitem [\protect \citeauthoryear {%
Berg%
, Hasenclever%
, Tomczak%
\BCBL {}\ \BBA {} Welling%
}{%
Berg%
\ \protect \BOthers {.}}{%
{\protect \APACyear {2018}}%
}]{%
berg2018sylvester}
\APACinsertmetastar {%
berg2018sylvester}%
\begin{APACrefauthors}%
Berg, R.v.d.%
, Hasenclever, L.%
, Tomczak, J.M.%
\BCBL {} Welling, M.%
\end{APACrefauthors}%
\unskip\
\newblock
\APACrefYearMonthDay{2018}{}{}.
\newblock
{\BBOQ}\APACrefatitle {{S}ylvester normalizing flows for variational inference}
  {{S}ylvester normalizing flows for variational inference}.{\BBCQ}
\newblock
 \APACrefbtitle {{Proceedings of the Conference on Uncertainty in Artificial
  Intelligence}.} {{Proceedings of the Conference on Uncertainty in Artificial
  Intelligence}.}
\PrintBackRefs{\CurrentBib}

\bibitem [\protect \citeauthoryear {%
Bevilacqua%
, Roumy%
, Guillemot%
\BCBL {}\ \BBA {} Morel%
}{%
Bevilacqua%
\ \protect \BOthers {.}}{%
{\protect \APACyear {2012}}%
}]{%
bevilacqua2012low}
\APACinsertmetastar {%
bevilacqua2012low}%
\begin{APACrefauthors}%
Bevilacqua, M.%
, Roumy, A.%
, Guillemot, C.%
\BCBL {} Morel, M\BHBI L.A.%
\end{APACrefauthors}%
\unskip\
\newblock
\APACrefYearMonthDay{2012}{}{}.
\newblock
{\BBOQ}\APACrefatitle {Low-Complexity Single-Image Super-Resolution based on
  Nonnegative Neighbor Embedding} {Low-complexity single-image super-resolution
  based on nonnegative neighbor embedding}.{\BBCQ}
\newblock
 \APACrefbtitle {{British Machine Vision Conference (BMVC)}.} {{British Machine
  Vision Conference (BMVC)}.}
\PrintBackRefs{\CurrentBib}

\bibitem [\protect \citeauthoryear {%
Blau%
, Mechrez%
, Timofte%
, Michaeli%
\BCBL {}\ \BBA {} Zelnik-Manor%
}{%
Blau%
\ \protect \BOthers {.}}{%
{\protect \APACyear {2018}}%
}]{%
blau20182018}
\APACinsertmetastar {%
blau20182018}%
\begin{APACrefauthors}%
Blau, Y.%
, Mechrez, R.%
, Timofte, R.%
, Michaeli, T.%
\BCBL {} Zelnik-Manor, L.%
\end{APACrefauthors}%
\unskip\
\newblock
\APACrefYearMonthDay{2018}{}{}.
\newblock
{\BBOQ}\APACrefatitle {The 2018 pirm challenge on perceptual image
  super-resolution} {The 2018 pirm challenge on perceptual image
  super-resolution}.{\BBCQ}
\newblock
 \APACrefbtitle {{European Conference on Computer Vision Workshops (ECCVW)}.}
  {{European Conference on Computer Vision Workshops (ECCVW)}.}
\PrintBackRefs{\CurrentBib}

\bibitem [\protect \citeauthoryear {%
Bruckstein%
, Elad%
\BCBL {}\ \BBA {} Kimmel%
}{%
Bruckstein%
\ \protect \BOthers {.}}{%
{\protect \APACyear {2003}}%
}]{%
bruckstein2003down}
\APACinsertmetastar {%
bruckstein2003down}%
\begin{APACrefauthors}%
Bruckstein, A.M.%
, Elad, M.%
\BCBL {} Kimmel, R.%
\end{APACrefauthors}%
\unskip\
\newblock
\APACrefYearMonthDay{2003}{}{}.
\newblock
{\BBOQ}\APACrefatitle {Down-scaling for better transform compression}
  {Down-scaling for better transform compression}.{\BBCQ}
\newblock
\APACjournalVolNumPages{IEEE Transactions on Image
  Processing}{12}{9}{1132--1144}.
\newblock

\newblock

\PrintBackRefs{\CurrentBib}

\bibitem [\protect \citeauthoryear {%
R.T.~Chen%
, Behrmann%
, Duvenaud%
\BCBL {}\ \BBA {} Jacobsen%
}{%
R.T.~Chen%
\ \protect \BOthers {.}}{%
{\protect \APACyear {2019}}%
}]{%
chen2019residual}
\APACinsertmetastar {%
chen2019residual}%
\begin{APACrefauthors}%
Chen, R.T.%
, Behrmann, J.%
, Duvenaud, D.K.%
\BCBL {} Jacobsen, J\BHBI H.%
\end{APACrefauthors}%
\unskip\
\newblock
\APACrefYearMonthDay{2019}{}{}.
\newblock
{\BBOQ}\APACrefatitle {Residual flows for invertible generative modeling}
  {Residual flows for invertible generative modeling}.{\BBCQ}
\newblock
 \APACrefbtitle {{Advances in Neural Information Processing Systems}.}
  {{Advances in Neural Information Processing Systems}.}
\PrintBackRefs{\CurrentBib}

\bibitem [\protect \citeauthoryear {%
Y.~Chen%
, Xiao%
, Dai%
\BCBL {}\ \BBA {} Xia%
}{%
Y.~Chen%
\ \protect \BOthers {.}}{%
{\protect \APACyear {2020}}%
}]{%
chen2020hrnet}
\APACinsertmetastar {%
chen2020hrnet}%
\begin{APACrefauthors}%
Chen, Y.%
, Xiao, X.%
, Dai, T.%
\BCBL {} Xia, S\BHBI T.%
\end{APACrefauthors}%
\unskip\
\newblock
\APACrefYearMonthDay{2020}{}{}.
\newblock
{\BBOQ}\APACrefatitle {Hrnet: Hamiltonian Rescaling Network for Image
  Downscaling} {Hrnet: Hamiltonian rescaling network for image
  downscaling}.{\BBCQ}
\newblock
 \APACrefbtitle {{Proceedings of the IEEE International Conference on Image
  Processing (ICIP)}.} {{Proceedings of the IEEE International Conference on
  Image Processing (ICIP)}.}
\PrintBackRefs{\CurrentBib}

\bibitem [\protect \citeauthoryear {%
K.L.~Cheng%
, Xie%
\BCBL {}\ \BBA {} Chen%
}{%
K.L.~Cheng%
\ \protect \BOthers {.}}{%
{\protect \APACyear {2021}}%
}]{%
cheng2021iicnet}
\APACinsertmetastar {%
cheng2021iicnet}%
\begin{APACrefauthors}%
Cheng, K.L.%
, Xie, Y.%
\BCBL {} Chen, Q.%
\end{APACrefauthors}%
\unskip\
\newblock
\APACrefYearMonthDay{2021}{}{}.
\newblock
{\BBOQ}\APACrefatitle {{IICNet: A Generic Framework for Reversible Image
  Conversion}} {{IICNet: A Generic Framework for Reversible Image
  Conversion}}.{\BBCQ}
\newblock
 \APACrefbtitle {{Proceedings of the IEEE International Conference on Computer
  Vision}.} {{Proceedings of the IEEE International Conference on Computer
  Vision}.}
\PrintBackRefs{\CurrentBib}

\bibitem [\protect \citeauthoryear {%
Z.~Cheng%
, Sun%
, Takeuchi%
\BCBL {}\ \BBA {} Katto%
}{%
Z.~Cheng%
\ \protect \BOthers {.}}{%
{\protect \APACyear {2020}}%
}]{%
cheng2020learned}
\APACinsertmetastar {%
cheng2020learned}%
\begin{APACrefauthors}%
Cheng, Z.%
, Sun, H.%
, Takeuchi, M.%
\BCBL {} Katto, J.%
\end{APACrefauthors}%
\unskip\
\newblock
\APACrefYearMonthDay{2020}{}{}.
\newblock
{\BBOQ}\APACrefatitle {Learned image compression with discretized gaussian
  mixture likelihoods and attention modules} {Learned image compression with
  discretized gaussian mixture likelihoods and attention modules}.{\BBCQ}
\newblock
 \APACrefbtitle {{Proceedings of the IEEE Conference on Computer Vision and
  Pattern Recognition}.} {{Proceedings of the IEEE Conference on Computer
  Vision and Pattern Recognition}.}
\PrintBackRefs{\CurrentBib}

\bibitem [\protect \citeauthoryear {%
Dai%
, Cai%
, Zhang%
, Xia%
\BCBL {}\ \BBA {} Zhang%
}{%
Dai%
\ \protect \BOthers {.}}{%
{\protect \APACyear {2019}}%
}]{%
dai2019second}
\APACinsertmetastar {%
dai2019second}%
\begin{APACrefauthors}%
Dai, T.%
, Cai, J.%
, Zhang, Y.%
, Xia, S\BHBI T.%
\BCBL {} Zhang, L.%
\end{APACrefauthors}%
\unskip\
\newblock
\APACrefYearMonthDay{2019}{}{}.
\newblock
{\BBOQ}\APACrefatitle {Second-order Attention Network for Single Image
  Super-Resolution} {Second-order attention network for single image
  super-resolution}.{\BBCQ}
\newblock
 \APACrefbtitle {{Proceedings of the IEEE Conference on Computer Vision and
  Pattern Recognition}.} {{Proceedings of the IEEE Conference on Computer
  Vision and Pattern Recognition}.}
\PrintBackRefs{\CurrentBib}

\bibitem [\protect \citeauthoryear {%
Deshpande%
, Lu%
, Yeh%
, Jin~Chong%
\BCBL {}\ \BBA {} Forsyth%
}{%
Deshpande%
\ \protect \BOthers {.}}{%
{\protect \APACyear {2017}}%
}]{%
deshpande2017learning}
\APACinsertmetastar {%
deshpande2017learning}%
\begin{APACrefauthors}%
Deshpande, A.%
, Lu, J.%
, Yeh, M\BHBI C.%
, Jin~Chong, M.%
\BCBL {} Forsyth, D.%
\end{APACrefauthors}%
\unskip\
\newblock
\APACrefYearMonthDay{2017}{}{}.
\newblock
{\BBOQ}\APACrefatitle {Learning diverse image colorization} {Learning diverse
  image colorization}.{\BBCQ}
\newblock
 \APACrefbtitle {{Proceedings of the IEEE Conference on Computer Vision and
  Pattern Recognition}.} {{Proceedings of the IEEE Conference on Computer
  Vision and Pattern Recognition}.}
\PrintBackRefs{\CurrentBib}

\bibitem [\protect \citeauthoryear {%
Dinh%
, Krueger%
\BCBL {}\ \BBA {} Bengio%
}{%
Dinh%
\ \protect \BOthers {.}}{%
{\protect \APACyear {2015}}%
}]{%
dinh2015nice}
\APACinsertmetastar {%
dinh2015nice}%
\begin{APACrefauthors}%
Dinh, L.%
, Krueger, D.%
\BCBL {} Bengio, Y.%
\end{APACrefauthors}%
\unskip\
\newblock
\APACrefYearMonthDay{2015}{}{}.
\newblock
{\BBOQ}\APACrefatitle {{NICE}: Non-linear independent components estimation}
  {{NICE}: Non-linear independent components estimation}.{\BBCQ}
\newblock
 \APACrefbtitle {{Workshop of the International Conference on Learning
  Representations}.} {{Workshop of the International Conference on Learning
  Representations}.}
\PrintBackRefs{\CurrentBib}

\bibitem [\protect \citeauthoryear {%
Dinh%
, Sohl-Dickstein%
\BCBL {}\ \BBA {} Bengio%
}{%
Dinh%
\ \protect \BOthers {.}}{%
{\protect \APACyear {2017}}%
}]{%
dinh2017density}
\APACinsertmetastar {%
dinh2017density}%
\begin{APACrefauthors}%
Dinh, L.%
, Sohl-Dickstein, J.%
\BCBL {} Bengio, S.%
\end{APACrefauthors}%
\unskip\
\newblock
\APACrefYearMonthDay{2017}{}{}.
\newblock
{\BBOQ}\APACrefatitle {Density estimation using real {NVP}} {Density estimation
  using real {NVP}}.{\BBCQ}
\newblock
 \APACrefbtitle {{Proceedings of the International Conference on Learning
  Representations}.} {{Proceedings of the International Conference on Learning
  Representations}.}
\PrintBackRefs{\CurrentBib}

\bibitem [\protect \citeauthoryear {%
Dong%
, Loy%
, He%
\BCBL {}\ \BBA {} Tang%
}{%
Dong%
\ \protect \BOthers {.}}{%
{\protect \APACyear {2015}}%
}]{%
dong2015image}
\APACinsertmetastar {%
dong2015image}%
\begin{APACrefauthors}%
Dong, C.%
, Loy, C.C.%
, He, K.%
\BCBL {} Tang, X.%
\end{APACrefauthors}%
\unskip\
\newblock
\APACrefYearMonthDay{2015}{}{}.
\newblock
{\BBOQ}\APACrefatitle {Image super-resolution using deep convolutional
  networks} {Image super-resolution using deep convolutional networks}.{\BBCQ}
\newblock
\APACjournalVolNumPages{IEEE Transactions on Pattern Analysis and Machine
  Intelligence}{38}{2}{295--307}.
\newblock

\newblock

\PrintBackRefs{\CurrentBib}

\bibitem [\protect \citeauthoryear {%
Franzen%
}{%
Franzen%
}{%
{\protect \APACyear {1999}}%
}]{%
franzen1999kodak}
\APACinsertmetastar {%
franzen1999kodak}%
\begin{APACrefauthors}%
Franzen, R.%
\end{APACrefauthors}%
\unskip\
\newblock
\APACrefYearMonthDay{1999}{}{}.
\newblock
{\BBOQ}\APACrefatitle {Kodak lossless true color image suite} {Kodak lossless
  true color image suite}.{\BBCQ}
\newblock
\APACjournalVolNumPages{{source: http://r0k.us/graphics/kodak}}{}{}{}.
\newblock

\newblock

\PrintBackRefs{\CurrentBib}

\bibitem [\protect \citeauthoryear {%
Freedman%
\ \BBA {} Fattal%
}{%
Freedman%
\ \BBA {} Fattal%
}{%
{\protect \APACyear {2011}}%
}]{%
freedman2011image}
\APACinsertmetastar {%
freedman2011image}%
\begin{APACrefauthors}%
Freedman, G.%
\BCBT {}\ \BBA {} Fattal, R.%
\end{APACrefauthors}%
\unskip\
\newblock
\APACrefYearMonthDay{2011}{}{}.
\newblock
{\BBOQ}\APACrefatitle {Image and video upscaling from local self-examples}
  {Image and video upscaling from local self-examples}.{\BBCQ}
\newblock
\APACjournalVolNumPages{ACM Transactions on Graphics (TOG)}{30}{2}{12}.
\newblock

\newblock

\PrintBackRefs{\CurrentBib}

\bibitem [\protect \citeauthoryear {%
Giachetti%
\ \BBA {} Asuni%
}{%
Giachetti%
\ \BBA {} Asuni%
}{%
{\protect \APACyear {2011}}%
}]{%
giachetti2011real}
\APACinsertmetastar {%
giachetti2011real}%
\begin{APACrefauthors}%
Giachetti, A.%
\BCBT {}\ \BBA {} Asuni, N.%
\end{APACrefauthors}%
\unskip\
\newblock
\APACrefYearMonthDay{2011}{}{}.
\newblock
{\BBOQ}\APACrefatitle {Real-time artifact-free image upscaling} {Real-time
  artifact-free image upscaling}.{\BBCQ}
\newblock
\APACjournalVolNumPages{IEEE Transactions on Image
  Processing}{20}{10}{2760--2768}.
\newblock

\newblock

\PrintBackRefs{\CurrentBib}

\bibitem [\protect \citeauthoryear {%
Glasner%
, Bagon%
\BCBL {}\ \BBA {} Irani%
}{%
Glasner%
\ \protect \BOthers {.}}{%
{\protect \APACyear {2009}}%
{\protect \APACexlab {{\protect \BCnt {1}}}}}]{%
irani2009super}
\APACinsertmetastar {%
irani2009super}%
\begin{APACrefauthors}%
Glasner, D.%
, Bagon, S.%
\BCBL {} Irani, M.%
\end{APACrefauthors}%
\unskip\
\newblock
\APACrefYearMonthDay{2009{\protect \BCnt {1}}}{}{}.
\newblock
{\BBOQ}\APACrefatitle {Super-resolution from a single image} {Super-resolution
  from a single image}.{\BBCQ}
\newblock
 \APACrefbtitle {{Proceedings of the IEEE International Conference on Computer
  Vision}.} {{Proceedings of the IEEE International Conference on Computer
  Vision}.}
\PrintBackRefs{\CurrentBib}

\bibitem [\protect \citeauthoryear {%
Glasner%
, Bagon%
\BCBL {}\ \BBA {} Irani%
}{%
Glasner%
\ \protect \BOthers {.}}{%
{\protect \APACyear {2009}}%
{\protect \APACexlab {{\protect \BCnt {2}}}}}]{%
glasner2009super}
\APACinsertmetastar {%
glasner2009super}%
\begin{APACrefauthors}%
Glasner, D.%
, Bagon, S.%
\BCBL {} Irani, M.%
\end{APACrefauthors}%
\unskip\
\newblock
\APACrefYearMonthDay{2009{\protect \BCnt {2}}}{}{}.
\newblock
{\BBOQ}\APACrefatitle {Super-resolution from a single image} {Super-resolution
  from a single image}.{\BBCQ}
\newblock
 \APACrefbtitle {{Proceedings of the IEEE International Conference on Computer
  Vision}.} {{Proceedings of the IEEE International Conference on Computer
  Vision}.}
\PrintBackRefs{\CurrentBib}

\bibitem [\protect \citeauthoryear {%
Goodfellow%
\ \protect \BOthers {.}}{%
Goodfellow%
\ \protect \BOthers {.}}{%
{\protect \APACyear {2014}}%
}]{%
goodfellow2014generative}
\APACinsertmetastar {%
goodfellow2014generative}%
\begin{APACrefauthors}%
Goodfellow, I.%
, Pouget-Abadie, J.%
, Mirza, M.%
, Xu, B.%
, Warde-Farley, D.%
, Ozair, S.%
\BDBL {}Bengio, Y.%
\end{APACrefauthors}%
\unskip\
\newblock
\APACrefYearMonthDay{2014}{}{}.
\newblock
{\BBOQ}\APACrefatitle {Generative adversarial nets} {Generative adversarial
  nets}.{\BBCQ}
\newblock
 \APACrefbtitle {{Advances in Neural Information Processing Systems}.}
  {{Advances in Neural Information Processing Systems}.}
\PrintBackRefs{\CurrentBib}

\bibitem [\protect \citeauthoryear {%
Grathwohl%
, Chen%
, Betterncourt%
, Sutskever%
\BCBL {}\ \BBA {} Duvenaud%
}{%
Grathwohl%
\ \protect \BOthers {.}}{%
{\protect \APACyear {2019}}%
}]{%
grathwohl2019ffjord}
\APACinsertmetastar {%
grathwohl2019ffjord}%
\begin{APACrefauthors}%
Grathwohl, W.%
, Chen, R.T.%
, Betterncourt, J.%
, Sutskever, I.%
\BCBL {} Duvenaud, D.%
\end{APACrefauthors}%
\unskip\
\newblock
\APACrefYearMonthDay{2019}{}{}.
\newblock
{\BBOQ}\APACrefatitle {{FFJORD}: Free-form continuous dynamics for scalable
  reversible generative models} {{FFJORD}: Free-form continuous dynamics for
  scalable reversible generative models}.{\BBCQ}
\newblock
 \APACrefbtitle {{Proceedings of the International Conference on Learning
  Representations}.} {{Proceedings of the International Conference on Learning
  Representations}.}
\PrintBackRefs{\CurrentBib}

\bibitem [\protect \citeauthoryear {%
Guo%
\ \protect \BOthers {.}}{%
Guo%
\ \protect \BOthers {.}}{%
{\protect \APACyear {2020}}%
}]{%
guo2020closed}
\APACinsertmetastar {%
guo2020closed}%
\begin{APACrefauthors}%
Guo, Y.%
, Chen, J.%
, Wang, J.%
, Chen, Q.%
, Cao, J.%
, Deng, Z.%
\BDBL {}Tan, M.%
\end{APACrefauthors}%
\unskip\
\newblock
\APACrefYearMonthDay{2020}{}{}.
\newblock
{\BBOQ}\APACrefatitle {Closed-loop matters: Dual regression networks for single
  image super-resolution} {Closed-loop matters: Dual regression networks for
  single image super-resolution}.{\BBCQ}
\newblock
 \APACrefbtitle {{Proceedings of the IEEE Conference on Computer Vision and
  Pattern Recognition}.} {{Proceedings of the IEEE Conference on Computer
  Vision and Pattern Recognition}.}
\PrintBackRefs{\CurrentBib}

\bibitem [\protect \citeauthoryear {%
Heusel%
, Ramsauer%
, Unterthiner%
, Nessler%
\BCBL {}\ \BBA {} Hochreiter%
}{%
Heusel%
\ \protect \BOthers {.}}{%
{\protect \APACyear {2017}}%
}]{%
heusel2017gans}
\APACinsertmetastar {%
heusel2017gans}%
\begin{APACrefauthors}%
Heusel, M.%
, Ramsauer, H.%
, Unterthiner, T.%
, Nessler, B.%
\BCBL {} Hochreiter, S.%
\end{APACrefauthors}%
\unskip\
\newblock
\APACrefYearMonthDay{2017}{}{}.
\newblock
{\BBOQ}\APACrefatitle {Gans trained by a two time-scale update rule converge to
  a local nash equilibrium} {Gans trained by a two time-scale update rule
  converge to a local nash equilibrium}.{\BBCQ}
\newblock
 \APACrefbtitle {Advances in Neural Information Processing Systems.} {Advances
  in neural information processing systems.}
\PrintBackRefs{\CurrentBib}

\bibitem [\protect \citeauthoryear {%
J\BHBI B.~Huang%
, Singh%
\BCBL {}\ \BBA {} Ahuja%
}{%
J\BHBI B.~Huang%
\ \protect \BOthers {.}}{%
{\protect \APACyear {2015}}%
}]{%
huang2015single}
\APACinsertmetastar {%
huang2015single}%
\begin{APACrefauthors}%
Huang, J\BHBI B.%
, Singh, A.%
\BCBL {} Ahuja, N.%
\end{APACrefauthors}%
\unskip\
\newblock
\APACrefYearMonthDay{2015}{}{}.
\newblock
{\BBOQ}\APACrefatitle {Single image super-resolution from transformed
  self-exemplars} {Single image super-resolution from transformed
  self-exemplars}.{\BBCQ}
\newblock
 \APACrefbtitle {{Proceedings of the IEEE Conference on Computer Vision and
  Pattern Recognition}.} {{Proceedings of the IEEE Conference on Computer
  Vision and Pattern Recognition}.}
\PrintBackRefs{\CurrentBib}

\bibitem [\protect \citeauthoryear {%
Y\BHBI C.~Huang%
\ \protect \BOthers {.}}{%
Y\BHBI C.~Huang%
\ \protect \BOthers {.}}{%
{\protect \APACyear {2021}}%
}]{%
huang2021video}
\APACinsertmetastar {%
huang2021video}%
\begin{APACrefauthors}%
Huang, Y\BHBI C.%
, Chen, Y\BHBI H.%
, Lu, C\BHBI Y.%
, Wang, H\BHBI P.%
, Peng, W\BHBI H.%
\BCBL {} Huang, C\BHBI C.%
\end{APACrefauthors}%
\unskip\
\newblock
\APACrefYearMonthDay{2021}{}{}.
\newblock
{\BBOQ}\APACrefatitle {Video Rescaling Networks with Joint Optimization
  Strategies for Downscaling and Upscaling} {Video rescaling networks with
  joint optimization strategies for downscaling and upscaling}.{\BBCQ}
\newblock
 \APACrefbtitle {{Proceedings of the IEEE Conference on Computer Vision and
  Pattern Recognition}.} {{Proceedings of the IEEE Conference on Computer
  Vision and Pattern Recognition}.}
\PrintBackRefs{\CurrentBib}

\bibitem [\protect \citeauthoryear {%
Hyvärinen%
\ \BBA {} Pajunen%
}{%
Hyvärinen%
\ \BBA {} Pajunen%
}{%
{\protect \APACyear {1999}}%
}]{%
HYVARINEN1999}
\APACinsertmetastar {%
HYVARINEN1999}%
\begin{APACrefauthors}%
Hyvärinen, A.%
\BCBT {}\ \BBA {} Pajunen, P.%
\end{APACrefauthors}%
\unskip\
\newblock
\APACrefYearMonthDay{1999}{}{}.
\newblock
{\BBOQ}\APACrefatitle {Nonlinear independent component analysis: Existence and
  uniqueness results} {Nonlinear independent component analysis: Existence and
  uniqueness results}.{\BBCQ}
\newblock
\APACjournalVolNumPages{Neural Networks}{12}{3}{429 - 439}.
\newblock

\newblock

\PrintBackRefs{\CurrentBib}

\bibitem [\protect \citeauthoryear {%
Jacobsen%
, Smeulders%
\BCBL {}\ \BBA {} Oyallon%
}{%
Jacobsen%
\ \protect \BOthers {.}}{%
{\protect \APACyear {2018}}%
}]{%
jacobsen2018revnet}
\APACinsertmetastar {%
jacobsen2018revnet}%
\begin{APACrefauthors}%
Jacobsen, J\BHBI H.%
, Smeulders, A.W.%
\BCBL {} Oyallon, E.%
\end{APACrefauthors}%
\unskip\
\newblock
\APACrefYearMonthDay{2018}{}{}.
\newblock
{\BBOQ}\APACrefatitle {i-RevNet: Deep Invertible Networks} {i-revnet: Deep
  invertible networks}.{\BBCQ}
\newblock
 \APACrefbtitle {{Proceedings of the International Conference on Learning
  Representations}.} {{Proceedings of the International Conference on Learning
  Representations}.}
\PrintBackRefs{\CurrentBib}

\bibitem [\protect \citeauthoryear {%
Jing%
, Deng%
, Xu%
, Wang%
\BCBL {}\ \BBA {} Guan%
}{%
Jing%
\ \protect \BOthers {.}}{%
{\protect \APACyear {2021}}%
}]{%
jing2021hinet}
\APACinsertmetastar {%
jing2021hinet}%
\begin{APACrefauthors}%
Jing, J.%
, Deng, X.%
, Xu, M.%
, Wang, J.%
\BCBL {} Guan, Z.%
\end{APACrefauthors}%
\unskip\
\newblock
\APACrefYearMonthDay{2021}{}{}.
\newblock
{\BBOQ}\APACrefatitle {HiNet: Deep Image Hiding by Invertible Network} {Hinet:
  Deep image hiding by invertible network}.{\BBCQ}
\newblock
 \APACrefbtitle {{Proceedings of the IEEE International Conference on Computer
  Vision}.} {{Proceedings of the IEEE International Conference on Computer
  Vision}.}
\PrintBackRefs{\CurrentBib}

\bibitem [\protect \citeauthoryear {%
Johnson%
, Alahi%
\BCBL {}\ \BBA {} Fei-Fei%
}{%
Johnson%
\ \protect \BOthers {.}}{%
{\protect \APACyear {2016}}%
}]{%
johnson2016perceptual}
\APACinsertmetastar {%
johnson2016perceptual}%
\begin{APACrefauthors}%
Johnson, J.%
, Alahi, A.%
\BCBL {} Fei-Fei, L.%
\end{APACrefauthors}%
\unskip\
\newblock
\APACrefYearMonthDay{2016}{}{}.
\newblock
{\BBOQ}\APACrefatitle {Perceptual losses for real-time style transfer and
  super-resolution} {Perceptual losses for real-time style transfer and
  super-resolution}.{\BBCQ}
\newblock
 \APACrefbtitle {{Proceedings of the European Conference on Computer Vision
  (ECCV)}.} {{Proceedings of the European Conference on Computer Vision
  (ECCV)}.}
\PrintBackRefs{\CurrentBib}

\bibitem [\protect \citeauthoryear {%
H.~Kim%
, Choi%
, Lim%
\BCBL {}\ \BBA {} Mu~Lee%
}{%
H.~Kim%
\ \protect \BOthers {.}}{%
{\protect \APACyear {2018}}%
}]{%
kim2018task}
\APACinsertmetastar {%
kim2018task}%
\begin{APACrefauthors}%
Kim, H.%
, Choi, M.%
, Lim, B.%
\BCBL {} Mu~Lee, K.%
\end{APACrefauthors}%
\unskip\
\newblock
\APACrefYearMonthDay{2018}{}{}.
\newblock
{\BBOQ}\APACrefatitle {Task-Aware Image Downscaling} {Task-aware image
  downscaling}.{\BBCQ}
\newblock
 \APACrefbtitle {{Proceedings of the European Conference on Computer Vision
  (ECCV)}.} {{Proceedings of the European Conference on Computer Vision
  (ECCV)}.}
\PrintBackRefs{\CurrentBib}

\bibitem [\protect \citeauthoryear {%
K.I.~Kim%
\ \BBA {} Kwon%
}{%
K.I.~Kim%
\ \BBA {} Kwon%
}{%
{\protect \APACyear {2010}}%
}]{%
kim2010single}
\APACinsertmetastar {%
kim2010single}%
\begin{APACrefauthors}%
Kim, K.I.%
\BCBT {}\ \BBA {} Kwon, Y.%
\end{APACrefauthors}%
\unskip\
\newblock
\APACrefYearMonthDay{2010}{}{}.
\newblock
{\BBOQ}\APACrefatitle {Single-image super-resolution using sparse regression
  and natural image prior} {Single-image super-resolution using sparse
  regression and natural image prior}.{\BBCQ}
\newblock
\APACjournalVolNumPages{IEEE Transactions on Pattern Analysis and Machine
  Intelligence}{32}{6}{1127--1133}.
\newblock

\newblock

\PrintBackRefs{\CurrentBib}

\bibitem [\protect \citeauthoryear {%
Kingma%
\ \BBA {} Ba%
}{%
Kingma%
\ \BBA {} Ba%
}{%
{\protect \APACyear {2015}}%
}]{%
kingma2014adam}
\APACinsertmetastar {%
kingma2014adam}%
\begin{APACrefauthors}%
Kingma, D.P.%
\BCBT {}\ \BBA {} Ba, J.%
\end{APACrefauthors}%
\unskip\
\newblock
\APACrefYearMonthDay{2015}{}{}.
\newblock
{\BBOQ}\APACrefatitle {Adam: A method for stochastic optimization} {Adam: A
  method for stochastic optimization}.{\BBCQ}
\newblock
 \APACrefbtitle {{Proceedings of the International Conference on Learning
  Representations}.} {{Proceedings of the International Conference on Learning
  Representations}.}
\PrintBackRefs{\CurrentBib}

\bibitem [\protect \citeauthoryear {%
Kingma%
\ \BBA {} Dhariwal%
}{%
Kingma%
\ \BBA {} Dhariwal%
}{%
{\protect \APACyear {2018}}%
}]{%
kingma2018glow}
\APACinsertmetastar {%
kingma2018glow}%
\begin{APACrefauthors}%
Kingma, D.P.%
\BCBT {}\ \BBA {} Dhariwal, P.%
\end{APACrefauthors}%
\unskip\
\newblock
\APACrefYearMonthDay{2018}{}{}.
\newblock
{\BBOQ}\APACrefatitle {Glow: Generative flow with invertible 1x1 convolutions}
  {Glow: Generative flow with invertible 1x1 convolutions}.{\BBCQ}
\newblock
 \APACrefbtitle {{Advances in Neural Information Processing Systems}.}
  {{Advances in Neural Information Processing Systems}.}
\PrintBackRefs{\CurrentBib}

\bibitem [\protect \citeauthoryear {%
Kingma%
\ \protect \BOthers {.}}{%
Kingma%
\ \protect \BOthers {.}}{%
{\protect \APACyear {2016}}%
}]{%
kingma2016improved}
\APACinsertmetastar {%
kingma2016improved}%
\begin{APACrefauthors}%
Kingma, D.P.%
, Salimans, T.%
, Jozefowicz, R.%
, Chen, X.%
, Sutskever, I.%
\BCBL {} Welling, M.%
\end{APACrefauthors}%
\unskip\
\newblock
\APACrefYearMonthDay{2016}{}{}.
\newblock
{\BBOQ}\APACrefatitle {Improved variational inference with inverse
  autoregressive flow} {Improved variational inference with inverse
  autoregressive flow}.{\BBCQ}
\newblock
 \APACrefbtitle {{Advances in Neural Information Processing Systems}.}
  {{Advances in Neural Information Processing Systems}.}
\PrintBackRefs{\CurrentBib}

\bibitem [\protect \citeauthoryear {%
Kobyzev%
, Prince%
\BCBL {}\ \BBA {} Brubaker%
}{%
Kobyzev%
\ \protect \BOthers {.}}{%
{\protect \APACyear {2020}}%
}]{%
kobyzev2020normalizing}
\APACinsertmetastar {%
kobyzev2020normalizing}%
\begin{APACrefauthors}%
Kobyzev, I.%
, Prince, S.%
\BCBL {} Brubaker, M.%
\end{APACrefauthors}%
\unskip\
\newblock
\APACrefYearMonthDay{2020}{}{}.
\newblock
{\BBOQ}\APACrefatitle {Normalizing flows: An introduction and review of current
  methods} {Normalizing flows: An introduction and review of current
  methods}.{\BBCQ}
\newblock
\APACjournalVolNumPages{IEEE Transactions on Pattern Analysis and Machine
  Intelligence}{}{}{}.
\newblock

\newblock

\PrintBackRefs{\CurrentBib}

\bibitem [\protect \citeauthoryear {%
Kopf%
, Shamir%
\BCBL {}\ \BBA {} Peers%
}{%
Kopf%
\ \protect \BOthers {.}}{%
{\protect \APACyear {2013}}%
}]{%
kopf2013content}
\APACinsertmetastar {%
kopf2013content}%
\begin{APACrefauthors}%
Kopf, J.%
, Shamir, A.%
\BCBL {} Peers, P.%
\end{APACrefauthors}%
\unskip\
\newblock
\APACrefYearMonthDay{2013}{}{}.
\newblock
{\BBOQ}\APACrefatitle {Content-adaptive image downscaling} {Content-adaptive
  image downscaling}.{\BBCQ}
\newblock
\APACjournalVolNumPages{ACM Transactions on Graphics (TOG)}{32}{6}{173}.
\newblock

\newblock

\PrintBackRefs{\CurrentBib}

\bibitem [\protect \citeauthoryear {%
Kumar%
\ \protect \BOthers {.}}{%
Kumar%
\ \protect \BOthers {.}}{%
{\protect \APACyear {2020}}%
}]{%
kumar2019videoflow}
\APACinsertmetastar {%
kumar2019videoflow}%
\begin{APACrefauthors}%
Kumar, M.%
, Babaeizadeh, M.%
, Erhan, D.%
, Finn, C.%
, Levine, S.%
, Dinh, L.%
\BCBL {} Kingma, D.%
\end{APACrefauthors}%
\unskip\
\newblock
\APACrefYearMonthDay{2020}{}{}.
\newblock
{\BBOQ}\APACrefatitle {VideoFlow: A Conditional Flow-Based Model for Stochastic
  Video Generation} {Videoflow: A conditional flow-based model for stochastic
  video generation}.{\BBCQ}
\newblock
 \APACrefbtitle {{Proceedings of the International Conference on Learning
  Representations}.} {{Proceedings of the International Conference on Learning
  Representations}.}
\PrintBackRefs{\CurrentBib}

\bibitem [\protect \citeauthoryear {%
Ledig%
\ \protect \BOthers {.}}{%
Ledig%
\ \protect \BOthers {.}}{%
{\protect \APACyear {2017}}%
}]{%
ledig2017photo}
\APACinsertmetastar {%
ledig2017photo}%
\begin{APACrefauthors}%
Ledig, C.%
, Theis, L.%
, Husz{\'a}r, F.%
, Caballero, J.%
, Cunningham, A.%
, Acosta, A.%
\BDBL {}others%
\end{APACrefauthors}%
\unskip\
\newblock
\APACrefYearMonthDay{2017}{}{}.
\newblock
{\BBOQ}\APACrefatitle {Photo-realistic single image super-resolution using a
  generative adversarial network} {Photo-realistic single image
  super-resolution using a generative adversarial network}.{\BBCQ}
\newblock
 \APACrefbtitle {{Proceedings of the IEEE Conference on Computer Vision and
  Pattern Recognition}.} {{Proceedings of the IEEE Conference on Computer
  Vision and Pattern Recognition}.}
\PrintBackRefs{\CurrentBib}

\bibitem [\protect \citeauthoryear {%
Levin%
, Lischinski%
\BCBL {}\ \BBA {} Weiss%
}{%
Levin%
\ \protect \BOthers {.}}{%
{\protect \APACyear {2004}}%
}]{%
levin2004colorization}
\APACinsertmetastar {%
levin2004colorization}%
\begin{APACrefauthors}%
Levin, A.%
, Lischinski, D.%
\BCBL {} Weiss, Y.%
\end{APACrefauthors}%
\unskip\
\newblock
\APACrefYearMonthDay{2004}{}{}.
\newblock
{\BBOQ}\APACrefatitle {Colorization using optimization} {Colorization using
  optimization}.{\BBCQ}
\newblock
 \APACrefbtitle {{ACM SIGGRAPH}.} {{ACM SIGGRAPH}.}
\PrintBackRefs{\CurrentBib}

\bibitem [\protect \citeauthoryear {%
M.~Li%
, Zuo%
, Gu%
, You%
\BCBL {}\ \BBA {} Zhang%
}{%
M.~Li%
\ \protect \BOthers {.}}{%
{\protect \APACyear {2020}}%
}]{%
li2020learning}
\APACinsertmetastar {%
li2020learning}%
\begin{APACrefauthors}%
Li, M.%
, Zuo, W.%
, Gu, S.%
, You, J.%
\BCBL {} Zhang, D.%
\end{APACrefauthors}%
\unskip\
\newblock
\APACrefYearMonthDay{2020}{}{}.
\newblock
{\BBOQ}\APACrefatitle {Learning content-weighted deep image compression}
  {Learning content-weighted deep image compression}.{\BBCQ}
\newblock
\APACjournalVolNumPages{IEEE Transactions on Pattern Analysis and Machine
  Intelligence}{}{}{}.
\newblock

\newblock

\PrintBackRefs{\CurrentBib}

\bibitem [\protect \citeauthoryear {%
Y.~Li%
\ \protect \BOthers {.}}{%
Y.~Li%
\ \protect \BOthers {.}}{%
{\protect \APACyear {2018}}%
}]{%
li2018learning}
\APACinsertmetastar {%
li2018learning}%
\begin{APACrefauthors}%
Li, Y.%
, Liu, D.%
, Li, H.%
, Li, L.%
, Li, Z.%
\BCBL {} Wu, F.%
\end{APACrefauthors}%
\unskip\
\newblock
\APACrefYearMonthDay{2018}{}{}.
\newblock
{\BBOQ}\APACrefatitle {Learning a convolutional neural network for image
  compact-resolution} {Learning a convolutional neural network for image
  compact-resolution}.{\BBCQ}
\newblock
\APACjournalVolNumPages{IEEE Transactions on Image
  Processing}{28}{3}{1092--1107}.
\newblock

\newblock

\PrintBackRefs{\CurrentBib}

\bibitem [\protect \citeauthoryear {%
Z.~Li%
, Li%
, Zhang%
, Wang%
\BCBL {}\ \BBA {} Xue%
}{%
Z.~Li%
\ \protect \BOthers {.}}{%
{\protect \APACyear {2019}}%
}]{%
li2019multi}
\APACinsertmetastar {%
li2019multi}%
\begin{APACrefauthors}%
Li, Z.%
, Li, S.%
, Zhang, N.%
, Wang, L.%
\BCBL {} Xue, Z.%
\end{APACrefauthors}%
\unskip\
\newblock
\APACrefYearMonthDay{2019}{}{}.
\newblock
{\BBOQ}\APACrefatitle {Multi-Scale Invertible Network for Image
  Super-Resolution} {Multi-scale invertible network for image
  super-resolution}.{\BBCQ}
\newblock
 \APACrefbtitle {{Proceedings of the ACM Multimedia Asia}.} {{Proceedings of
  the ACM Multimedia Asia}.}
\PrintBackRefs{\CurrentBib}

\bibitem [\protect \citeauthoryear {%
Lim%
, Son%
, Kim%
, Nah%
\BCBL {}\ \BBA {} Mu~Lee%
}{%
Lim%
\ \protect \BOthers {.}}{%
{\protect \APACyear {2017}}%
}]{%
lim2017enhanced}
\APACinsertmetastar {%
lim2017enhanced}%
\begin{APACrefauthors}%
Lim, B.%
, Son, S.%
, Kim, H.%
, Nah, S.%
\BCBL {} Mu~Lee, K.%
\end{APACrefauthors}%
\unskip\
\newblock
\APACrefYearMonthDay{2017}{}{}.
\newblock
{\BBOQ}\APACrefatitle {Enhanced deep residual networks for single image
  super-resolution} {Enhanced deep residual networks for single image
  super-resolution}.{\BBCQ}
\newblock
 \APACrefbtitle {{Proceedings of the IEEE Conference on Computer Vision and
  Pattern Recognition Workshops}.} {{Proceedings of the IEEE Conference on
  Computer Vision and Pattern Recognition Workshops}.}
\PrintBackRefs{\CurrentBib}

\bibitem [\protect \citeauthoryear {%
Lin%
\ \BBA {} Dong%
}{%
Lin%
\ \BBA {} Dong%
}{%
{\protect \APACyear {2006}}%
}]{%
lin2006adaptive}
\APACinsertmetastar {%
lin2006adaptive}%
\begin{APACrefauthors}%
Lin, W.%
\BCBT {}\ \BBA {} Dong, L.%
\end{APACrefauthors}%
\unskip\
\newblock
\APACrefYearMonthDay{2006}{}{}.
\newblock
{\BBOQ}\APACrefatitle {Adaptive downsampling to improve image compression at
  low bit rates} {Adaptive downsampling to improve image compression at low bit
  rates}.{\BBCQ}
\newblock
\APACjournalVolNumPages{IEEE Transactions on Image
  Processing}{15}{9}{2513--2521}.
\newblock

\newblock

\PrintBackRefs{\CurrentBib}

\bibitem [\protect \citeauthoryear {%
C.~Liu%
, Tang%
, Qin%
, Wang%
\BCBL {}\ \BBA {} Liu%
}{%
C.~Liu%
\ \protect \BOthers {.}}{%
{\protect \APACyear {2021}}%
}]{%
liu2021generative}
\APACinsertmetastar {%
liu2021generative}%
\begin{APACrefauthors}%
Liu, C.%
, Tang, H.%
, Qin, T.%
, Wang, J.%
\BCBL {} Liu, T\BHBI Y.%
\end{APACrefauthors}%
\unskip\
\newblock
\APACrefYearMonthDay{2021}{}{}.
\newblock
{\BBOQ}\APACrefatitle {On the Generative Utility of Cyclic Conditionals} {On
  the generative utility of cyclic conditionals}.{\BBCQ}
\newblock
 \APACrefbtitle {Advances in Neural Information Processing Systems.} {Advances
  in neural information processing systems.}
\PrintBackRefs{\CurrentBib}

\bibitem [\protect \citeauthoryear {%
J.~Liu%
, He%
\BCBL {}\ \BBA {} Lau%
}{%
J.~Liu%
\ \protect \BOthers {.}}{%
{\protect \APACyear {2017}}%
}]{%
liu2017l_}
\APACinsertmetastar {%
liu2017l_}%
\begin{APACrefauthors}%
Liu, J.%
, He, S.%
\BCBL {} Lau, R.W.%
\end{APACrefauthors}%
\unskip\
\newblock
\APACrefYearMonthDay{2017}{}{}.
\newblock
{\BBOQ}\APACrefatitle {$ L\_ $\{$0$\}$ $-regularized image downscaling} {$ l\_
  $\{$0$\}$ $-regularized image downscaling}.{\BBCQ}
\newblock
\APACjournalVolNumPages{IEEE Transactions on Image
  Processing}{27}{3}{1076--1085}.
\newblock

\newblock

\PrintBackRefs{\CurrentBib}

\bibitem [\protect \citeauthoryear {%
Q.~Liu%
, Liu%
, Xie%
, Wang%
\BCBL {}\ \BBA {} Liang%
}{%
Q.~Liu%
\ \protect \BOthers {.}}{%
{\protect \APACyear {2015}}%
}]{%
liu2015gcsdecolor}
\APACinsertmetastar {%
liu2015gcsdecolor}%
\begin{APACrefauthors}%
Liu, Q.%
, Liu, P.X.%
, Xie, W.%
, Wang, Y.%
\BCBL {} Liang, D.%
\end{APACrefauthors}%
\unskip\
\newblock
\APACrefYearMonthDay{2015}{}{}.
\newblock
{\BBOQ}\APACrefatitle {GcsDecolor: gradient correlation similarity for
  efficient contrast preserving decolorization} {Gcsdecolor: gradient
  correlation similarity for efficient contrast preserving
  decolorization}.{\BBCQ}
\newblock
\APACjournalVolNumPages{IEEE Transactions on Image
  Processing}{24}{9}{2889--2904}.
\newblock

\newblock

\PrintBackRefs{\CurrentBib}

\bibitem [\protect \citeauthoryear {%
Y.~Liu%
\ \protect \BOthers {.}}{%
Y.~Liu%
\ \protect \BOthers {.}}{%
{\protect \APACyear {2021}}%
}]{%
liu2021invertible}
\APACinsertmetastar {%
liu2021invertible}%
\begin{APACrefauthors}%
Liu, Y.%
, Qin, Z.%
, Anwar, S.%
, Ji, P.%
, Kim, D.%
, Caldwell, S.%
\BCBL {} Gedeon, T.%
\end{APACrefauthors}%
\unskip\
\newblock
\APACrefYearMonthDay{2021}{}{}.
\newblock
{\BBOQ}\APACrefatitle {Invertible Denoising Network: A Light Solution for Real
  Noise Removal} {Invertible denoising network: A light solution for real noise
  removal}.{\BBCQ}
\newblock
 \APACrefbtitle {{Proceedings of the IEEE Conference on Computer Vision and
  Pattern Recognition}.} {{Proceedings of the IEEE Conference on Computer
  Vision and Pattern Recognition}.}
\PrintBackRefs{\CurrentBib}

\bibitem [\protect \citeauthoryear {%
C.~Lu%
, Chen%
, Li%
, Wang%
\BCBL {}\ \BBA {} Zhu%
}{%
C.~Lu%
\ \protect \BOthers {.}}{%
{\protect \APACyear {2021}}%
}]{%
lu2021implicit}
\APACinsertmetastar {%
lu2021implicit}%
\begin{APACrefauthors}%
Lu, C.%
, Chen, J.%
, Li, C.%
, Wang, Q.%
\BCBL {} Zhu, J.%
\end{APACrefauthors}%
\unskip\
\newblock
\APACrefYearMonthDay{2021}{}{}.
\newblock
{\BBOQ}\APACrefatitle {Implicit Normalizing Flows} {Implicit normalizing
  flows}.{\BBCQ}
\newblock
 \APACrefbtitle {International Conference on Learning Representations.}
  {International conference on learning representations.}
\PrintBackRefs{\CurrentBib}

\bibitem [\protect \citeauthoryear {%
S\BHBI P.~Lu%
, Wang%
, Zhong%
\BCBL {}\ \BBA {} Rosin%
}{%
S\BHBI P.~Lu%
\ \protect \BOthers {.}}{%
{\protect \APACyear {2021}}%
}]{%
lu2021large}
\APACinsertmetastar {%
lu2021large}%
\begin{APACrefauthors}%
Lu, S\BHBI P.%
, Wang, R.%
, Zhong, T.%
\BCBL {} Rosin, P.L.%
\end{APACrefauthors}%
\unskip\
\newblock
\APACrefYearMonthDay{2021}{}{}.
\newblock
{\BBOQ}\APACrefatitle {Large-Capacity Image Steganography Based on Invertible
  Neural Networks} {Large-capacity image steganography based on invertible
  neural networks}.{\BBCQ}
\newblock
 \APACrefbtitle {{Proceedings of the IEEE Conference on Computer Vision and
  Pattern Recognition}.} {{Proceedings of the IEEE Conference on Computer
  Vision and Pattern Recognition}.}
\PrintBackRefs{\CurrentBib}

\bibitem [\protect \citeauthoryear {%
Lugmayr%
, Danelljan%
, Van~Gool%
\BCBL {}\ \BBA {} Timofte%
}{%
Lugmayr%
\ \protect \BOthers {.}}{%
{\protect \APACyear {2020}}%
}]{%
lugmayr2020srflow}
\APACinsertmetastar {%
lugmayr2020srflow}%
\begin{APACrefauthors}%
Lugmayr, A.%
, Danelljan, M.%
, Van~Gool, L.%
\BCBL {} Timofte, R.%
\end{APACrefauthors}%
\unskip\
\newblock
\APACrefYearMonthDay{2020}{}{}.
\newblock
{\BBOQ}\APACrefatitle {Srflow: Learning the super-resolution space with
  normalizing flow} {Srflow: Learning the super-resolution space with
  normalizing flow}.{\BBCQ}
\newblock
 \APACrefbtitle {{Proceedings of the European Conference on Computer Vision
  (ECCV)}.} {{Proceedings of the European Conference on Computer Vision
  (ECCV)}.}
\PrintBackRefs{\CurrentBib}

\bibitem [\protect \citeauthoryear {%
Martin%
, Fowlkes%
, Tal%
, Malik%
\BCBL {}\ \protect \BOthers {.}}{%
Martin%
\ \protect \BOthers {.}}{%
{\protect \APACyear {2001}}%
}]{%
martin2001database}
\APACinsertmetastar {%
martin2001database}%
\begin{APACrefauthors}%
Martin, D.%
, Fowlkes, C.%
, Tal, D.%
, Malik, J.%
\BCBL {}\ \BOthersPeriod {.}\end{APACrefauthors}%
\unskip\
\newblock
\APACrefYearMonthDay{2001}{}{}.
\newblock
{\BBOQ}\APACrefatitle {A database of human segmented natural images and its
  application to evaluating segmentation algorithms and measuring ecological
  statistics} {A database of human segmented natural images and its application
  to evaluating segmentation algorithms and measuring ecological
  statistics}.{\BBCQ}
\newblock
 \APACrefbtitle {{Proceedings of the IEEE International Conference on Computer
  Vision}.} {{Proceedings of the IEEE International Conference on Computer
  Vision}.}
\PrintBackRefs{\CurrentBib}

\bibitem [\protect \citeauthoryear {%
Minnen%
, Ball{\'e}%
\BCBL {}\ \BBA {} Toderici%
}{%
Minnen%
\ \protect \BOthers {.}}{%
{\protect \APACyear {2018}}%
}]{%
minnen2018joint}
\APACinsertmetastar {%
minnen2018joint}%
\begin{APACrefauthors}%
Minnen, D.%
, Ball{\'e}, J.%
\BCBL {} Toderici, G.D.%
\end{APACrefauthors}%
\unskip\
\newblock
\APACrefYearMonthDay{2018}{}{}.
\newblock
{\BBOQ}\APACrefatitle {Joint autoregressive and hierarchical priors for learned
  image compression} {Joint autoregressive and hierarchical priors for learned
  image compression}.{\BBCQ}
\newblock
 \APACrefbtitle {{Advances in Neural Information Processing Systems}.}
  {{Advances in Neural Information Processing Systems}.}
\PrintBackRefs{\CurrentBib}

\bibitem [\protect \citeauthoryear {%
Mitchell%
\ \BBA {} Netravali%
}{%
Mitchell%
\ \BBA {} Netravali%
}{%
{\protect \APACyear {1988}}%
}]{%
mitchell1988reconstruction}
\APACinsertmetastar {%
mitchell1988reconstruction}%
\begin{APACrefauthors}%
Mitchell, D.P.%
\BCBT {}\ \BBA {} Netravali, A.N.%
\end{APACrefauthors}%
\unskip\
\newblock
\APACrefYearMonthDay{1988}{}{}.
\newblock
{\BBOQ}\APACrefatitle {Reconstruction filters in computer-graphics}
  {Reconstruction filters in computer-graphics}.{\BBCQ}
\newblock
 \APACrefbtitle {{ACM Siggraph Computer Graphics}} {{ACM Siggraph Computer
  Graphics}}\ (\BVOL\ 22-4, \BPGS\ 221--228).
\PrintBackRefs{\CurrentBib}

\bibitem [\protect \citeauthoryear {%
Oeztireli%
\ \BBA {} Gross%
}{%
Oeztireli%
\ \BBA {} Gross%
}{%
{\protect \APACyear {2015}}%
}]{%
oeztireli2015perceptually}
\APACinsertmetastar {%
oeztireli2015perceptually}%
\begin{APACrefauthors}%
Oeztireli, A.C.%
\BCBT {}\ \BBA {} Gross, M.%
\end{APACrefauthors}%
\unskip\
\newblock
\APACrefYearMonthDay{2015}{}{}.
\newblock
{\BBOQ}\APACrefatitle {Perceptually based downscaling of images} {Perceptually
  based downscaling of images}.{\BBCQ}
\newblock
\APACjournalVolNumPages{ACM Transactions on Graphics (TOG)}{34}{4}{77}.
\newblock

\newblock

\PrintBackRefs{\CurrentBib}

\bibitem [\protect \citeauthoryear {%
Ren%
, Padilla%
\BCBL {}\ \BBA {} Malof%
}{%
Ren%
\ \protect \BOthers {.}}{%
{\protect \APACyear {2020}}%
}]{%
NEURIPS2020_007ff380}
\APACinsertmetastar {%
NEURIPS2020_007ff380}%
\begin{APACrefauthors}%
Ren, S.%
, Padilla, W.%
\BCBL {} Malof, J.%
\end{APACrefauthors}%
\unskip\
\newblock
\APACrefYearMonthDay{2020}{}{}.
\newblock
{\BBOQ}\APACrefatitle {Benchmarking Deep Inverse Models over time, and the
  Neural-Adjoint method} {Benchmarking deep inverse models over time, and the
  neural-adjoint method}.{\BBCQ}
\newblock
 \APACrefbtitle {{Advances in Neural Information Processing Systems}.}
  {{Advances in Neural Information Processing Systems}.}
\PrintBackRefs{\CurrentBib}

\bibitem [\protect \citeauthoryear {%
Rezende%
\ \BBA {} Mohamed%
}{%
Rezende%
\ \BBA {} Mohamed%
}{%
{\protect \APACyear {2015}}%
}]{%
rezende2015variational}
\APACinsertmetastar {%
rezende2015variational}%
\begin{APACrefauthors}%
Rezende, D.%
\BCBT {}\ \BBA {} Mohamed, S.%
\end{APACrefauthors}%
\unskip\
\newblock
\APACrefYearMonthDay{2015}{}{}.
\newblock
{\BBOQ}\APACrefatitle {Variational Inference with Normalizing Flows}
  {Variational inference with normalizing flows}.{\BBCQ}
\newblock
 \APACrefbtitle {{Proceedings of the International Conference on Machine
  Learning}.} {{Proceedings of the International Conference on Machine
  Learning}.}
\PrintBackRefs{\CurrentBib}

\bibitem [\protect \citeauthoryear {%
Rippel%
\ \BBA {} Bourdev%
}{%
Rippel%
\ \BBA {} Bourdev%
}{%
{\protect \APACyear {2017}}%
}]{%
rippel2017real}
\APACinsertmetastar {%
rippel2017real}%
\begin{APACrefauthors}%
Rippel, O.%
\BCBT {}\ \BBA {} Bourdev, L.%
\end{APACrefauthors}%
\unskip\
\newblock
\APACrefYearMonthDay{2017}{}{}.
\newblock
{\BBOQ}\APACrefatitle {Real-time adaptive image compression} {Real-time
  adaptive image compression}.{\BBCQ}
\newblock
 \APACrefbtitle {{Proceedings of the International Conference on Machine
  Learning}.} {{Proceedings of the International Conference on Machine
  Learning}.}
\PrintBackRefs{\CurrentBib}

\bibitem [\protect \citeauthoryear {%
Schulter%
, Leistner%
\BCBL {}\ \BBA {} Bischof%
}{%
Schulter%
\ \protect \BOthers {.}}{%
{\protect \APACyear {2015}}%
}]{%
schulter2015fast}
\APACinsertmetastar {%
schulter2015fast}%
\begin{APACrefauthors}%
Schulter, S.%
, Leistner, C.%
\BCBL {} Bischof, H.%
\end{APACrefauthors}%
\unskip\
\newblock
\APACrefYearMonthDay{2015}{}{}.
\newblock
{\BBOQ}\APACrefatitle {Fast and accurate image upscaling with super-resolution
  forests} {Fast and accurate image upscaling with super-resolution
  forests}.{\BBCQ}
\newblock
 \APACrefbtitle {{Proceedings of the IEEE Conference on Computer Vision and
  Pattern Recognition}.} {{Proceedings of the IEEE Conference on Computer
  Vision and Pattern Recognition}.}
\PrintBackRefs{\CurrentBib}

\bibitem [\protect \citeauthoryear {%
Shannon%
}{%
Shannon%
}{%
{\protect \APACyear {1949}}%
}]{%
shannon1949communication}
\APACinsertmetastar {%
shannon1949communication}%
\begin{APACrefauthors}%
Shannon, C.E.%
\end{APACrefauthors}%
\unskip\
\newblock
\APACrefYearMonthDay{1949}{}{}.
\newblock
{\BBOQ}\APACrefatitle {Communication in the presence of noise} {Communication
  in the presence of noise}.{\BBCQ}
\newblock
\APACjournalVolNumPages{Proceedings of the IRE}{37}{1}{10--21}.
\newblock

\newblock

\PrintBackRefs{\CurrentBib}

\bibitem [\protect \citeauthoryear {%
Shen%
, Xue%
\BCBL {}\ \BBA {} Wang%
}{%
Shen%
\ \protect \BOthers {.}}{%
{\protect \APACyear {2011}}%
}]{%
shen2011down}
\APACinsertmetastar {%
shen2011down}%
\begin{APACrefauthors}%
Shen, M.%
, Xue, P.%
\BCBL {} Wang, C.%
\end{APACrefauthors}%
\unskip\
\newblock
\APACrefYearMonthDay{2011}{}{}.
\newblock
{\BBOQ}\APACrefatitle {Down-sampling based video coding using super-resolution
  technique} {Down-sampling based video coding using super-resolution
  technique}.{\BBCQ}
\newblock
\APACjournalVolNumPages{IEEE Transactions on Circuits and Systems for Video
  Technology}{21}{6}{755--765}.
\newblock

\newblock

\PrintBackRefs{\CurrentBib}

\bibitem [\protect \citeauthoryear {%
Sneyers%
\ \BBA {} Wuille%
}{%
Sneyers%
\ \BBA {} Wuille%
}{%
{\protect \APACyear {2016}}%
}]{%
sneyers2016flif}
\APACinsertmetastar {%
sneyers2016flif}%
\begin{APACrefauthors}%
Sneyers, J.%
\BCBT {}\ \BBA {} Wuille, P.%
\end{APACrefauthors}%
\unskip\
\newblock
\APACrefYearMonthDay{2016}{}{}.
\newblock
{\BBOQ}\APACrefatitle {FLIF: Free lossless image format based on MANIAC
  compression} {Flif: Free lossless image format based on maniac
  compression}.{\BBCQ}
\newblock
 \APACrefbtitle {{Proceedings of the IEEE International Conference on Image
  Processing (ICIP)}.} {{Proceedings of the IEEE International Conference on
  Image Processing (ICIP)}.}
\PrintBackRefs{\CurrentBib}

\bibitem [\protect \citeauthoryear {%
Sullivan%
, Ohm%
, Han%
\BCBL {}\ \BBA {} Wiegand%
}{%
Sullivan%
\ \protect \BOthers {.}}{%
{\protect \APACyear {2013}}%
}]{%
2013Overview}
\APACinsertmetastar {%
2013Overview}%
\begin{APACrefauthors}%
Sullivan, G.J.%
, Ohm, J.R.%
, Han, W.J.%
\BCBL {} Wiegand, T.%
\end{APACrefauthors}%
\unskip\
\newblock
\APACrefYearMonthDay{2013}{}{}.
\newblock
{\BBOQ}\APACrefatitle {Overview of the High Efficiency Video Coding (HEVC)
  Standard} {Overview of the high efficiency video coding (hevc)
  standard}.{\BBCQ}
\newblock
\APACjournalVolNumPages{IEEE Transactions on Circuits and Systems for Video
  Technology}{22}{12}{1649-1668}.
\newblock

\newblock

\PrintBackRefs{\CurrentBib}

\bibitem [\protect \citeauthoryear {%
Sun%
\ \BBA {} Chen%
}{%
Sun%
\ \BBA {} Chen%
}{%
{\protect \APACyear {2020}}%
}]{%
sun2020learned}
\APACinsertmetastar {%
sun2020learned}%
\begin{APACrefauthors}%
Sun, W.%
\BCBT {}\ \BBA {} Chen, Z.%
\end{APACrefauthors}%
\unskip\
\newblock
\APACrefYearMonthDay{2020}{}{}.
\newblock
{\BBOQ}\APACrefatitle {Learned image downscaling for upscaling using content
  adaptive resampler} {Learned image downscaling for upscaling using content
  adaptive resampler}.{\BBCQ}
\newblock
\APACjournalVolNumPages{IEEE Transactions on Image
  Processing}{29}{}{4027--4040}.
\newblock

\newblock

\PrintBackRefs{\CurrentBib}

\bibitem [\protect \citeauthoryear {%
Teshima%
\ \protect \BOthers {.}}{%
Teshima%
\ \protect \BOthers {.}}{%
{\protect \APACyear {2020}}%
}]{%
NEURIPS2020_2290a738}
\APACinsertmetastar {%
NEURIPS2020_2290a738}%
\begin{APACrefauthors}%
Teshima, T.%
, Ishikawa, I.%
, Tojo, K.%
, Oono, K.%
, Ikeda, M.%
\BCBL {} Sugiyama, M.%
\end{APACrefauthors}%
\unskip\
\newblock
\APACrefYearMonthDay{2020}{}{}.
\newblock
{\BBOQ}\APACrefatitle {Coupling-based Invertible Neural Networks Are Universal
  Diffeomorphism Approximators} {Coupling-based invertible neural networks are
  universal diffeomorphism approximators}.{\BBCQ}
\newblock
 \APACrefbtitle {Advances in Neural Information Processing Systems.} {Advances
  in neural information processing systems.}
\PrintBackRefs{\CurrentBib}

\bibitem [\protect \citeauthoryear {%
Tian%
\ \protect \BOthers {.}}{%
Tian%
\ \protect \BOthers {.}}{%
{\protect \APACyear {2021}}%
}]{%
tian2021self}
\APACinsertmetastar {%
tian2021self}%
\begin{APACrefauthors}%
Tian, Y.%
, Lu, G.%
, Min, X.%
, Che, Z.%
, Zhai, G.%
, Guo, G.%
\BCBL {} Gao, Z.%
\end{APACrefauthors}%
\unskip\
\newblock
\APACrefYearMonthDay{2021}{}{}.
\newblock
{\BBOQ}\APACrefatitle {Self-Conditioned Probabilistic Learning of Video
  Rescaling} {Self-conditioned probabilistic learning of video
  rescaling}.{\BBCQ}
\newblock
 \APACrefbtitle {{Proceedings of the IEEE International Conference on Computer
  Vision}.} {{Proceedings of the IEEE International Conference on Computer
  Vision}.}
\PrintBackRefs{\CurrentBib}

\bibitem [\protect \citeauthoryear {%
van~der Ouderaa%
\ \BBA {} Worrall%
}{%
van~der Ouderaa%
\ \BBA {} Worrall%
}{%
{\protect \APACyear {2019}}%
}]{%
van2019reversible}
\APACinsertmetastar {%
van2019reversible}%
\begin{APACrefauthors}%
van~der Ouderaa, T.F.%
\BCBT {}\ \BBA {} Worrall, D.E.%
\end{APACrefauthors}%
\unskip\
\newblock
\APACrefYearMonthDay{2019}{}{}.
\newblock
{\BBOQ}\APACrefatitle {Reversible GANs for Memory-efficient Image-to-Image
  Translation} {Reversible gans for memory-efficient image-to-image
  translation}.{\BBCQ}
\newblock
 \APACrefbtitle {{Proceedings of the IEEE Conference on Computer Vision and
  Pattern Recognition}.} {{Proceedings of the IEEE Conference on Computer
  Vision and Pattern Recognition}.}
\PrintBackRefs{\CurrentBib}

\bibitem [\protect \citeauthoryear {%
X.~Wang%
\ \protect \BOthers {.}}{%
X.~Wang%
\ \protect \BOthers {.}}{%
{\protect \APACyear {2018}}%
}]{%
wang2018esrgan}
\APACinsertmetastar {%
wang2018esrgan}%
\begin{APACrefauthors}%
Wang, X.%
, Yu, K.%
, Wu, S.%
, Gu, J.%
, Liu, Y.%
, Dong, C.%
\BDBL {}Loy, C.C.%
\end{APACrefauthors}%
\unskip\
\newblock
\APACrefYearMonthDay{2018}{}{}.
\newblock
{\BBOQ}\APACrefatitle {ESRGAN: Enhanced super-resolution generative adversarial
  networks} {Esrgan: Enhanced super-resolution generative adversarial
  networks}.{\BBCQ}
\newblock
 \APACrefbtitle {{European Conference on Computer Vision Workshops (ECCVW)}.}
  {{European Conference on Computer Vision Workshops (ECCVW)}.}
\PrintBackRefs{\CurrentBib}

\bibitem [\protect \citeauthoryear {%
Y.~Wang%
, Xiao%
, Liu%
, Zheng%
\BCBL {}\ \BBA {} Liu%
}{%
Y.~Wang%
\ \protect \BOthers {.}}{%
{\protect \APACyear {2020}}%
}]{%
wang2020modeling}
\APACinsertmetastar {%
wang2020modeling}%
\begin{APACrefauthors}%
Wang, Y.%
, Xiao, M.%
, Liu, C.%
, Zheng, S.%
\BCBL {} Liu, T\BHBI Y.%
\end{APACrefauthors}%
\unskip\
\newblock
\APACrefYearMonthDay{2020}{}{}.
\newblock
{\BBOQ}\APACrefatitle {Modeling Lost Information in Lossy Image Compression}
  {Modeling lost information in lossy image compression}.{\BBCQ}
\newblock
\APACjournalVolNumPages{{arXiv preprint arXiv:2006.11999}}{}{}{}.
\newblock

\newblock

\PrintBackRefs{\CurrentBib}

\bibitem [\protect \citeauthoryear {%
Z.~Wang%
, Bovik%
, Sheikh%
, Simoncelli%
\BCBL {}\ \protect \BOthers {.}}{%
Z.~Wang%
\ \protect \BOthers {.}}{%
{\protect \APACyear {2004}}%
}]{%
wang2004image}
\APACinsertmetastar {%
wang2004image}%
\begin{APACrefauthors}%
Wang, Z.%
, Bovik, A.C.%
, Sheikh, H.R.%
, Simoncelli, E.P.%
\BCBL {}\ \BOthersPeriod {.}\end{APACrefauthors}%
\unskip\
\newblock
\APACrefYearMonthDay{2004}{}{}.
\newblock
{\BBOQ}\APACrefatitle {Image quality assessment: from error visibility to
  structural similarity} {Image quality assessment: from error visibility to
  structural similarity}.{\BBCQ}
\newblock
\APACjournalVolNumPages{IEEE Transactions on Image
  Processing}{13}{4}{600--612}.
\newblock

\newblock

\PrintBackRefs{\CurrentBib}

\bibitem [\protect \citeauthoryear {%
Weber%
, Waechter%
, Amend%
, Guthe%
\BCBL {}\ \BBA {} Goesele%
}{%
Weber%
\ \protect \BOthers {.}}{%
{\protect \APACyear {2016}}%
}]{%
weber2016rapid}
\APACinsertmetastar {%
weber2016rapid}%
\begin{APACrefauthors}%
Weber, N.%
, Waechter, M.%
, Amend, S.C.%
, Guthe, S.%
\BCBL {} Goesele, M.%
\end{APACrefauthors}%
\unskip\
\newblock
\APACrefYearMonthDay{2016}{}{}.
\newblock
{\BBOQ}\APACrefatitle {Rapid, detail-preserving image downscaling} {Rapid,
  detail-preserving image downscaling}.{\BBCQ}
\newblock
\APACjournalVolNumPages{ACM Transactions on Graphics (TOG)}{35}{6}{205}.
\newblock

\newblock

\PrintBackRefs{\CurrentBib}

\bibitem [\protect \citeauthoryear {%
Wu%
, Zhang%
\BCBL {}\ \BBA {} Wang%
}{%
Wu%
\ \protect \BOthers {.}}{%
{\protect \APACyear {2009}}%
}]{%
wu2009low}
\APACinsertmetastar {%
wu2009low}%
\begin{APACrefauthors}%
Wu, X.%
, Zhang, X.%
\BCBL {} Wang, X.%
\end{APACrefauthors}%
\unskip\
\newblock
\APACrefYearMonthDay{2009}{}{}.
\newblock
{\BBOQ}\APACrefatitle {Low bit-rate image compression via adaptive
  down-sampling and constrained least squares upconversion} {Low bit-rate image
  compression via adaptive down-sampling and constrained least squares
  upconversion}.{\BBCQ}
\newblock
\APACjournalVolNumPages{IEEE Transactions on Image
  Processing}{18}{3}{552--561}.
\newblock

\newblock

\PrintBackRefs{\CurrentBib}

\bibitem [\protect \citeauthoryear {%
Xia%
, Liu%
\BCBL {}\ \BBA {} Wong%
}{%
Xia%
\ \protect \BOthers {.}}{%
{\protect \APACyear {2018}}%
}]{%
xia2018invertible}
\APACinsertmetastar {%
xia2018invertible}%
\begin{APACrefauthors}%
Xia, M.%
, Liu, X.%
\BCBL {} Wong, T\BHBI T.%
\end{APACrefauthors}%
\unskip\
\newblock
\APACrefYearMonthDay{2018}{}{}.
\newblock
{\BBOQ}\APACrefatitle {Invertible grayscale} {Invertible grayscale}.{\BBCQ}
\newblock
\APACjournalVolNumPages{ACM Transactions on Graphics (TOG)}{37}{6}{1--10}.
\newblock

\newblock

\PrintBackRefs{\CurrentBib}

\bibitem [\protect \citeauthoryear {%
Xiao%
\ \protect \BOthers {.}}{%
Xiao%
\ \protect \BOthers {.}}{%
{\protect \APACyear {2020}}%
}]{%
xiao2020invertible}
\APACinsertmetastar {%
xiao2020invertible}%
\begin{APACrefauthors}%
Xiao, M.%
, Zheng, S.%
, Liu, C.%
, Wang, Y.%
, He, D.%
, Ke, G.%
\BDBL {}Liu, T\BHBI Y.%
\end{APACrefauthors}%
\unskip\
\newblock
\APACrefYearMonthDay{2020}{}{}.
\newblock
{\BBOQ}\APACrefatitle {Invertible image rescaling} {Invertible image
  rescaling}.{\BBCQ}
\newblock
 \APACrefbtitle {{Proceedings of the European Conference on Computer Vision
  (ECCV)}.} {{Proceedings of the European Conference on Computer Vision
  (ECCV)}.}
\PrintBackRefs{\CurrentBib}

\bibitem [\protect \citeauthoryear {%
Xie%
, Cheng%
\BCBL {}\ \BBA {} Chen%
}{%
Xie%
\ \protect \BOthers {.}}{%
{\protect \APACyear {2021}}%
}]{%
xie2021enhanced}
\APACinsertmetastar {%
xie2021enhanced}%
\begin{APACrefauthors}%
Xie, Y.%
, Cheng, K.L.%
\BCBL {} Chen, Q.%
\end{APACrefauthors}%
\unskip\
\newblock
\APACrefYearMonthDay{2021}{}{}.
\newblock
{\BBOQ}\APACrefatitle {Enhanced invertible encoding for learned image
  compression} {Enhanced invertible encoding for learned image
  compression}.{\BBCQ}
\newblock
 \APACrefbtitle {{Proceedings of the 29th ACM International Conference on
  Multimedia}.} {{Proceedings of the 29th ACM International Conference on
  Multimedia}.}
\PrintBackRefs{\CurrentBib}

\bibitem [\protect \citeauthoryear {%
J.~Xing%
, Hu%
\BCBL {}\ \BBA {} Wong%
}{%
J.~Xing%
\ \protect \BOthers {.}}{%
{\protect \APACyear {2022}}%
}]{%
xing2022scale}
\APACinsertmetastar {%
xing2022scale}%
\begin{APACrefauthors}%
Xing, J.%
, Hu, W.%
\BCBL {} Wong, T\BHBI T.%
\end{APACrefauthors}%
\unskip\
\newblock
\APACrefYearMonthDay{2022}{}{}.
\newblock
{\BBOQ}\APACrefatitle {Scale-arbitrary Invertible Image Downscaling}
  {Scale-arbitrary invertible image downscaling}.{\BBCQ}
\newblock
\APACjournalVolNumPages{{arXiv preprint arXiv:2201.12576}}{}{}{}.
\newblock

\newblock

\PrintBackRefs{\CurrentBib}

\bibitem [\protect \citeauthoryear {%
Y.~Xing%
, Qian%
\BCBL {}\ \BBA {} Chen%
}{%
Y.~Xing%
\ \protect \BOthers {.}}{%
{\protect \APACyear {2021}}%
}]{%
xing2021invertible}
\APACinsertmetastar {%
xing2021invertible}%
\begin{APACrefauthors}%
Xing, Y.%
, Qian, Z.%
\BCBL {} Chen, Q.%
\end{APACrefauthors}%
\unskip\
\newblock
\APACrefYearMonthDay{2021}{}{}.
\newblock
{\BBOQ}\APACrefatitle {Invertible image signal processing} {Invertible image
  signal processing}.{\BBCQ}
\newblock
 \APACrefbtitle {{Proceedings of the IEEE Conference on Computer Vision and
  Pattern Recognition}.} {{Proceedings of the IEEE Conference on Computer
  Vision and Pattern Recognition}.}
\PrintBackRefs{\CurrentBib}

\bibitem [\protect \citeauthoryear {%
Yang%
, Wright%
, Huang%
\BCBL {}\ \BBA {} Ma%
}{%
Yang%
\ \protect \BOthers {.}}{%
{\protect \APACyear {2010}}%
}]{%
yang2010image}
\APACinsertmetastar {%
yang2010image}%
\begin{APACrefauthors}%
Yang, J.%
, Wright, J.%
, Huang, T.S.%
\BCBL {} Ma, Y.%
\end{APACrefauthors}%
\unskip\
\newblock
\APACrefYearMonthDay{2010}{}{}.
\newblock
{\BBOQ}\APACrefatitle {Image super-resolution via sparse representation} {Image
  super-resolution via sparse representation}.{\BBCQ}
\newblock
\APACjournalVolNumPages{IEEE Transactions on Image
  Processing}{19}{11}{2861--2873}.
\newblock

\newblock

\PrintBackRefs{\CurrentBib}

\bibitem [\protect \citeauthoryear {%
Ye%
, Du%
, Deng%
\BCBL {}\ \BBA {} He%
}{%
Ye%
\ \protect \BOthers {.}}{%
{\protect \APACyear {2020}}%
}]{%
ye2020invertible}
\APACinsertmetastar {%
ye2020invertible}%
\begin{APACrefauthors}%
Ye, T.%
, Du, Y.%
, Deng, J.%
\BCBL {} He, S.%
\end{APACrefauthors}%
\unskip\
\newblock
\APACrefYearMonthDay{2020}{}{}.
\newblock
{\BBOQ}\APACrefatitle {Invertible Grayscale via Dual Features Ensemble}
  {Invertible grayscale via dual features ensemble}.{\BBCQ}
\newblock
\APACjournalVolNumPages{IEEE Access}{8}{}{89670--89679}.
\newblock

\newblock

\PrintBackRefs{\CurrentBib}

\bibitem [\protect \citeauthoryear {%
Yeo%
, Do%
\BCBL {}\ \BBA {} Han%
}{%
Yeo%
\ \protect \BOthers {.}}{%
{\protect \APACyear {2017}}%
}]{%
yeo2017will}
\APACinsertmetastar {%
yeo2017will}%
\begin{APACrefauthors}%
Yeo, H.%
, Do, S.%
\BCBL {} Han, D.%
\end{APACrefauthors}%
\unskip\
\newblock
\APACrefYearMonthDay{2017}{}{}.
\newblock
{\BBOQ}\APACrefatitle {How will Deep Learning Change Internet Video Delivery?}
  {How will deep learning change internet video delivery?}{\BBCQ}
\newblock
 \APACrefbtitle {{Proceedings of the 16th ACM Workshop on Hot Topics in
  Networks}.} {{Proceedings of the 16th ACM Workshop on Hot Topics in
  Networks}.}
\PrintBackRefs{\CurrentBib}

\bibitem [\protect \citeauthoryear {%
Yeo%
, Jung%
, Kim%
, Shin%
\BCBL {}\ \BBA {} Han%
}{%
Yeo%
\ \protect \BOthers {.}}{%
{\protect \APACyear {2018}}%
}]{%
yeo2018neural}
\APACinsertmetastar {%
yeo2018neural}%
\begin{APACrefauthors}%
Yeo, H.%
, Jung, Y.%
, Kim, J.%
, Shin, J.%
\BCBL {} Han, D.%
\end{APACrefauthors}%
\unskip\
\newblock
\APACrefYearMonthDay{2018}{}{}.
\newblock
{\BBOQ}\APACrefatitle {Neural adaptive content-aware internet video delivery}
  {Neural adaptive content-aware internet video delivery}.{\BBCQ}
\newblock
 \APACrefbtitle {{13th $\{$USENIX$\}$ Symposium on Operating Systems Design and
  Implementation ($\{$OSDI$\}$ 18)}.} {{13th $\{$USENIX$\}$ Symposium on
  Operating Systems Design and Implementation ($\{$OSDI$\}$ 18)}.}
\PrintBackRefs{\CurrentBib}

\bibitem [\protect \citeauthoryear {%
Zeyde%
, Elad%
\BCBL {}\ \BBA {} Protter%
}{%
Zeyde%
\ \protect \BOthers {.}}{%
{\protect \APACyear {2010}}%
}]{%
zeyde2010single}
\APACinsertmetastar {%
zeyde2010single}%
\begin{APACrefauthors}%
Zeyde, R.%
, Elad, M.%
\BCBL {} Protter, M.%
\end{APACrefauthors}%
\unskip\
\newblock
\APACrefYearMonthDay{2010}{}{}.
\newblock
{\BBOQ}\APACrefatitle {On single image scale-up using sparse-representations}
  {On single image scale-up using sparse-representations}.{\BBCQ}
\newblock
 \APACrefbtitle {{International Conference on Curves and Surfaces}.}
  {{International Conference on Curves and Surfaces}.}
\PrintBackRefs{\CurrentBib}

\bibitem [\protect \citeauthoryear {%
R.~Zhang%
, Isola%
\BCBL {}\ \BBA {} Efros%
}{%
R.~Zhang%
\ \protect \BOthers {.}}{%
{\protect \APACyear {2016}}%
}]{%
zhang2016colorful}
\APACinsertmetastar {%
zhang2016colorful}%
\begin{APACrefauthors}%
Zhang, R.%
, Isola, P.%
\BCBL {} Efros, A.A.%
\end{APACrefauthors}%
\unskip\
\newblock
\APACrefYearMonthDay{2016}{}{}.
\newblock
{\BBOQ}\APACrefatitle {Colorful image colorization} {Colorful image
  colorization}.{\BBCQ}
\newblock
 \APACrefbtitle {{Proceedings of the European Conference on Computer Vision
  (ECCV)}.} {{Proceedings of the European Conference on Computer Vision
  (ECCV)}.}
\PrintBackRefs{\CurrentBib}

\bibitem [\protect \citeauthoryear {%
R.~Zhang%
, Isola%
, Efros%
, Shechtman%
\BCBL {}\ \BBA {} Wang%
}{%
R.~Zhang%
\ \protect \BOthers {.}}{%
{\protect \APACyear {2018}}%
}]{%
zhang2018perceptual}
\APACinsertmetastar {%
zhang2018perceptual}%
\begin{APACrefauthors}%
Zhang, R.%
, Isola, P.%
, Efros, A.A.%
, Shechtman, E.%
\BCBL {} Wang, O.%
\end{APACrefauthors}%
\unskip\
\newblock
\APACrefYearMonthDay{2018}{}{}.
\newblock
{\BBOQ}\APACrefatitle {The Unreasonable Effectiveness of Deep Features as a
  Perceptual Metric} {The unreasonable effectiveness of deep features as a
  perceptual metric}.{\BBCQ}
\newblock
 \APACrefbtitle {{Proceedings of the IEEE Conference on Computer Vision and
  Pattern Recognition}.} {{Proceedings of the IEEE Conference on Computer
  Vision and Pattern Recognition}.}
\PrintBackRefs{\CurrentBib}

\bibitem [\protect \citeauthoryear {%
R.~Zhang%
\ \protect \BOthers {.}}{%
R.~Zhang%
\ \protect \BOthers {.}}{%
{\protect \APACyear {2017}}%
}]{%
zhang2017real}
\APACinsertmetastar {%
zhang2017real}%
\begin{APACrefauthors}%
Zhang, R.%
, Zhu, J\BHBI Y.%
, Isola, P.%
, Geng, X.%
, Lin, A.S.%
, Yu, T.%
\BCBL {} Efros, A.A.%
\end{APACrefauthors}%
\unskip\
\newblock
\APACrefYearMonthDay{2017}{}{}.
\newblock
{\BBOQ}\APACrefatitle {Real-time user-guided image colorization with learned
  deep priors} {Real-time user-guided image colorization with learned deep
  priors}.{\BBCQ}
\newblock
\APACjournalVolNumPages{ACM Transactions on Graphics (TOG)}{36}{4}{1--11}.
\newblock

\newblock

\PrintBackRefs{\CurrentBib}

\bibitem [\protect \citeauthoryear {%
Y.~Zhang%
, Li%
\BCBL {}\ \protect \BOthers {.}}{%
Y.~Zhang%
, Li%
\BCBL {}\ \protect \BOthers {.}}{%
{\protect \APACyear {2018}}%
}]{%
zhang2018image}
\APACinsertmetastar {%
zhang2018image}%
\begin{APACrefauthors}%
Zhang, Y.%
, Li, K.%
, Li, K.%
, Wang, L.%
, Zhong, B.%
\BCBL {} Fu, Y.%
\end{APACrefauthors}%
\unskip\
\newblock
\APACrefYearMonthDay{2018}{}{}.
\newblock
{\BBOQ}\APACrefatitle {Image super-resolution using very deep residual channel
  attention networks} {Image super-resolution using very deep residual channel
  attention networks}.{\BBCQ}
\newblock
 \APACrefbtitle {{Proceedings of the European Conference on Computer Vision
  (ECCV)}.} {{Proceedings of the European Conference on Computer Vision
  (ECCV)}.}
\PrintBackRefs{\CurrentBib}

\bibitem [\protect \citeauthoryear {%
Y.~Zhang%
, Tian%
, Kong%
, Zhong%
\BCBL {}\ \BBA {} Fu%
}{%
Y.~Zhang%
, Tian%
\BCBL {}\ \protect \BOthers {.}}{%
{\protect \APACyear {2018}}%
}]{%
zhang2018residual}
\APACinsertmetastar {%
zhang2018residual}%
\begin{APACrefauthors}%
Zhang, Y.%
, Tian, Y.%
, Kong, Y.%
, Zhong, B.%
\BCBL {} Fu, Y.%
\end{APACrefauthors}%
\unskip\
\newblock
\APACrefYearMonthDay{2018}{}{}.
\newblock
{\BBOQ}\APACrefatitle {Residual dense network for image super-resolution}
  {Residual dense network for image super-resolution}.{\BBCQ}
\newblock
 \APACrefbtitle {{Proceedings of the IEEE Conference on Computer Vision and
  Pattern Recognition}.} {{Proceedings of the IEEE Conference on Computer
  Vision and Pattern Recognition}.}
\PrintBackRefs{\CurrentBib}

\bibitem [\protect \citeauthoryear {%
Zhao%
, Liu%
, Xiao%
, Lun%
\BCBL {}\ \BBA {} Lam%
}{%
Zhao%
\ \protect \BOthers {.}}{%
{\protect \APACyear {2021}}%
}]{%
zhao2021invertible}
\APACinsertmetastar {%
zhao2021invertible}%
\begin{APACrefauthors}%
Zhao, R.%
, Liu, T.%
, Xiao, J.%
, Lun, D.P.%
\BCBL {} Lam, K\BHBI M.%
\end{APACrefauthors}%
\unskip\
\newblock
\APACrefYearMonthDay{2021}{}{}.
\newblock
{\BBOQ}\APACrefatitle {Invertible image decolorization} {Invertible image
  decolorization}.{\BBCQ}
\newblock
\APACjournalVolNumPages{IEEE Transactions on Image
  Processing}{30}{}{6081--6095}.
\newblock

\newblock

\PrintBackRefs{\CurrentBib}

\bibitem [\protect \citeauthoryear {%
Zhong%
, Shen%
, Yang%
, Lin%
\BCBL {}\ \BBA {} Zhang%
}{%
Zhong%
\ \protect \BOthers {.}}{%
{\protect \APACyear {2018}}%
}]{%
zhong2018joint}
\APACinsertmetastar {%
zhong2018joint}%
\begin{APACrefauthors}%
Zhong, Z.%
, Shen, T.%
, Yang, Y.%
, Lin, Z.%
\BCBL {} Zhang, C.%
\end{APACrefauthors}%
\unskip\
\newblock
\APACrefYearMonthDay{2018}{}{}.
\newblock
{\BBOQ}\APACrefatitle {Joint sub-bands learning with clique structures for
  wavelet domain super-resolution} {Joint sub-bands learning with clique
  structures for wavelet domain super-resolution}.{\BBCQ}
\newblock
 \APACrefbtitle {{Advances in Neural Information Processing Systems}.}
  {{Advances in Neural Information Processing Systems}.}
\PrintBackRefs{\CurrentBib}

\bibitem [\protect \citeauthoryear {%
Zhu%
\ \protect \BOthers {.}}{%
Zhu%
\ \protect \BOthers {.}}{%
{\protect \APACyear {2019}}%
}]{%
zhu2019residual}
\APACinsertmetastar {%
zhu2019residual}%
\begin{APACrefauthors}%
Zhu, X.%
, Li, Z.%
, Zhang, X\BHBI Y.%
, Li, C.%
, Liu, Y.%
\BCBL {} Xue, Z.%
\end{APACrefauthors}%
\unskip\
\newblock
\APACrefYearMonthDay{2019}{}{}.
\newblock
{\BBOQ}\APACrefatitle {Residual invertible spatio-temporal network for video
  super-resolution} {Residual invertible spatio-temporal network for video
  super-resolution}.{\BBCQ}
\newblock
 \APACrefbtitle {{Proceedings of the AAAI Conference on Artificial
  Intelligence}.} {{Proceedings of the AAAI Conference on Artificial
  Intelligence}.}
\PrintBackRefs{\CurrentBib}

\end{thebibliography}


\clearpage

\begin{appendices}

\begin{table*} [htbp]
	\centering
	\footnotesize
	\caption{Quantitative evaluation results (PSNR / SSIM) of different 4$\times$ image downscaling and upscaling methods on benchmark datasets: Set5, Set14, BSD100, Urban100, and DIV2K validation set. For our model, differences on average PSNR / SSIM of different samples for z are less than 0.02. We report the mean result. The best result is in red, while the second is in blue.}
	\begin{tabular}{c|c|c|c|c|c|c|c}
		
		\hline
		Downscaling \& Upscaling  &  Scale & Param  &  Set5  &  Set14  &  BSD100  &  Urban100  &  DIV2K \\
	
		\hline  
		\hline
	
		Bicubic \& Bicubic & 4$\times$ & / & 28.42 / 0.8104 & 26.00 / 0.7027 & 25.96 / 0.6675 & 23.14 / 0.6577 & 26.66 / 0.8521 \\

		\hline
		Bicubic \& SRCNN & 4$\times$ & 57.3K &  30.48 / 0.8628  &  27.50 / 0.7513  &  26.90 / 0.7101  &  24.52 / 0.7221 &   --  \\
		
		\hline
		Bicubic \& EDSR & 4$\times$ & 43.1M & 32.62 / 0.8984 & 28.94 / 0.7901 & 27.79 / 0.7437 & 26.86 / 0.8080 & 29.38 / 0.9032 \\

		\hline
		Bicubic \& RDN & 4$\times$ & 22.3M &  32.47 / 0.8990  &  28.81 / 0.7871  &  27.72 / 0.7419  &  26.61 / 0.8028 &   --  \\
		
		\hline
		Bicubic \& RCAN & 4$\times$ & 15.6M & 32.63 / 0.9002 & 28.87 / 0.7889 & 27.77 / 0.7436 & 26.82 / 0.8087 & 30.77 / 0.8460 \\
		
		\hline
		Bicubic \& ESRGAN & 4$\times$ & 16.3M & 32.74 / 0.9012 & 29.00 / 0.7915 & 27.84 / 0.7455 & 27.03 / 0.8152 & 30.92 / 0.8486 \\
		
		\hline
		Bicubic \& SAN & 4$\times$ & 15.7M &  32.64 / 0.9003  &  28.92 / 0.7888  &  27.78 / 0.7436  &  26.79 / 0.8068  &  --  \\
		
		\hline
		TAD \& TAU & 4$\times$ & -- & 31.81 /  --  & 28.63 /  --   & 28.51 /  --   & 26.63 /  --   & 31.16 /  --   \\
		
		\hline
		CAR \& EDSR & 4$\times$ & 52.8M & \textcolor{blue}{33.88} / \textcolor{blue}{0.9174} & \textcolor{blue}{30.31} / 0.8382 & \textcolor{blue}{29.15} / 0.8001 & \textcolor{blue}{29.28} / \textcolor{blue}{0.8711} & \textcolor{blue}{32.82} / 0.8837 \\
		
		\hline
		IRN (ours) & 4$\times$ & 4.35M & \textcolor{red}{36.19} / \textcolor{red}{0.9451} & \textcolor{red}{32.67} / \textcolor{red}{0.9015} & \textcolor{red}{31.64} / \textcolor{red}{0.8826} & \textcolor{red}{31.41} / \textcolor{red}{0.9157} & \textcolor{red}{35.07} / \textcolor{red}{0.9318} \\
        
		\hline
		IRN+ (ours) & 4$\times$ & 4.35M & 33.59 / 0.9147 & 29.97 / \textcolor{blue}{0.8444} & 28.94 / \textcolor{blue}{0.8189} & 28.24 / 0.8684 & 32.24 / \textcolor{blue}{0.8921} \\
		
        \hline
	    
	\end{tabular}
	
    \label{quantitative results of IRN+}
    
\end{table*}

\section{Quantitive results of IRN+}

IRN+ aims at producing more realistic images by minimizing the distribution difference, not exactly matching details of original images as IRN does. The difference will lead to lower PSNR and SSIM, which is the same as GAN-based super-resolution methods. Despite the difference, IRN+ still outperforms most methods in PSNR and SSIM as shown in Table.~\ref{quantitative results of IRN+}, demonstrating the good similarity between the reconstructed images and original HR images.

\section{Different samples of $z$}
\begin{figure*} [htbp]
    \centering
    \subfigure[]{
    \includegraphics[scale=0.1]{attachment/0810_GT_frame.png}
    \label{ground truth}
    }
    \subfigure[]{
    \includegraphics[scale=0.1]{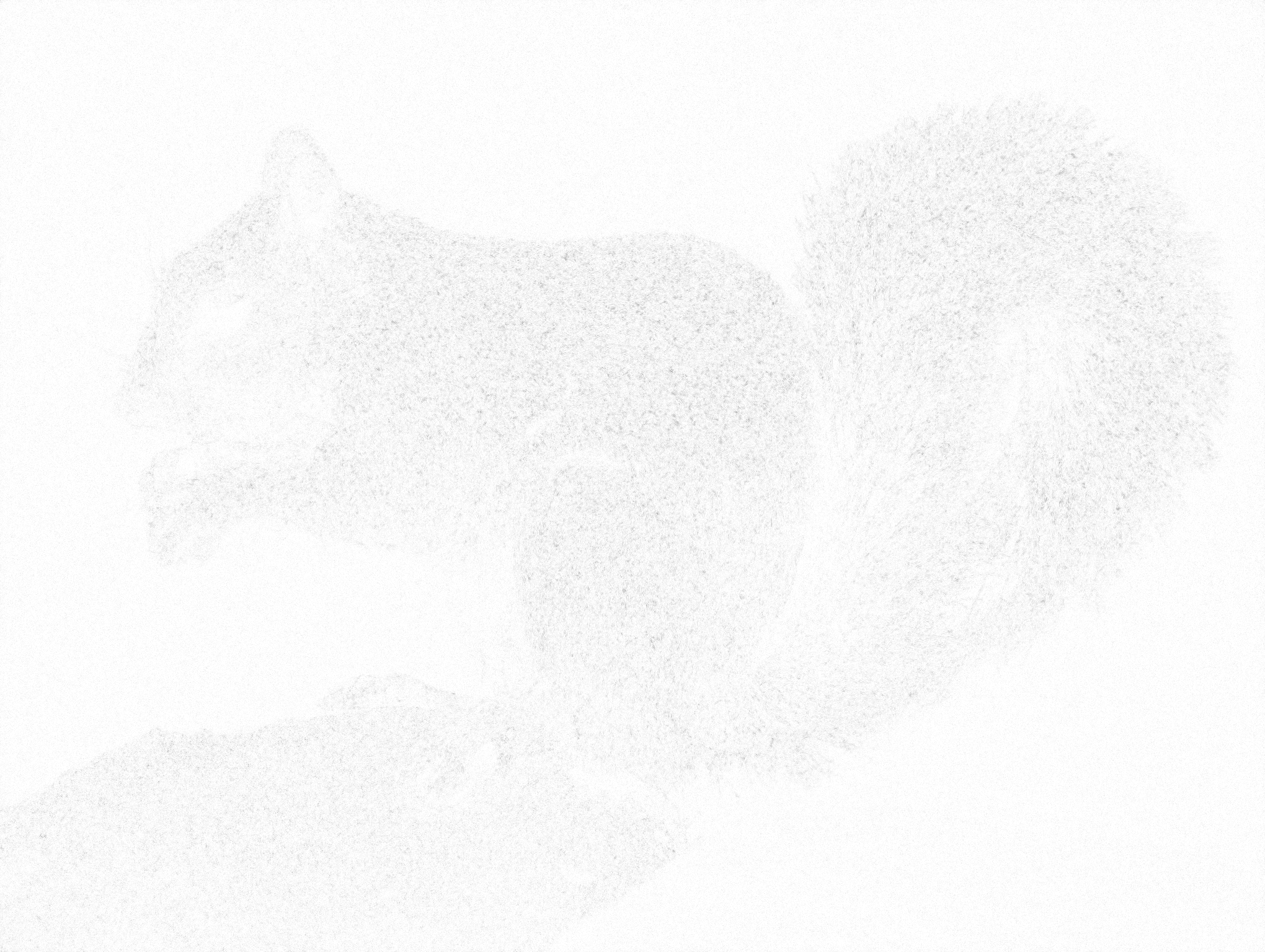}
    \label{sample0&sample1}
    }
    \subfigure[]{
    \includegraphics[scale=0.1]{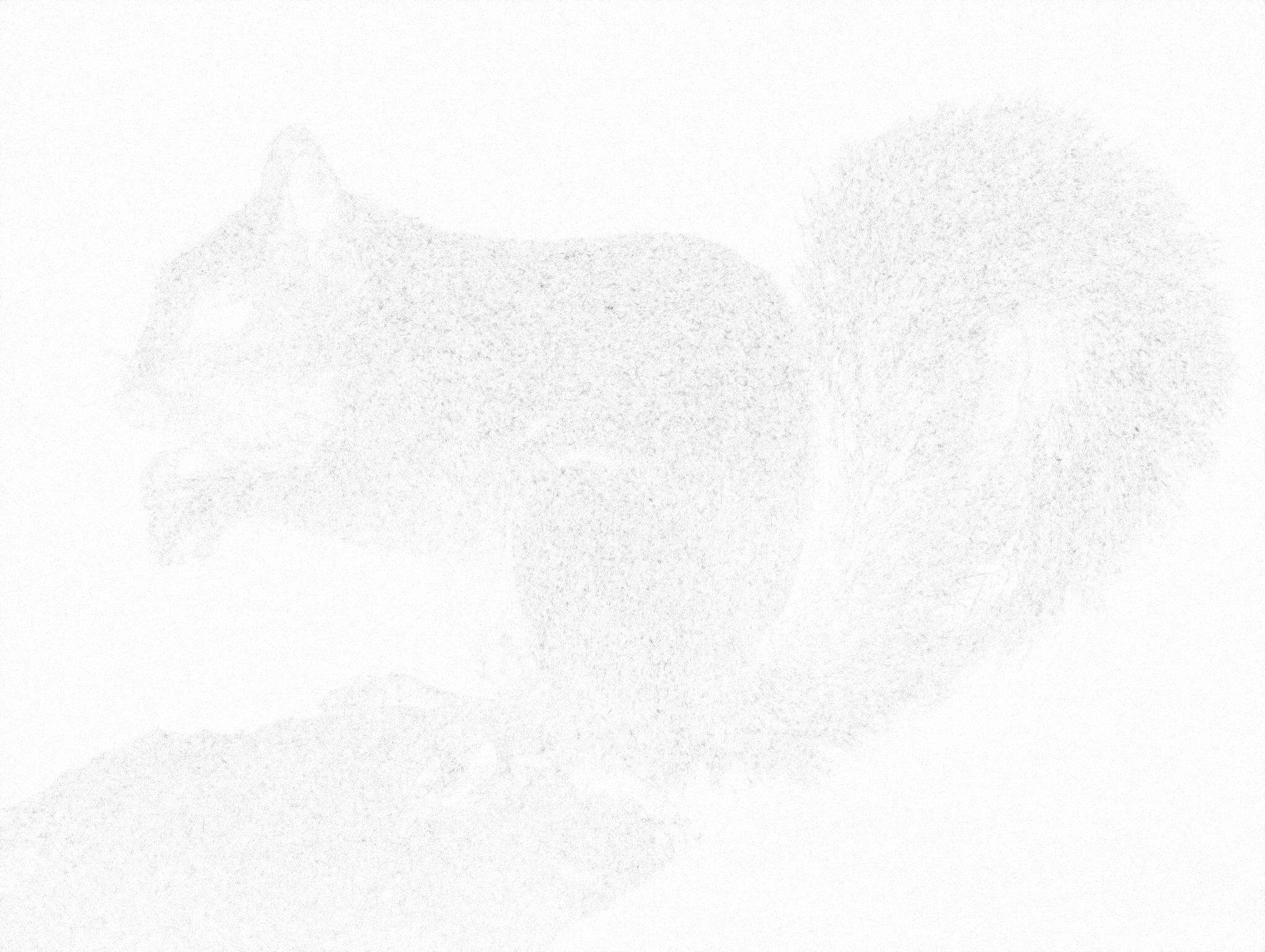}
    \label{sample0&sample2}
    }
    \subfigure[]{
    \includegraphics[scale=0.1]{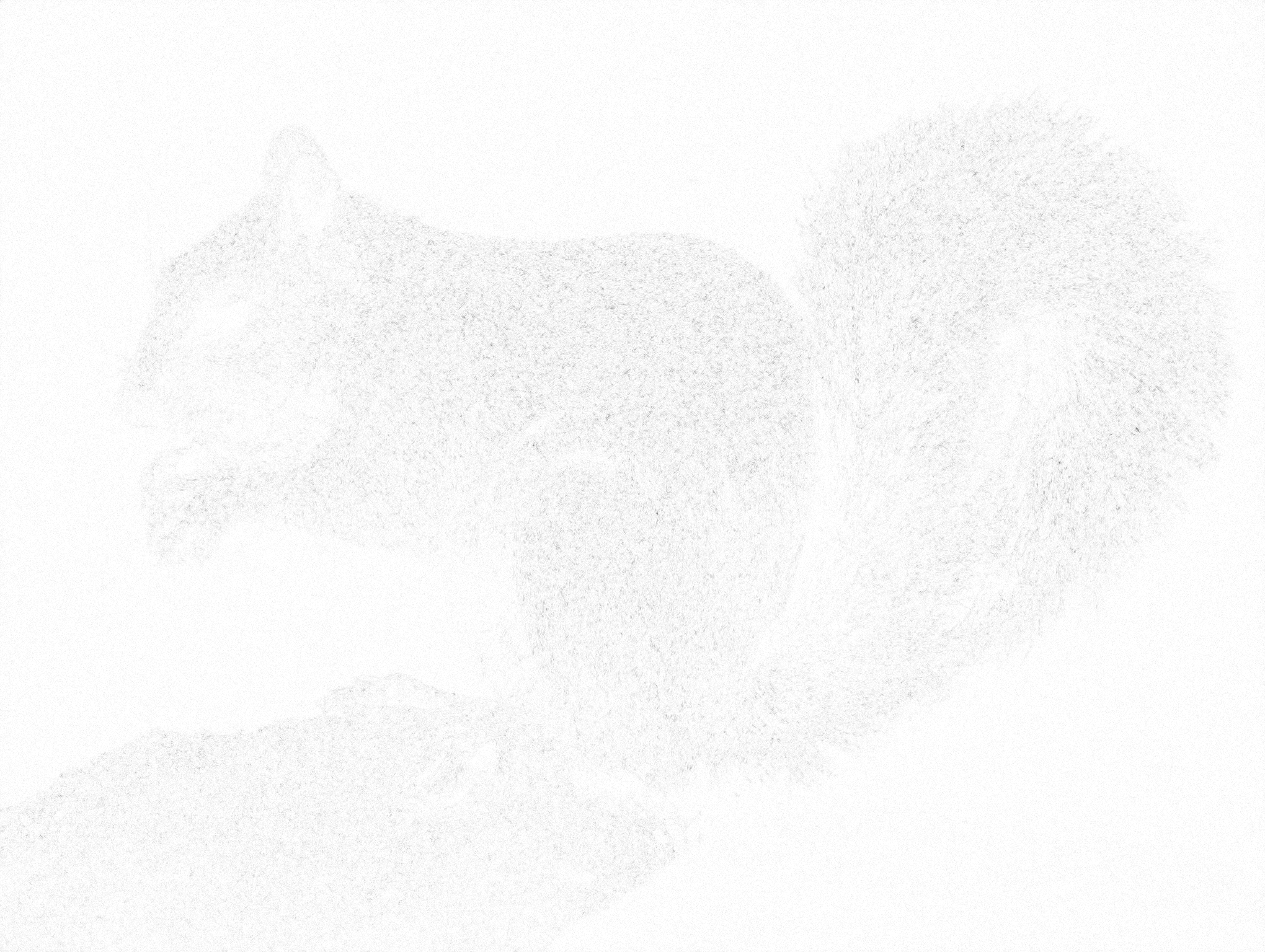}
    \label{sample0&sample3}
    }
    \subfigure[]{
    \includegraphics[scale=1]{attachment/grey_0810_diff01_patch_1.png}
    \label{detail1}
    }
    \hspace{2mm}
    \subfigure[]{
    \includegraphics[scale=1]{attachment/grey_0810_diff02_patch_1.png}
    \label{detail2}
    }
    \hspace{2mm}
    \subfigure[]{
    \includegraphics[scale=1]{attachment/grey_0810_diff03_patch_1.png}
    \label{detail3}
    }
    \caption{Difference between upscaled images by different samples of $z$. (a): Original image. (b-d): Residual of three randomly upscaled images with another sample (averaged over the three channels). (e-g): Detailed difference of (b-d). The darker the larger difference. To ensure the visual perception, we set rebalance factor by 20.
    }
    \label{fig:different samples}
\end{figure*}

As shown in Fig.~\ref{fig:different samples}, there is only a tiny noisy distinction in high-frequency areas without typical textures, which can hardly be perceived when combined with low-frequency contents. Different samples lead to different but perceptually meaningless noisy distinctions.

\section{More qualitative results}

\begin{figure*} [htbp]
    \centering
    \includegraphics[scale=0.135]{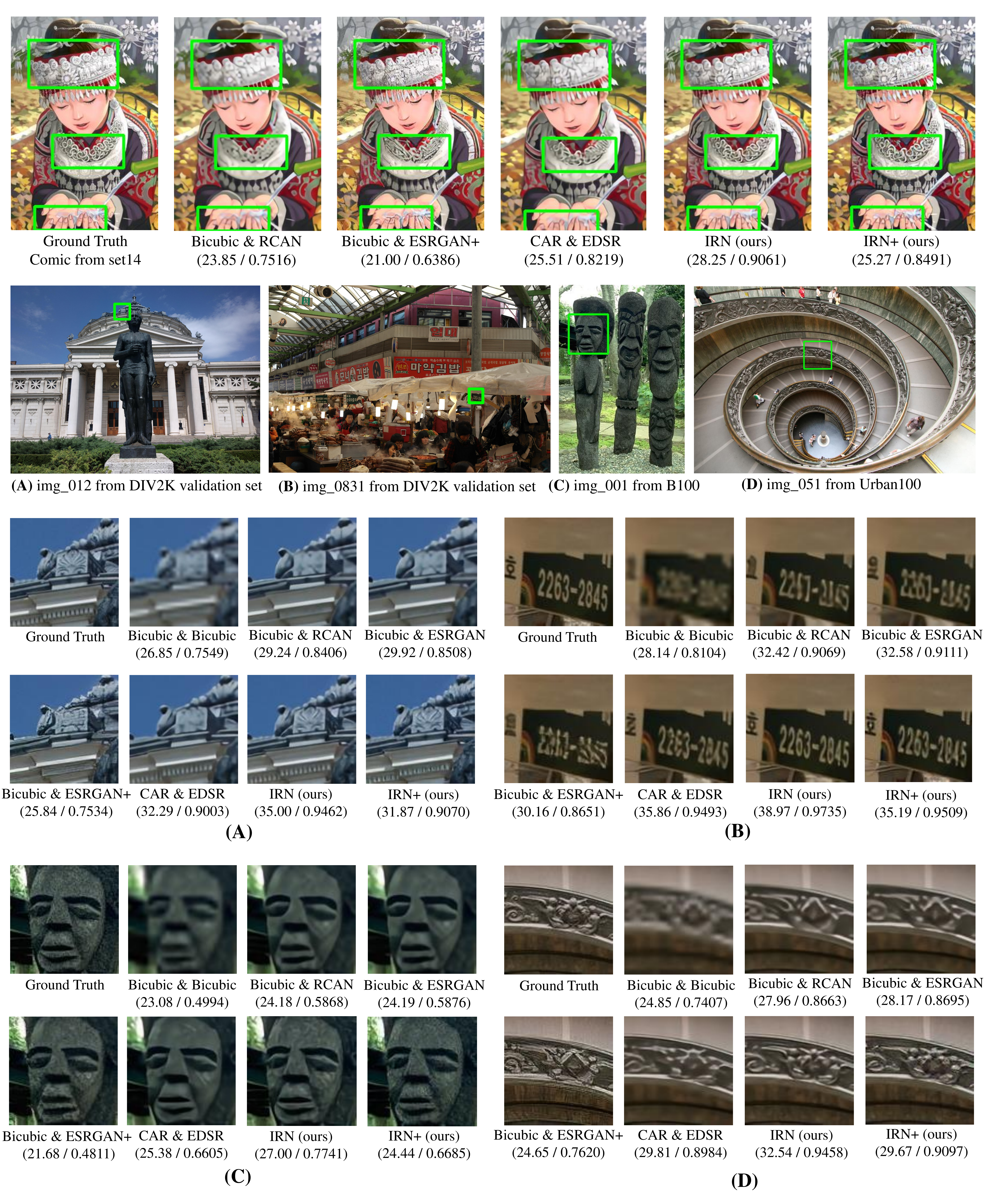}
    \caption{More qualitative results of upscaling the $4\times$ downscaled images on Set14, BSD100, Urban100 and DIV2K validation datasets.}
    \label{fig:qualitative results}
\end{figure*}

\begin{figure*} [htbp]
    \centering
    \includegraphics[scale=0.135]{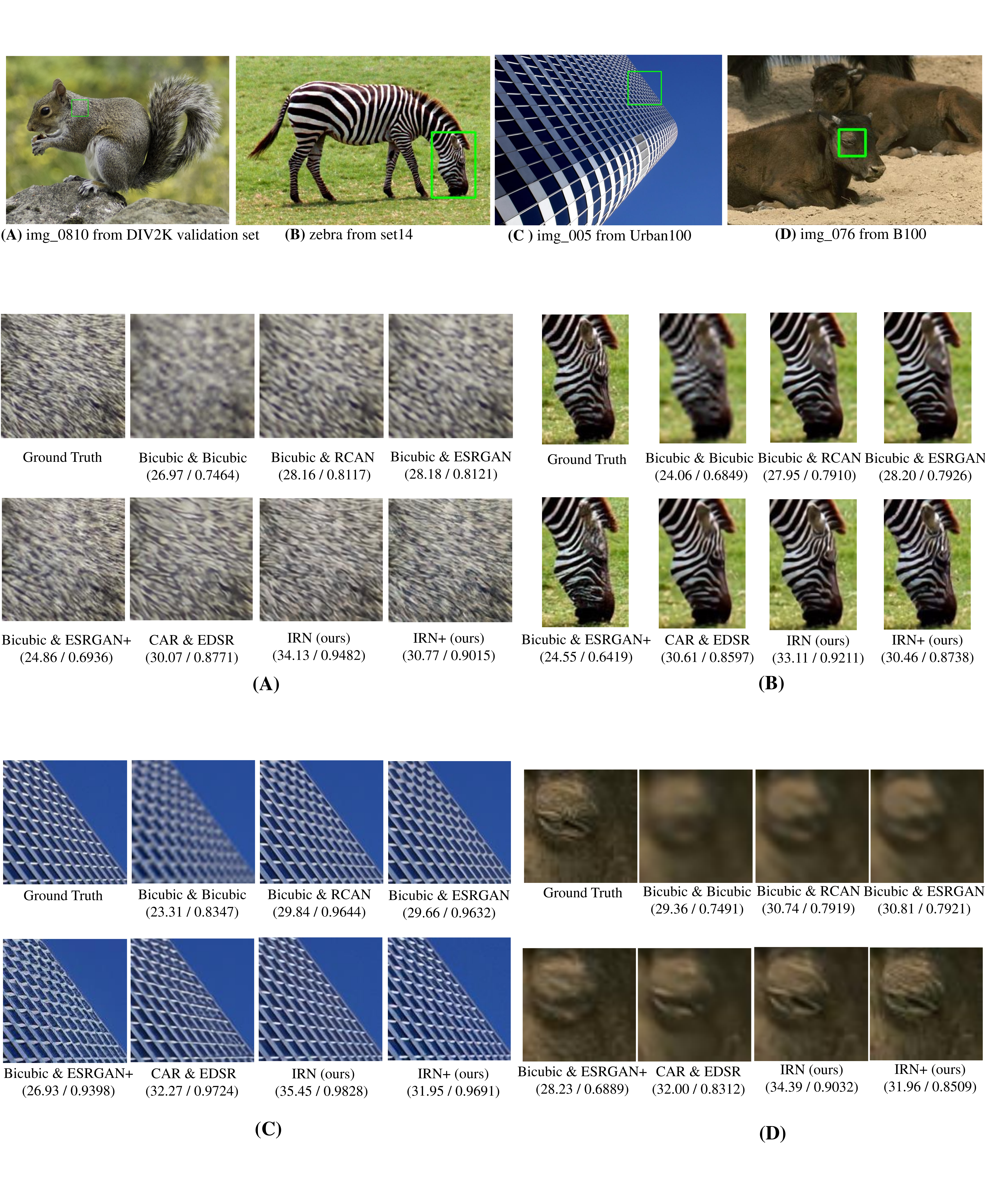}
    \caption{More qualitative results of upscaling the $4\times$ downscaled images on Set14, BSD100, Urban100 and DIV2K validation datasets.}
    \label{fig:qualitative results1}
\end{figure*}

\begin{figure*} [htbp]
    \centering
    \includegraphics[scale=0.135]{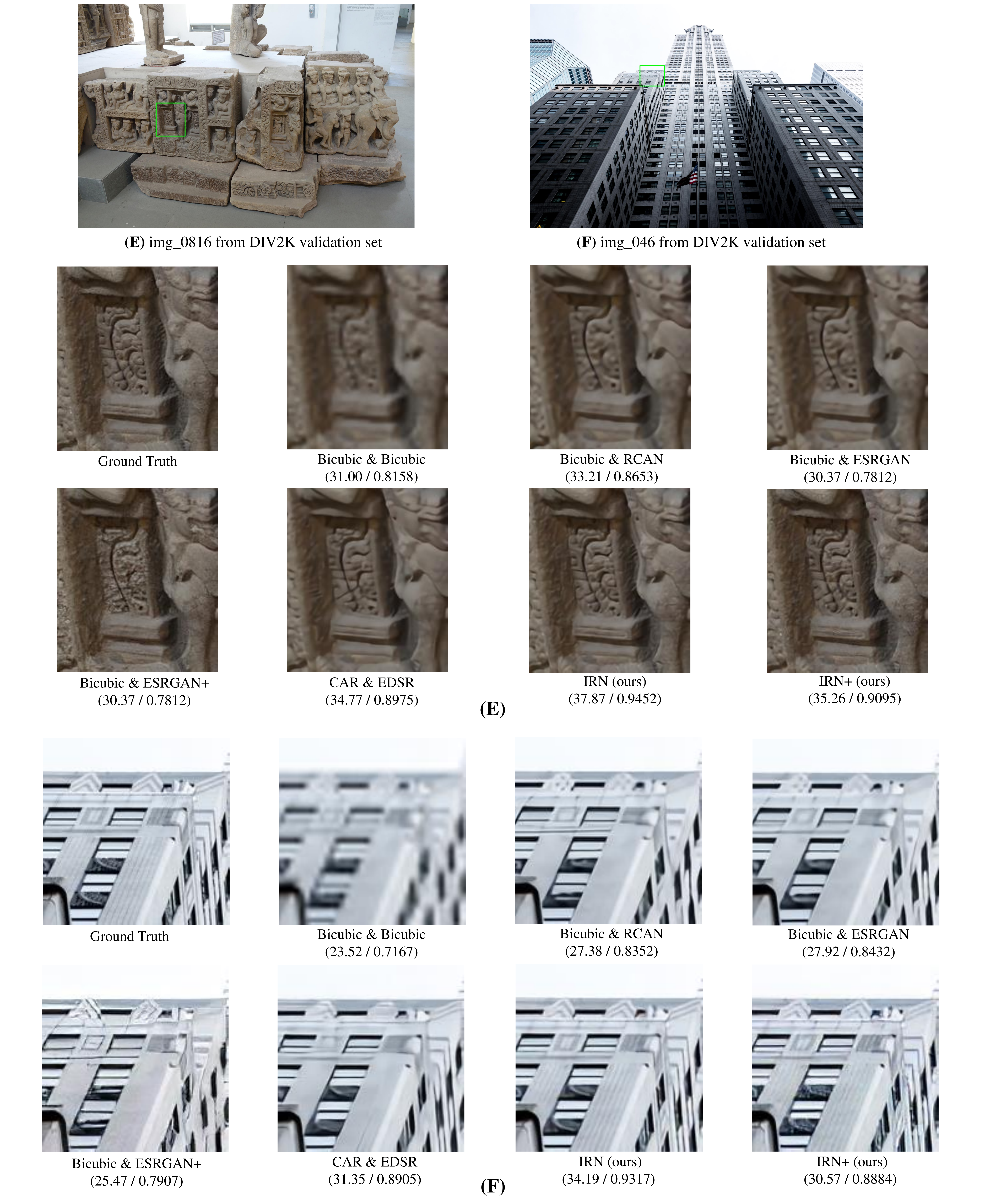}
    \caption{More qualitative results of upscaling the $4\times$ downscaled images on DIV2K validation dataset.}
    \label{fig:qualitative results2}
\end{figure*}

\begin{figure*} [htbp]
    \centering
    \includegraphics[scale=0.135]{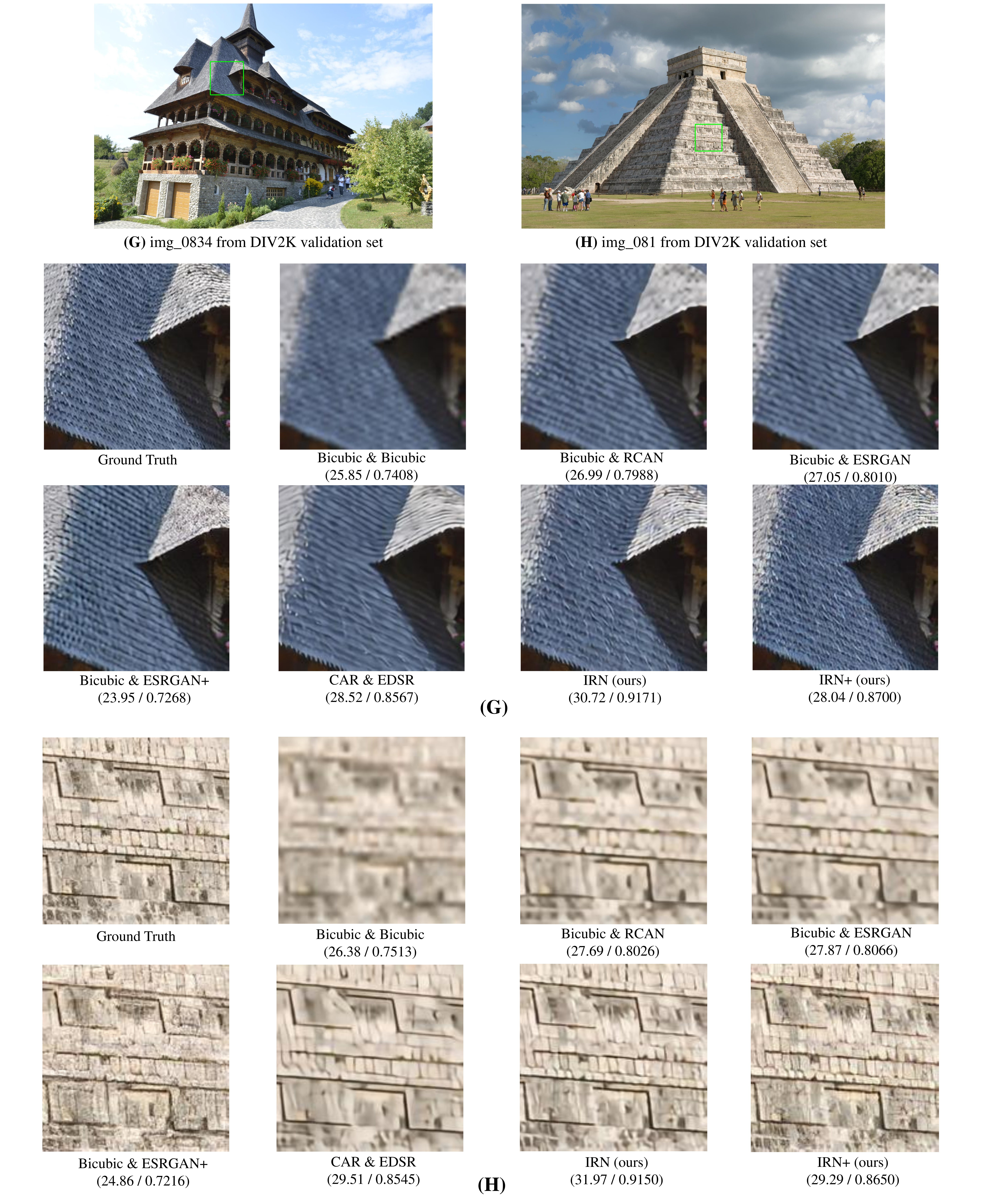}
    \caption{More qualitative results of upscaling the $4\times$ downscaled images on DIV2K validation dataset.}
    \label{fig:qualitative results3}
\end{figure*}

As shown in Fig.\ref{fig:qualitative results},\ref{fig:qualitative results1},\ref{fig:qualitative results2},\ref{fig:qualitative results3}, images reconstructed by IRN and IRN+ significantly outperforms previous both PSNR-oriented and perceptual-driven methods in visual quality and similarity to original images. IRN can reconstruct rich details including detailed lines and textures, which contributes to the pleasing perception. IRN+ further produces sharper and more realistic images as a result of the distribution matching objective.

\section{Evaluation on downscaled images}

\begin{figure*} [htbp]
    \centering
    \subfigure[]{
    \includegraphics[scale=0.5]{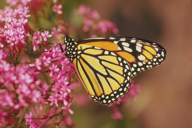}
    \label{img1_lrgt}
    }
    \subfigure[]{
    \includegraphics[scale=0.5]{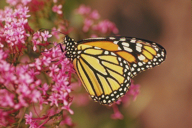}
    \label{img1_lr}
    }
    \subfigure[]{
    \includegraphics[scale=0.2]{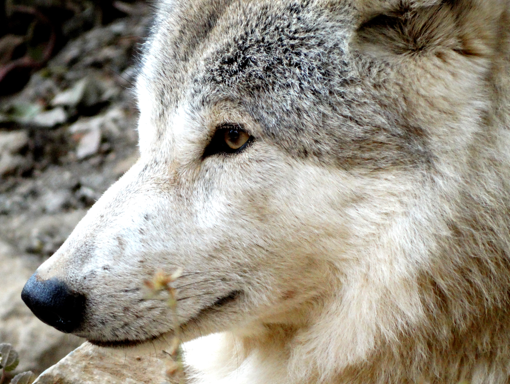}
    \label{img2_lrgt}
    }
    \subfigure[]{
    \includegraphics[scale=0.2]{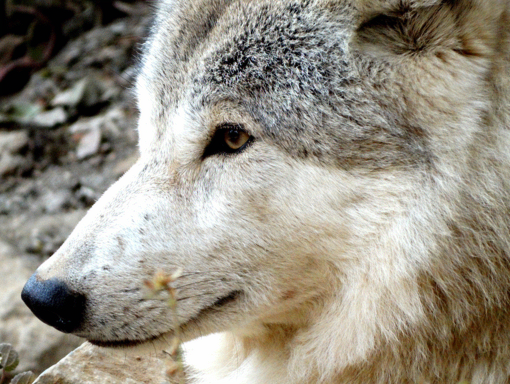}
    \label{img2_lr}
    }
    \\
    \subfigure[]{
    \includegraphics[scale=0.375]{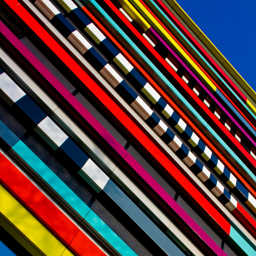}
    \label{img3_lrgt}
    }
    \subfigure[]{
    \includegraphics[scale=0.375]{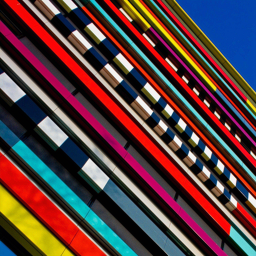}
    \label{img3_lr}
    }
    \subfigure[]{
    \includegraphics[scale=0.2]{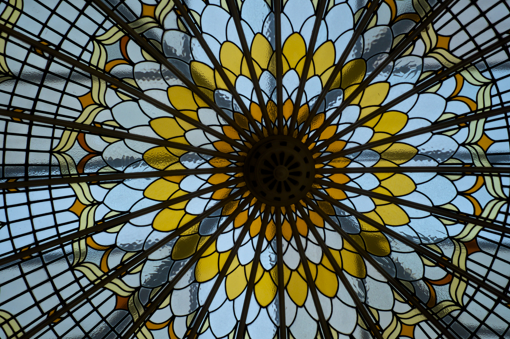}
    \label{img4_lrgt}
    }
    \subfigure[]{
    \includegraphics[scale=0.2]{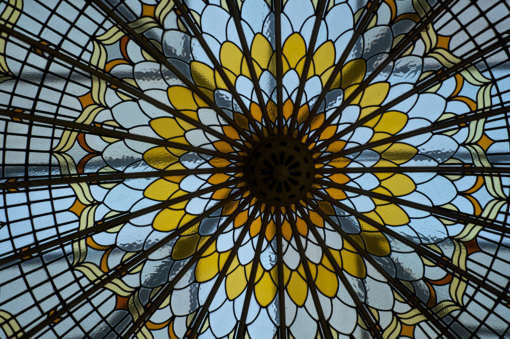}
    \label{img4_lr}
    }
    \\
    \subfigure[]{
    \includegraphics[scale=1.2]{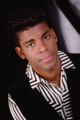}
    \label{img5_lrgt}
    }
    \subfigure[]{
    \includegraphics[scale=1.2]{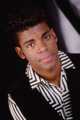}
    \label{img5_lr}
    }
    \subfigure[]{
    \includegraphics[scale=0.825]{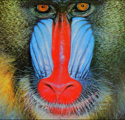}
    \label{img6_lrgt}
    }
    \subfigure[]{
    \includegraphics[scale=0.825]{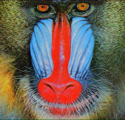}
    \label{img6_lr}
    }
    \\
    \subfigure[]{
    \includegraphics[scale=0.41]{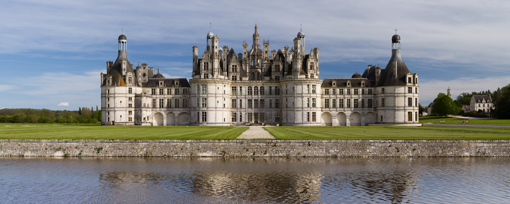}
    \label{img7_lrgt}
    }
    \subfigure[]{
    \includegraphics[scale=0.41]{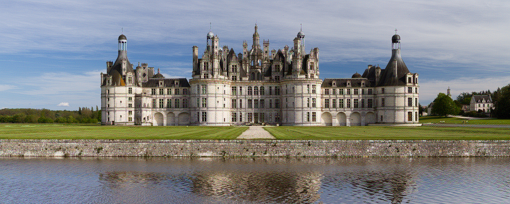}
    \label{img7_lr}
    }
    \caption{Demonstration of the downscaled images from Set14, B100, Urban100, and DIV2K validation set. Left column (a,c,e,g,i,k,m): Image downscaled by Bicubic. Right column (b,d,f,h,j,l,n): Image downscaled by IRN. They share a similar visual perception.}
    \label{fig:downscaled images}
\end{figure*}

As shown in Fig.~\ref{fig:downscaled images}, images downscaled by IRN share a similar visual perception with images downscaled by bicubic.

\end{appendices}

\end{document}